\documentclass{aa}
\usepackage{amsmath}
\usepackage{breqn}
\usepackage{natbib}
\usepackage{hyperref}
\hypersetup{colorlinks=true, citecolor=blue}
\usepackage{graphicx}
\usepackage{txfonts}
\bibpunct{(}{)}{;}{a}{}{,}

\begin{document}
  \title{Towards understanding the relation between the gas and the attenuation in galaxies at kpc scales}
  \author{M. Boquien\inst{1} \and A. Boselli\inst{1} \and V. Buat\inst{1} \and M. Baes\inst{2} \and G. Bendo\inst{3} \and S. Boissier\inst{1} \and L. Ciesla\inst{1} \and A. Cooray\inst{4} \and L. Cortese\inst{5} \and S. Eales\inst{6} \and J. Koda\inst{7} \and V. Lebouteiller\inst{8} \and I. de Looze\inst{2} \and M. W. L. Smith\inst{6} \and L. Spinoglio\inst{9} \and C. D. Wilson\inst{10}}
  \authorrunning{Boquien et al.}
  \institute{
       Aix Marseille Universit\'e, CNRS, LAM (Laboratoire d'Astrophysique de Marseille) UMR 7326, 13388, Marseille, France \email{mederic.boquien@oamp.fr}
  \and Sterrenkundig Observatorium, Universiteit Gent, Krijgslaan 281 S9, B-9000 Gent, Belgium
  \and UK ALMA Regional Centre Node, Jordell Bank Center for Astrophysics, School of Physics and Astronomy, University of Manchester, Oxford Road, Manchester M13 9PL, UK
  \and Department of Physics \& Astronomy, University of California, Irvine, CA 92697, USA
  \and European Southern Observatory, Karl Schwarzschild Str. 2, 85748, Garching bei M\"unchen, Germany
  \and School of Physics \& Astronomy, Cardiff University, The Parade, Cardiff CF24 3AA, UK
  \and Department of Physics and Astronomy, Stony Brook University, Stony Brook, NY 11794--3800, USA
  \and Laboratoire AIM, CEA/DSM-CNRS-Universit\'e Paris Diderot DAPNIA/Service d'Astrophysique, B\^at. 709, CEA-Saclay, 91191, Gif--sur--Yvette Cedex, France
  \and Istituto di Astrofisica e Planetologia Spaziali, INAF-IAPS, Via Fosso del Cavaliere 100, I-00133 Roma, Italy
  \and Department of Physics and Astronomy, McMaster University, Hamilton, Ontario L8S 4M1 Canada
  }
  \date{}
 \abstract
  {Understanding the relation of the attenuation with the gas is fundamental for interpreting the appearance of galaxies. This can now be done down to a local scale in the local Universe, thanks to the high spatial resolution achievable from the FUV to the FIR. More importantly, this relation is also crucial for predicting the emission of galaxies. This is essential in semi--analytic models that link dark matter from large cosmological simulations, to the baryonic matter which we can observe directly.}
  {The aim of the present paper is to provide new and more detailed relations at the kpc scale between the gas surface density and the face--on optical depth directly calibrated on galaxies, in order to compute the attenuation not only for semi--analytic models but also observationally as new and upcoming radio observatories are able to trace gas ever farther in the Universe.}
  {We have selected a sample of 4 nearby resolved galaxies and a sample of 27 unresolved galaxies from the \textit{Herschel Reference Survey} and the \textit{Very Nearby Galaxies Survey}, for which we have a large set of multi--wavelength data from the FUV to the FIR including metallicity gradients for resolved galaxies, along with radio HI and CO observations. For each pixel in resolved galaxies and for each galaxy in the unresolved sample, we compute the face--on optical depth from the attenuation determined with the CIGALE Spectral Energy Distribution fitting code and an assumed geometry. We determine the gas surface density from HI and CO observations with a metallicity--dependent $X_{CO}$ factor.}
  {We provide new, simple to use, relations to determine the face--on optical depth from the gas surface density, taking the metallicity into account, which proves to be crucial for a proper estimate. The method used to determine the gas surface density or the face--on optical depth has little impact on the relations except for galaxies that have an inclination over 50$^\circ$. Finally, we provide detailed instructions on how to compute the attenuation practically from the gas surface density taking into account possible information on the metallicity.}
  {Examination of the influence of these new relations on simulated FUV and IR luminosity functions shows a clear impact compared to older oft--used relations, which in turn could affect the conclusions drawn from studies based on large scale cosmological simulations.}
  \maketitle

\section{Introduction\label{sec:introduction}}

Understanding galaxy formation and evolution is one of the central questions in modern astrophysics. Major efforts in terms of observations and theory have been made over the last few decades in order to gain insight into the physical processes driving the transformation of the protogalaxies seen at high redshift into the diverse range of galaxies we see in the local Universe. Large scale numerical simulations have become increasingly important tools over the last 10 years with remarkable breakthroughs. With recent advances in computing power, cosmological simulations, which were initially limited to massive structures, now have a mass resolution sufficient to resolve objects down to dwarf galaxies in large volumes \citep[e.g.][]{guo2011a} yielding unprecedented insight into the formation and evolution of galaxies across a large dynamical range of types and masses.

What determines the formation and evolution of galaxies is the transformation of baryonic matter. Yet the computational cost of an extensive physical model of the processes affecting baryonic matter (gas heating and cooling, star formation, feedback, active nuclei, metal enrichment, etc.) at the sub--grid level remains particularly high. Even large hydrodynamical simulations, such as the OverWhelmingly Large Simulations project \citep{schaye2010a}, that model in detail the physics of the baryons are either limited to significantly smaller volumes or require a dazzling amount of computing power: $10^5$ cores and $65.5\times10^9$ particles in the case of the {\it MassiveBlack} simulation \citep{dimatteo2012a} for instance. Large scale cosmological simulations such as the Millennium simulation \citep{springel2005a}, which are limited to dark matter, are then coerced to employ coarser, yet physically sound, ``recipes''. Still, the baryonic matter is key to constrain the formation and evolution of galaxies as it is directly traced through high redshift surveys. The connection between the dark matter traced by N--body simulations and the baryonic one traced by observations can be made through SAMs (semi--analytic models) such as the ones developed by \cite{croton2006a} or \cite{delucia2006a} for instance, among many others. Reassuringly, comparisons between different SAMs show that they are globally consistent but some systematic uncertainties can nevertheless be found \citep{fontanot2009b}. To constrain models, simulated observations can be computed from the combination of numerical simulations with SAMs. These artificial catalogues can then be compared with observations \citep[e.g.][]{henriques2012a}. 

The attenuation due to the dust affects dramatically the observed or predicted fluxes, colours, and Spectral Energy Distributions (SEDs) of galaxies which are reddened. Computing the attenuation in galaxies is an important challenge observationally, and one of the most delicate parts in SAMs \citep{fontanot2009a} along with the computation of the amount of cold gas \citep{obreschkow2009b}. As the radiation from different stellar populations propagates through clouds of gas and dust, it finds itself partly absorbed and reprocessed towards longer wavelengths. Because the energetic radiation from the youngest stellar populations is the most affected by the presence of dust, the most widely used star formation tracers are strongly impacted, either positively (the infrared) or negatively (the ultraviolet or H$\alpha$). This heavily affects our ability to trace star formation and by extension our understanding of galaxy formation and evolution across cosmic times.

In galaxies, the observed attenuation is actually the consequence of an intricate interplay between 1) the relative weights of the different stellar populations across the SED, 2) the relative geometry of the dust and the stars, and 3) the line--of--sight of the observation. Radiation transfer codes can be used to model in detail the emerging light of dusty galaxies \citep{baes2001a,tuffs2004a,bianchi2008a,popescu2011a}. Unfortunately their high computational cost makes them unsuitable for cosmological simulations. In SAMs, different methods to compute the attenuation yield different predicted counts in ultraviolet (UV) Luminosity Functions (LFs) by a factor $\sim2$ \citep{finlator2006a,fontanot2009a}. This is worrisome as the UV is one of the most widely used star formation tracers \citep{kennicutt1998a,kennicutt2012a} and is observable by ground--based surveys for galaxies with $z\gtrsim1$. One of the methods followed to compute the attenuation in SAMs revolves around variations of the following: 1) link the face--on optical depth to the gas surface density which is a known quantity, and 2) assuming a given extinction curve, geometry, and line--of--sight, retrieve the V--band attenuation from the face--on optical depth \citep[e.g.][]{guiderdoni1987a,devriendt1999a,delucia2007a}. Even though galaxies have a complex geometry, such methods are generally calibrated on objects in the Milky Way. In our galaxy, the reddening can be linked with the gas column density \citep{burstein1978a,burstein1982a,schlegel1998a}. However, studies carried out by \cite{xu1997a} and \cite{boissier2004a,boissier2007a} on different samples of galaxies actually found only weak relations between the face--on optical depth and the gas surface density. This sheds doubt on the reliability of such methods, and rises the issue of the potential impact on our understanding of galaxy formation and evolution. The main difficulties in performing studies on galaxies essentially revolved until recently around the coarse resolution available to trace both the gas and the attenuation in galaxies. The increasing availability of high resolution maps of the gas through HI and CO observations, and multi--wavelengths observations from the FUV to the FIR now allow us to gain a much needed insight into the relation between the gas and the attenuation in galaxies. Beyond the direct impact of such relations on SAMs, with the inception of ALMA and other millimetre instruments, we can trace the gas content of galaxies across the Universe both indirectly through the emission of the dust and directly through molecular lines such as CO. The gas can then be used to correct the SED for the attenuation. Therefore it is now timely to reexamine at last the relation between the attenuation and the gas surface density in galaxies. 

The overarching aim of this paper is to determine whether there is a physical relation between the face--on optical depth and the gas surface density in galaxies at kpc scales and if so, to provide new, well calibrated, but still simple and easy to use relations to compute the attenuation for observations and for models. To do so we carry out a study on 2 samples made of nearby resolved and unresolved galaxies defined in Sect.~\ref{sec:sample-data}. The sensitive and delicate issue of the determination of the attenuation and of the gas mass surface density is detailed in Sect.~\ref{sec:att-gas-determination}. We provide a first overview of the relation between the attenuation and the difference gas phases in Sect.~\ref{sec:first-overview}. We then derive new relations between the face--on optical depth and the gas mass surface density for resolved galaxies (Sect.~\ref{sec:rel-gas-tau}). In Sect.~\ref{sec:recipe} we describe how to apply these new relations to obtain a reliable estimate of the attenuation. We then examine the case and unresolved galaxies (Sect.~\ref{sec:comp-int}). We compare the new relations we have obtained m to standard relations from the literature in Sect.~\ref{sec:comp-literature}. The impact of the new relations on LFs is discussed in Sect.~\ref{sec:impact}, and finally we conclude in Sect.~\ref{sec:conclusion}.

\section{Sample and data\label{sec:sample-data}}

\subsection{Sample selection\label{ssec:sample}}

\subsubsection{Multi--wavelength requirements\label{sssec:requirements}}

We want to determine the relation between the attenuation and the gas in nearby galaxies. To reach this goal we need to determine these two quantities on a resolved basis in galaxies as well as for unresolved galaxies.

The attenuation which is due to absorption and scattering of the radiation out and into the line--of--sight can be determined in several ways. A widespread method to correct for the attenuation in nearby galaxies is to compute the ionised, gas--phase attenuation from the Balmer decrement between the H$\alpha$ and H$\beta$ lines \citep[e.g.][and many others]{lequeux1979a,kennicutt1992a}. This is best performed using spectra, and would ideally require integral field unit spectroscopy of the disk to map the attenuation reliably when considering resolved galaxies. Narrow--band observations of galaxies can be adversely affected by underlying absorption features, especially for the H$\beta$ line. In any case this method requires stellar populations no older than typically 10~Myr. Another method is to combine extinguished and unextinguished star formation tracers such as UV or H$\alpha$ with 24~$\mu$m for instance. However, such methods often rely on a statistical analysis over a sample of galaxies \citep{kennicutt2009a,calzetti2010a} or a sample of star forming regions \citep{calzetti2005a,calzetti2007a,kennicutt2007a,relano2007a,li2010a} rather than being specifically tailored to each individual region. With the availability of large multi--wavelength datasets from the FUV to the FIR at sufficient resolution to resolve nearby galaxies, it is now possible to determine the attenuation accurately from the continuum emission by modelling the SED not only of entire galaxies but also on a pixel--by--pixel basis, naturally taking into account the local physical parameters \citep[e.g.][]{boquien2012a}. \cite{ly2012a} showed that at least in the case of unresolved galaxies, the use of SED fitting to determine the attenuation is as robust as using the Balmer decrement from spectra. We therefore determine the attenuation from a full SED modelling.

The gas is overwhelmingly made of hydrogen either in atomic or molecular form. The atomic gas is easily detectable through its 21~cm spin--flip transition. The detection of molecular gas however is more difficult due to the intrinsic nature of H$_2$, which accounts for the bulk the molecular gas mass. The most widespread method is to observe a rotational transition of the CO molecule which is a good tracer of molecular gas even though it is dependent on the temperature or the density of the gas \citep{maloney1988a}, and its metallicity \citep{wilson1995a,israel1997a,israel2000a,barone2000a,israel2003a,strong2004a,israel2005a}.

Following the multi--wavelength requirements we have exposed, in order to examine the relation between the gas and the attenuation at local and global scales in galaxies, we need 1) high resolution HI and CO data to trace the atomic and molecular gas, and 2) high resolution multi--wavelength data from the FUV to the FIR to determine the attenuation. For unresolved galaxies the wavelengths requirements are the same.

\subsubsection{Targets and data}

We carry out the current study in the context of the {\it Herschel} SPIRE guaranteed time consortium. We rely on 2 surveys dedicated to the observation of galaxies in the nearby Universe: the {\it Herschel} Reference Survey \citep[HRS,][]{boselli2010a} and the Very Nearby Galaxies Survey\footnote{\url{http://herschel.esac.esa.int/Docs/GTKP/KPGT_accepted.html\#KPGT_cwilso01_1}} (VNGS). We use these surveys as a baseline to guide our selection of a resolved sample (hereafter sample A), and an unresolved sample (hereafter sample B).

For sample A we select all late--type galaxies from these 2 surveys that are not dominated by a strong active nucleus, and that have high resolution data available in all bands at a sufficient depth: UV, optical, NIR, and FIR to model the SED accurately and determine the local physical parameters (including the attenuation), and also HI and CO to trace the atomic and molecular gas. To decide whether a galaxy is sufficiently resolved or not, which depends on its size, distance, and inclination, we rely on visual inspection of the data at 350~$\mu$m. We drop the 500~$\mu$m band because its resolution is coarser and it does not add value to compute the attenuation. If structures are visible, the galaxy is deemed resolved for our purpose. We carry out the analysis at a resolution of 30\arcsec, only slightly coarser than that of the 350~$\mu$m band. The shallow depth of GALEX UV data for some galaxies causes further reduction in the number of elements in the sample. The final set of photometric bands is the following: FUV, NUV (GALEX), $u'$, $g'$, $r'$, $i'$, $z'$ (SDSS), J, H, Ks (2MASS), 70~$\mu$m ({\it Spitzer}/MIPS), 250~$\mu$m, and 350~$\mu$m ({\it Herschel}/SPIRE), in addition to HI and CO maps. Despite the tremendous development of millimetre and radio imaging, high quality data covering the entire body of nearby galaxies at a resolution better than 30\arcsec\ remain rare. This is the most stringent criterion limiting the number of galaxies in the sample. We rely on two large surveys for HI data using the Very Large Array (VLA): VIVA \citep{chung2009a} and THINGS \citep{walter2008a}. For CO, we rely on the HERACLES survey \citep{bigiel2008a} carried out at the Institut de Radioastronomie Millim\'etrique (IRAM), which covers a sample of nearby galaxies in the CO(2--1) transition at 6\arcsec\ to map the molecular gas. In conclusion, sample A is constituted of 4 galaxies: NGC~2403, NGC~4254, NGC~4321, and NGC~5194 (Table~\ref{tab:sampleA}). 
\begin{table*}
\caption{Sample A. The physical parameters of the galaxies are provided by \cite{boselli2010a,ciesla2012a} (see references therein), or obtained from the NASA Extragalactic Database.}
\label{tab:sampleA}
\centering
\begin{tabular}{c c c c c c c c c}
 \hline\hline
NGC \# & RA       & Dec      & Type & K         & $a$          & $b/a$  & Inclination &Distance\\
       &(J2000)   &(J2000)   &      & (Vega mag)&(kpc)&        & (degrees)   &(Mpc)\\
\hline
2403 & 07 36 51.40 & +65 36 09.2 & SAB(s)cd; HII        & 6.19 & 10.7& 0.50 &60&3.37\\
4254 & 12 18 49.63 & +14 24 59.4 & Sa(s)c               & 6.93 & 30.4& 0.91 &24&17.0\\
4321 & 12 22 54.90 & +15 49 20.6 & SAB(s)bc; LINER; HII & 6.59 & 45.1& 0.89 &27&17.0\\
5194 & 13 59 52.71 & +47 11 42.6 & SA(s)bc pec; HIISy2.5& 5.50 & 11.3& 0.83 &34&6.92\\
\hline
\end{tabular}
\end{table*}

For sample B, we have applied the same basic criteria on the original HRS sample, substituting the requirement on the resolution for a similar requirement on the inclination $b/a>0.33$ in order to limit radiation transfer effects. We use the following combination of photometric bands which maximises the size of the sample: FUV, NUV, U, $g'$, V, $r'$, $i'$, J, H, Ks, 60~$\mu$m (IRAS), 250~$\mu$m, and 350~$\mu$m, in addition to HI and CO measurements. The requirement on the availability of both HI and CO is very stringent, along with the low--inclination requirement, both eliminating a large number of candidates. Sample B is made of 27 galaxies (Table~\ref{tab:sampleB}).

\begin{table*}
\caption{Sample B. The physical parameters of the galaxies are provided by \cite{boselli2010a} (see references therein), or obtained from the NASA Extragalactic Database.}
\label{tab:sampleB}
\centering
\begin{tabular}{c c c c c c c c c c c c c c}
 \hline\hline
HRS \# & NGC \# & RA      & Dec    & Type & K         & $a$    & $b/a$ & Inclination & Distance\\
       &        & (J2000) &(J2000) &      & (Vega mag)&(kpc)   &       &  (degrees)   &(Mpc)\\
\hline
015&3338&10 42 07.54&+13 44 49.2  &SA(s)c               & 8.13 &29.8& 0.62 &52&17.4\\
077&4030&12 00 23.64&$-$01 06 00.0&SA(s)bc;HII          & 7.33 &23.6& 0.72 &44&19.5\\
081&4045&12 02 42.26&+01 58 36.4  &SAB(r)a;HII          & 8.75 &14.8& 0.74 &42&17.0\\
084&4067&12 04 11.55&+10 51 15.8  &SA(s)b:              & 9.90 &5.9 & 0.74 &42&17.0\\
089&4178&12 12 46.45&+10 51 57.5  &SB(rs)dm;HII         & 9.58 &26.5& 0.35 &70&17.0\\
096&4212&12 15 39.36&+13 54 05.4  &SAc:;HII             & 8.38 &17.8& 0.56 &56&17.0\\
100&4237&12 17 11.42&+15 19 26.3  &SAB(rs)bc;HII        &10.03 &10.0& 0.58 &55&17.0\\
102&4254&12 18 49.63&+14 24 59.4  &SA(s)c               & 6.93 &30.4& 0.91 &24&17.0\\
114&4303&12 21 54.90&+04 28 25.1  &SAB(rs)bc;HII;Sy2    & 6.84 &32.6& 0.81 &36&17.0\\
122&4321&12 22 54.90&+15 49 20.6  &SAB(s)bc;LINER;HII   & 6.59 &45.1& 0.89 &27&17.0\\
142&4393&12 25 25.50&+16 28 12.0  &Sa? pec;HII          & 9.49 &12.9& 0.50 &60&17.0\\
154&--  &12 26 47.23&+08 53 04.6  &SAB(s)cd             &10.71 &23.1& 1.00 &0 &23.0\\
157&4420&12 26 58.48&+02 29 39.7  &SB(r)bc:             & 9.66 &10.0& 0.42 &65&17.0\\
171&4451&12 28 40.55&+09 15 32.2  &Sbc:                 & 9.99 &13.2& 0.49 &61&23.0\\
196&4519&12 33 30.25&+08 39 17.1  &SB(rs)d              & 9.56 &17.8& 0.72 &44&17.0\\
203&4532&12 34 19.33&+06 28 03.7  &IBm;HII              & 9.48 &12.9& 0.38 &68&17.0\\
204&4535&12 34 20.31&+08 11 51.9  &SAB(s)c;HII          & 7.38 &41.2& 0.89 &27&17.0\\
205&4536&12 34 27.13&+02 11 16.4  &SAB(rs)bc;HII;Sbrst  & 7.52 &35.8& 0.45 &63&17.0\\
206&--  &12 34 39.42&+07 09 36.0  &SBm pec;BCD          &11.01 &9.9 & 0.70 &46&17.0\\
208&4548&12 35 26.43&+14 29 46.8  &SBb(rs);LINER;Sy     & 7.12 &29.7& 0.83 &34&17.0\\
221&4580&12 37 48.40&+05 22 06.4  &SAB(rs)a pec         & 8.77 &10.7& 0.74 &42&17.0\\
246&4651&12 43 42.63&+16 23 36.2  &SA(rs)c;LINER        & 8.03 &19.3& 0.71 &45&17.0\\
247&4654&12 43 56.58&+13 07 36.0  &SAB(rs)cd;HII        & 7.74 &24.6& 0.52 &59&17.0\\
256&4691&12 48 13.63&$-$03 19 57.8&(R)SB(s)0/a pec;HII  & 8.54 &12.1& 0.81 &36&14.8\\
287&4904&13 00 58.67&$-$00 01 38.8&SB(s)cd;Sbrst        & 9.50 &11.9& 0.71 &45&17.0\\
293&5147&13 26 19.71&+02 06 02.7  &SB(s)dm              & 9.73 &8.1 & 0.81 &36&14.5\\
307&5364&13 56 12.00&+05 00 52.1  &SA(rs)bc pec;HII     & 7.80 &32.5& 0.65 &49&16.5\\
\hline
\end{tabular}
\end{table*}

Finally we summarise the origin of the data used in this study in Table~\ref{tab:data}. 
\begin{table*}
\caption{Origin of the selected data.}
\label{tab:data}
\centering
\begin{tabular}{c c c c c c c c c}
 \hline\hline
Galaxy/sample & UV & Optical & NIR & FIR & HI & CO\\\hline
NGC~2403 & GALEX/GI3-50& SDSS & 2MASS & SAG2/VNGS & VLA/THINGS & IRAM/HERACLES\\
NGC~4254 & GALEX/GI2-17& "    & "     & B08+SAG2/HRS+HeViCS  & VLA/VIVA   & "\\
NGC~4321 & GALEX/NGA   & "    & "     & "         & "          & "\\
NGC~5194 & GALEX/GI3-50& "    & "     & SAG2/VNGS & VLA/THINGS & "\\
Sample B & GALEX/GUViCS& SDSS+SAG2/HRS & 2MASS & IRAS+SAG2/HRS & Arecibo & Kitt Peak 12m\\
\hline
\end{tabular}
\tablebib{2MASS: \cite{skrutskie2006a}; Arecibo: Boselli et al., in prep.; B08: \cite{bendo2008a}; GALEX: \cite{martin2005a}; GUViCS: \cite{boselli2011a,cortese2012b}; HeViCS: \cite{davies2012a}; IRAM/HERACLES: \cite{leroy2009a}; IRAS: \cite{neugebauer1984a,sanders2003a}; Kitt Peak 12m: Boselli et al., in prep.; NGA: \cite{gildepaz2007a}; SAG2/HRS: \cite{boselli2010a,ciesla2012a}; SAG2/VNGS+HeViCS \textit{Spitzer}/MIPS: \cite{bendo2012b} SDSS: \cite{abazajian2009a}; VLA/THINGS: \cite{walter2008a}; VLA/VIVA: \cite{chung2009a}.}
\tablefoot{The resolution of each dataset is the following. GALEX: 6\arcsec; SDSS: 1.4\arcsec; 2MASS: 2.5\arcsec; \textit{Spitzer}/MIPS: 18\arcsec; \textit{Herschel}/SPIRE: 18--25\arcsec; VLA:6--25\arcsec; IRAM: 6\arcsec.}
\end{table*}

\subsection{Data processing}

The data processing requirements are different for both samples. We must ensure that sample A allows us to perform a pixel--by--pixel analysis across the entire SED. Conversely, for sample B we can use directly the fluxes that have been published in the literature so no extra processing is required.

For galaxies in sample A, the processing of the FUV to FIR bands, including HI and CO maps is performed similarly to \cite{boquien2012a}. In a nutshell: background and foreground objects are masked; UV and optical data are corrected for the foreground Galactic attenuation using the attenuation value from \cite{schlegel1998a} combined with a \cite{cardelli1989a} extinction curve including the \cite{odonnell1994a} update; using the convolution kernels published by \cite{aniano2011a}, images are convolved to a Gaussian with a 30\arcsec\ Full Width Half Maximum (from 0.5~kpc for NGC~2403, up to 2.5~kpc for NGC~4254 and NGC~4321) corresponding to the lowest resolution of the entire dataset; they are projected on the same reference grid with a pixel size of 10\arcsec; the background is subtracted; finally pixels under a 3--$\sigma$ threshold in any band, are discarded. The standard deviation of each image is computed taking into account both the variation of the background and the pixel--to--pixel noise. The entire procedure is described in greater detail in the aforementioned article.

\section{Attenuation and gas surface mass density\label{sec:att-gas-determination}}

\subsection{Determination of the attenuation\label{ssec:AFUV}}

As mentioned in Sect.~\ref{sssec:requirements}, to determine the attenuation, we model the SED for each pixel in each galaxy (sample A), or of unresolved galaxies (sample B). We present here how this modelling is performed, and what is the expected accuracy. The impact of using alternative methods such as the relation between the IR--to--UV ratio and the attenuation is discussed in appendix \ref{sec:afuv-measured-with-irx}.

\subsubsection{CIGALE}

We use the SED modelling code CIGALE \citep{burgarella2005a,noll2009a} which handles both the emission of the stars in the UV--optical domain, and the absorption and reemission of this radiation by the dust at longer wavelength through an energy balance requirement. Several recent works have made use of it to determine the physical parameters of low and high--redshift galaxies \citep{buat2011b,buat2011a,burgarella2011a,giovannoli2011a} as well as subregions within galaxies \citep{boquien2012a}. CIGALE has been amply described in the aforementioned articles. We succinctly remind its main characteristics here.

\subsubsection{Physical parameters\label{sssec:phys-params}}

CIGALE handles a large number of input parameters that can be fine--tuned to achieve different goals. This means that we have to select the parameters space carefully to constrain the attenuation precisely. We use the set of parameters defined in \cite{boquien2012a} as a baseline. We briefly describe the choice of these parameters hereafter.

The stellar component which dominates the UV, optical, and NIR domains is made of 2 populations: the first one modelling the old stellar population, and the second one modelling the most recent burst of star formation. Both populations follow a \cite{kroupa2001a} initial mass function with a standard metallicity of $Z=0.02$. We take the latter parameter as fixed because our samples are made of objects that have typically a solar neighbourhood metallicity and there is no object with an extreme metallicity that would induce a strong effect on the blue end of the stellar spectrum. The relative masses of these 2 populations can vary freely to take into account the strength of the last star formation episode compared to the older stellar population. The emission of the stellar populations in the UV and in the optical is attenuated following a bump--less starburst--like attenuation law \citep{calzetti1994a,calzetti2000a} of varying slope, parameterised by a $\left(\lambda/\lambda_0\right)^\delta$ multiplying factor, with $\lambda$ the wavelength, $\lambda_0$ the normalisation wavelength, and $\delta$ the slope--modifying parameter. Allowing for a differential attenuation between the older and younger stellar populations, the energy is absorbed by the dust following the previously defined attenuation curve and is re--emitted in the IR, conserving the total energy. The dust emission is modelled using the star--forming galaxies templates from \cite{dale2002a}, parameterised by the IR power--law slope $\alpha$. We should mention that because the computation of the IR luminosity is largely independent from the chosen library, this choice of the template library has no impact on our results. In Tab.~\ref{tab:parameters}, we present the list and the values of the parameters used.

\begin{table*}
\caption{List of CIGALE parameters.}
\label{tab:parameters}
\centering
\begin{tabular}{l c c}
 \hline\hline
Parameter&Unit&Range\\\hline
Metallicity&Z&0.02\\
E--folding time of the old population&Gyr&1 2 3 4 5 6 7\\
Age of the old population&Gyr&13\\
E--folding time of the young population&Myr&5 10 20 50 100 200\\
Age of the young population&Myr&5 10 20 50 100 200\\
Mass fraction of the young to old stellar population&&0.0001 0.001 0.01 0.1 0.999\\
Initial mass function&&\cite{kroupa2001a}\\
Slope modifying parameter $\delta$ &&-0.4 -0.3 -0.2 -0.1 0.0\\
V--band attenuation of the young population&mag&0.05 0.15 0.30 0.45 0.60 0.75 0.90 1.05 1.20 1.35 1.50\\
Reduction of the attenuation for the old population&mag&0.25 0.50 0.75\\
IR power--law slope $\alpha$ \citep{dale2002a}&&1.0 1.5 2.0 2.5 3.0 3.5 4.0\\\hline
\end{tabular}
\end{table*}

\begin{figure*}[!htbp]
\centering
\includegraphics[width=\columnwidth]{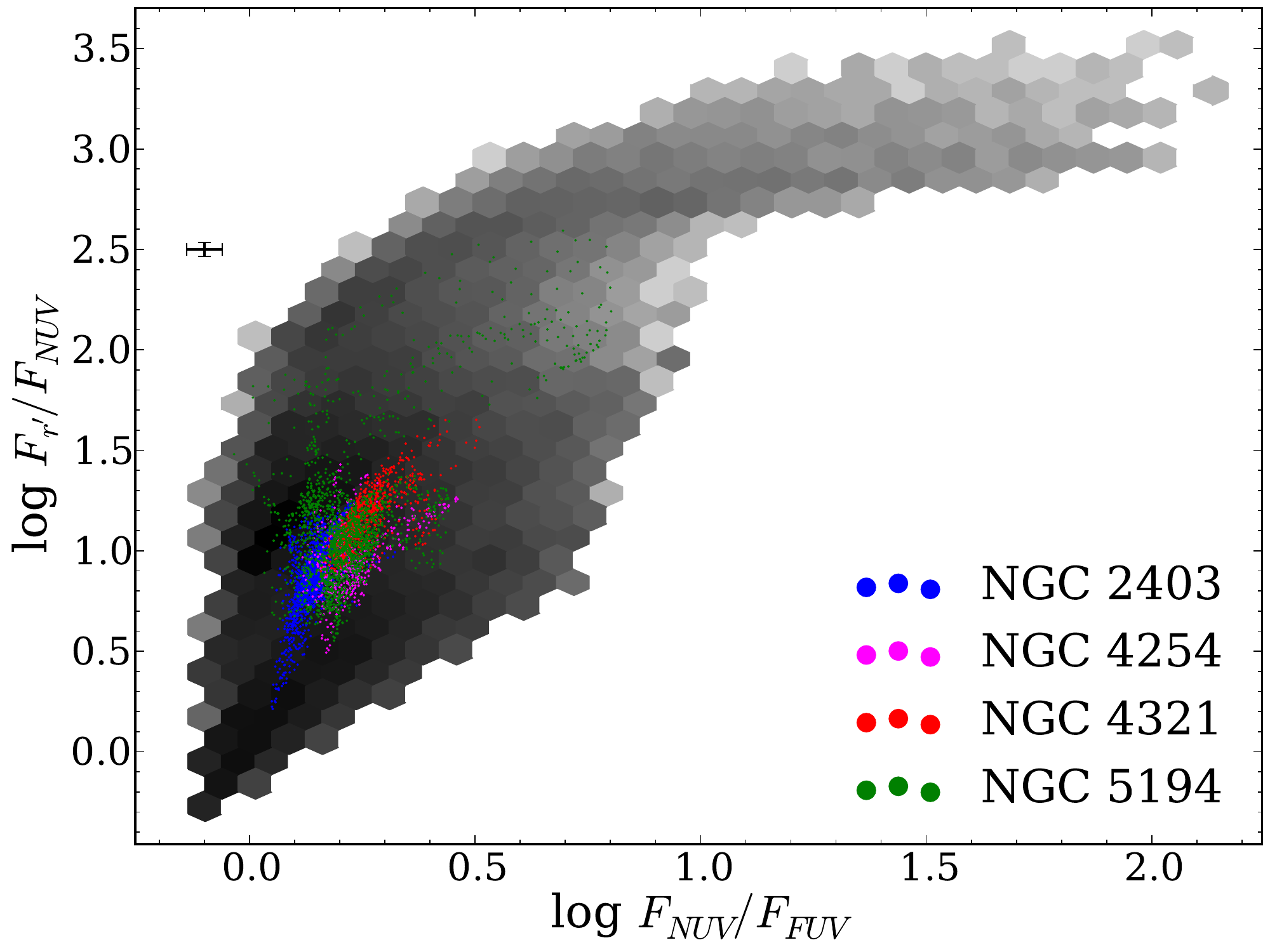}
\includegraphics[width=\columnwidth]{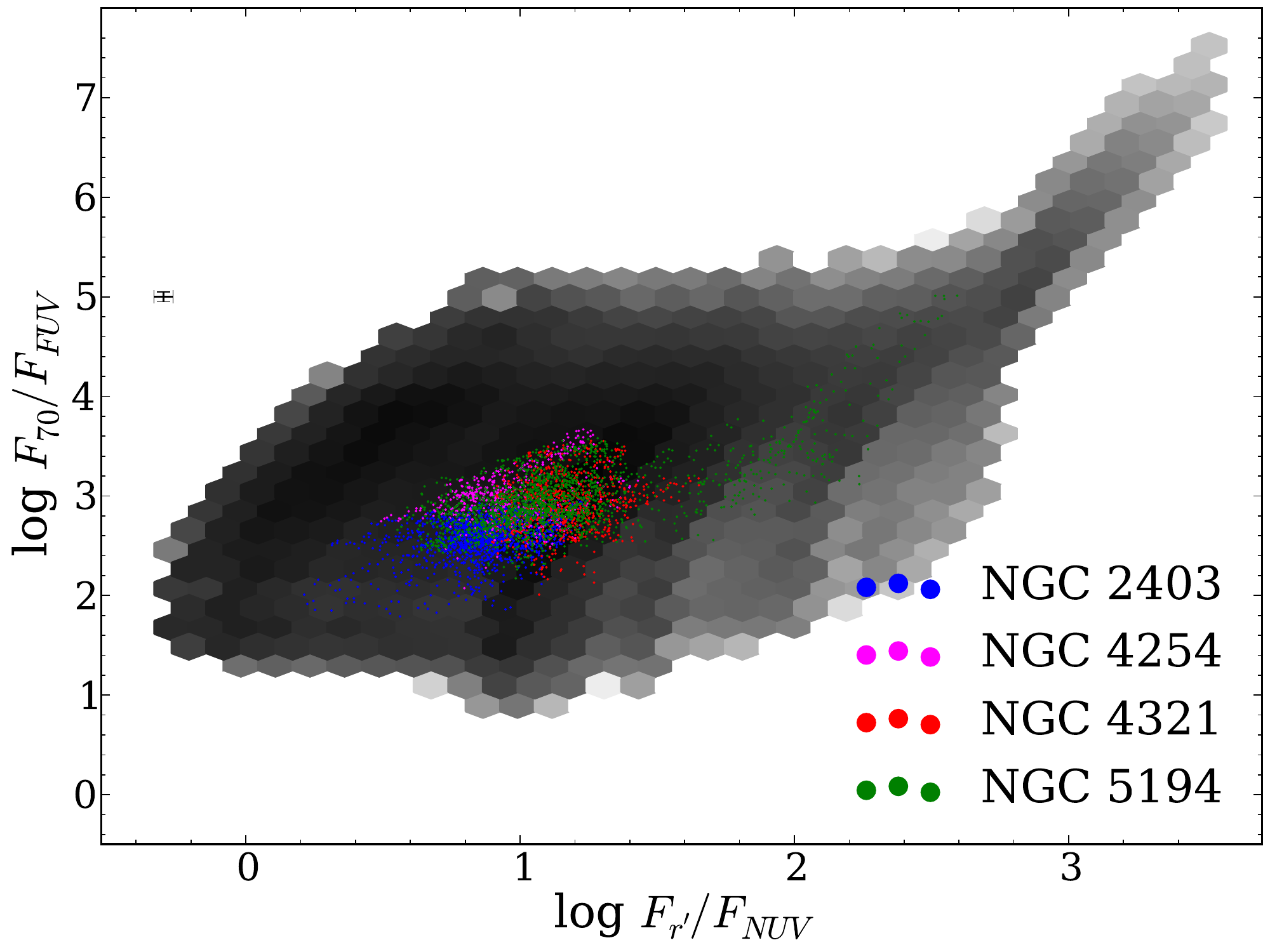}
\caption{Comparison between observed and modelled flux ratios. The shade of the grey hexagons represents the logarithm of the density of models, with a darker shade indicating a higher number of models in the bin. The observed fluxes are indicated by circles, the colour depending on the galaxy. The median error bars including systematic uncertainties are shown in the upper--left side of the plots. The left plot shows the $r'/NUV$ versus $NUV/FUV$ flux ratios and the right plot shows the $70/FUV$ versus $r'/NUV$ flux ratios.}
\label{fig:priors}
\end{figure*}

Using this space of parameters, we have generated a total of 1455300 models with CIGALE. We present in Fig.~\ref{fig:priors} a comparison between observed and modelled flux ratios constraining star formation, the attenuation, and the properties of the IR emission.
We see that these models and the observations overlap, showing that this set of CIGALE models is adequate to reproduce the observations. In the case of NGC~5194, we notice that there is a tail towards high $r'/NUV$ and $70/FUV$ flux ratios. These regions correspond to the northern end of the galaxy that is contaminated by NGC~5195, a peculiar SB0 galaxy \citep{sandage1994a}, with which NGC~5194 is interacting.

\begin{figure*}[!htbp]
\centering
\includegraphics[width=\columnwidth]{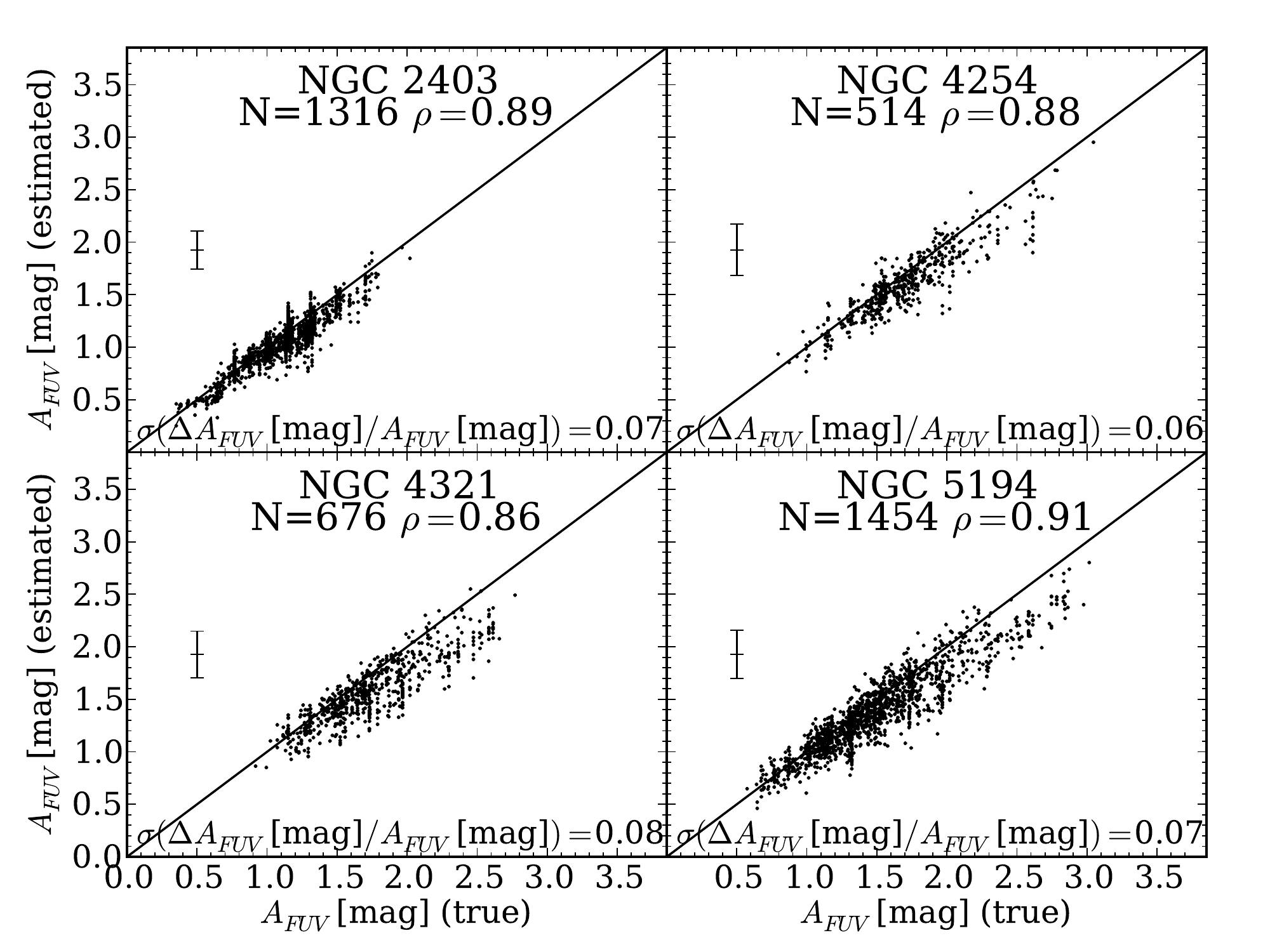}
\includegraphics[width=\columnwidth]{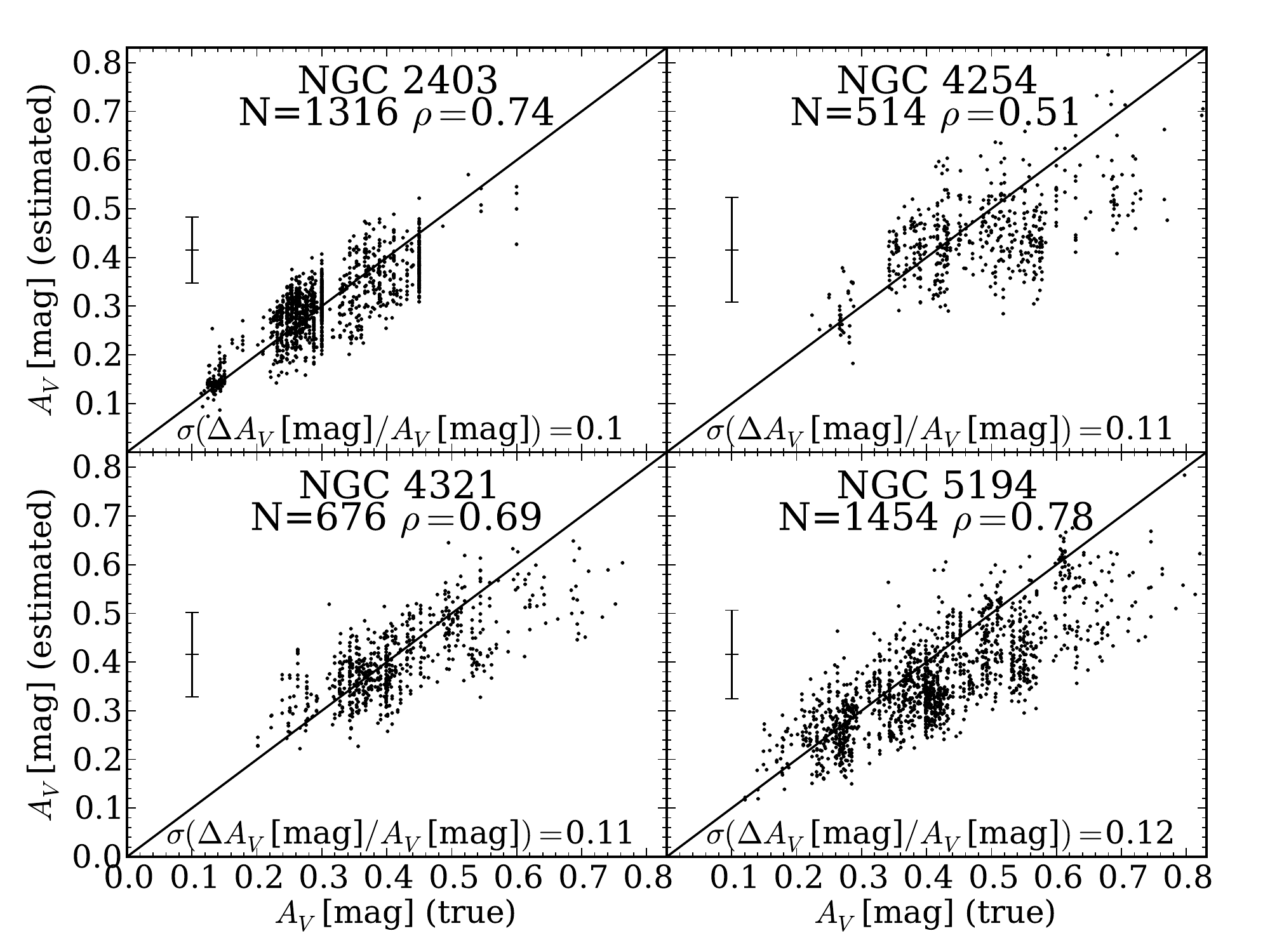}
\caption{Attenuation derived from the analysis of the probability distribution function performed with CIGALE versus the true value for artificial regions in galaxies, in order to test the accuracy of the modelling. In each quadrant we indicate the number of data points N, and the Spearman correlation coefficient $\rho$ at the top and the scatter of the relative difference in the attenuation at the bottom. The solid black line indicates the 1--to--1 relation. The error bar on the left indicates the median uncertainty as computed from the analysis of the probability distribution function. For consistency with the analysis carried out later in this paper, we have only selected datapoints with a good best fit ($\chi^2<2$). The left panel describes the FUV band attenuation whereas the right one describes the V band attenuation.}
\label{fig:mock}
\end{figure*}

Even if the models can reproduce the observations, it is not known how accurately the intrinsic physical parameters can be retrieved from such modelling with this combination of bands. Ensuring that the attenuation is reliably determined is especially important to understand its relation with the gas surface density. To do so, we follow the procedure outlined in \cite{giovannoli2011a} and \cite{boquien2012a} for instance. In a first step, we fit the SED of each pixel in each galaxy in sample A. Using the best fits, we then create a catalogue of artificial SEDs and their associated parameters which are perfectly known. Subsequently to model uncertainties on the observed fluxes, we add a random flux to each band in each artificial SED, following a Gaussian distribution whose width is determined from the standard deviation of the original data. This allows us to obtain an artificial catalogue of simulated observations of perfectly known regions in galaxies that are at the same time representative of the original sample. Finally, we compare the parameters of the artificial galaxies determined through the analysis of the probability distribution function by CIGALE, with the intrinsic, true parameters. The analysis of the results shows that the determination of the FUV attenuation is more reliable than the V--band attenuation (Fig.~\ref{fig:mock}).
We note however a slight underestimate of the attenuation for the highest values ($A_{FUV}\gtrsim$2). These biases remain no larger than the typical uncertainty which is computed from the probability function. As such, they should not affect the results of the paper. We see that the relative scatter of the attenuation in the FUV band is typically 30\% to 50\% smaller than for the V band, while the Spearman correlation coefficients are higher. For this reason, and because FUV more closely traces star formation than optical bands, we will derive relations between the gas surface density and the FUV attenuation rather than the V--band attenuation as is customary. We provide a systematic method to extend the FUV attenuation to all wavelengths in Sect.~\ref{sec:recipe}.

\subsection{Determination of the atomic and molecular gas mass surface density\label{ssec:NH}}

\subsubsection{Sample A}

The atomic and molecular gas components can be traced indirectly through the emission of the dust, or in a more standard way through HI and CO observations. We adopt the latter method as a baseline in this paper as it allows us to ensure that the determination of the gas surface density is entirely independent from the determination of the attenuation, which would not be the case with the former method. The impact of using the dust to trace the gas is discussed in appendix \ref{sec:gas-measured-with-dust}.

As long as HI remains optically thin, the conversion from HI emission to surface mass density is straightforward and reliable. For sample A, we convert the maps from brightness temperature to gas mass surface density using the following relation (see appendix \ref{sec:computation-HI} for the derivation):

\begin{equation}
\Sigma_{HI}=\frac{8.85}{\Delta\alpha\Delta\delta}\sum_i S_i\Delta V,
\end{equation}
with $\Sigma_{HI}$ the atomic gas surface density in M$_\odot$~pc$^{-2}$, $\Delta\alpha$ and $\Delta\delta$ the major and minor axes of the beam in arcsec, $S_i$ the HI flux density in Jy per beam in channel $i$, and $\Delta V$ the channel width in m~s$^{-1}$. We have to note however that self--absorption is not always negligible. A recent study by \cite{braun2009a} on the inclined galaxy M31 found that the HI mass had to be increased globally by 30\% to correct for self--absorption, and even more locally. The lower inclination of our galaxies and the relatively coarser resolution should limit the problem of HI self--absorption and its spatial variation.

The rotational transitions of CO are in theory good tracers of the molecular gas because its critical density is low enough to be excited collisionally in the bulk of the molecular gas. The relation between CO emission and the amount of molecular gas has been extensively studied in the literature but remains subject of significant uncertainties that depend on the metallicity, the temperature, the density, etc. Ideally, the combination of several CO transitions permits to constrain the physical properties of the CO emitting regions to trace the molecular gas as accurately as possible. Unfortunately such a large dataset is not available for the current sample. While CO(1--0) maps are often the most commonly available, for sample A we use the CO(2--1) maps from the HERACLES survey \citep{leroy2009a} because they provide simultaneously a good resolution and a good sensitivity. We discuss in appendix \ref{sec:impact-CO-transitions} the impact of using different CO transitions. To convert the CO emission to molecular gas surface density, in a first approach, we will use a standard $X_{CO}$ conversion factor valid for galaxies that have a metallicity similar to that of the Milky Way \citep{strong1988a}:
\begin{equation}
X_{CO}=2.3\times10^{20}~\mathrm{cm^{-2}~\left(K~km~s^{-1}\right)^{-1}},
\end{equation}
the molecular gas surface density coming directly:
\begin{equation}
\Sigma_{H_2}=2m_p\times X_{CO}\times I_{CO},\label{eqn:NH2}
\end{equation}
with $m_p$ the mass of a proton, and $I_{CO}$, the CO line intensity. Even though there are more recent estimates, it is well within the typical range of $X_{CO}$ found in the local group ($1\text{--}4\times10^{20}$~$\mathrm{cm^{-2}~\left(K~km~s^{-1}\right)^{-1}}$). As we will see later the choice of this value has been made for consistency with \cite{boselli2002a}. As mentioned before, one of the main parameters affecting the  $X_{CO}$ factor is the metallicity, with the conversion factor being higher in less metallic environments owing to complex factors linked to differences in dust shielding, lower CO abundance etc. The impact of such a dependence on the metallicity is discussed in Sect.~\ref{ssec:metal}.

The $X_{CO}$ factor is generally defined for the CO(1--0) line. As shown by \cite{braine1992a}, the typical CO(2--1)/CO(1--0) ratio for star forming galaxies does not show large variations from one galaxy to another: $0.89\pm0.06$. We therefore divide CO(2--1) maps by this factor before computing the molecular gas surface density with Eq.~\ref{eqn:NH2}. We should note that this ratio could be lower in the interarm regions (in the range 0.4--0.6) compared to the arms (in the range 0.8--1.0) as can be seen in NGC~5194 for instance \citep{koda2013a}.

Finally, to account for the presence of Helium, we multiply the gas masses by a factor 1.38. This value is derived assuming $X=0.7154$ and $Y=0.2703$ \citep{asplund2009a}. All quoted gas masses will implicitly be inclusive of Helium.

\subsubsection{Sample B\label{sssec:NH-sample-B}}

For unresolved galaxies, the computation of the mean gas surface density is more complex as it depends on the large scale distribution of the gas and the intrinsic characteristics of the observations (beam size, pointing strategy, etc.). The atomic gas is generally distributed on a large scale length and can extend well beyond the optical disk, fueling star forming regions seen in the UV with GALEX \citep{thilker2005a,thilker2007b}. Conversely on a broad scale the molecular gas is mostly concentrated towards the centre of the galaxy with an exponentially declining profile. High--resolution observations reveal however much more complex profiles with important variations in the azimuthal distribution \citep{regan2001a}. Similarly to \cite{xu1997a}, we consider different distributions to compute the mean gas surface density. When considering the atomic hydrogen only, we assume the gas has an exponential distribution and we average over a radius $r_{HI}$ corresponding to 3 scale--lengths, $r_{HI}=3\times0.61\times r_{25}$ \citep{bigiel2012a}, with $r_{25}$, the optical radius of the galaxy at 25~mag~arcsec$^{-2}$. When considering the molecular hydrogen, which is traced through CO(1--0) for all galaxies in sample B, we also assume an exponential distribution and we similarly average over 3 scale--lengths, $r_{H_2}=3\times0.2\times r_{25}$ \citep{leroy2008a,lisenfeld2011a}. When considering both we simply sum the atomic and molecular gas surface densities. We caution that these are only coarse estimates of the actual distribution of the gas and variations from galaxy to galaxy are to be expected.

Similarly to sample A, all masses are corrected to take into account the presence of Helium.

\subsection{Other gas components}

We have considered here only atomic gas traced through the 21~cm line and molecular gas traced through CO. However, that may not account for the total mass of gas in the galaxy as it does not take into account the ``dark gas'' and the ionised gas.

\subsubsection{Dark gas}

Besides the molecular gas traced by CO, there can also be so--called ``dark gas'' which is not traced by CO and therefore may be missed. This gas can be found in large quantities in relatively low metallicity galaxies such as the Large Magellanic Cloud for instance \citep{galliano2011a}, and also in a handful of high redshift objects \citep{maiolino2009a}, perhaps also due to the lower metallicity of these objects compared to local spiral galaxies. However our sample is made of relatively metal--rich spiral galaxies for which CO should be a more reliable tracer of the total molecular content. Still, some dark gas can also be found in the Milky Way \citep{grenier2005a,langer2010a,pineda2010a,velusamy2010a,planck2011a}. However, theoretical models suggest that for a given mean attenuation, the fraction of dark gas remains constant \citep{wolfire2010a}. Considering this, we have decided not to explicitly take into account this component. If necessary, it can be taken into account afterwards, rescaling the derived relations.

One way to naturally take into account the presence of this dark gas is to compute the gas surface density indirectly from the emission of the dust. We discuss this in detail in appendix \ref{sec:gas-measured-with-dust}.

\subsubsection{Ionised gas}

The last component we discuss here is the ionised gas. It can be found both in the disk in regions that have recently seen star formation and in the warm medium surrounding galaxies \citep[e.g.][]{ferriere2001a,haffner2009a}. Even though some dust has been detected in the warm medium \citep{lagache1999a,lagache2000a}, its contribution to the optical depth is difficult to ascertain. Consequently, we choose here to consider only the gas ionised by young, massive stars in the disk. This may induce an increase in the uncertainties on the determination of the final attenuation. However, regions with recent star formation are also regions where the atomic and molecular gas have the highest surface density, which makes their relative importance uncertain. Studies show that the emission of the warm gas is more closely associated to star formation than the emission of the cold gas \citep{bendo2010b,boquien2011b,bendo2012a}. In nearby galaxies, the warm to cold dust mass ratio has been shown to vary from $\sim100$ to a few $1000$s \citep[e.g.][]{willmer2009a,kramer2010a}, suggesting that only a small fraction of the dust actually lies in ionised gas. To ensure this, we have computed the approximate mass of ionised gas in each pixel. We have first computed the ionising photon luminosity following the case B recombination \citep{osterbrock1989a}:

\begin{equation}
 Q(H^0)=\frac{4}{3}\pi R^3n_e^2\alpha_B~\left[\mathrm{photons~s^{-1}}\right],
\end{equation}
with $R$ the radius of a spherical HII region, $n_e$ the mean electron density and $\alpha_B$ the case B recombination coefficient for the H$\beta$ line. Doing so we have assumed a constant density of ionised hydrogen within the radius $R$ and a fully neutral hydrogen beyond. It easily comes that:

\begin{equation}
 n_e=\sqrt{\frac{Q\left(H^0\right)}{\frac{4}{3}\pi R^3\alpha_B}}~\left[\mathrm{m^{-3}}\right].
\end{equation}
Expressing $Q\left(H^0\right)$ as a function of the H$\beta$ luminosity:
\begin{equation}
 n_e=\sqrt{\frac{L\left(H\beta\right)}{\frac{4}{3}\pi R^3\alpha_B h\nu}}~\left[\mathrm{m^{-3}}\right],
\end{equation}
with $h$ the Planck constant and $\nu$ the frequency of the H$\beta$ line. Accounting for He the total gas mass associated with the HII region can then be easily calculated:
\begin{equation}
 M_{HII}=1.38m_H\sqrt{\frac{\frac{4}{3}\pi R^3L\left(H\beta\right)}{\alpha_B h\nu}}~\left[\mathrm{M_\odot}\right],
\end{equation}
with $m_H$ the mass of an hydrogen atom.

We compute the H$\beta$ luminosity from the SFR. First we convert the SFR to an H$\alpha$ luminosity. Then we consider a temperature of 10000~K, which yields $L\left(H\alpha\right)/L\left(H\beta\right)=2.87$. If we assumed that each 10\arcsec\ pixel contains typically an HII region with a radius of 50~pc, we find that $\left<\Sigma_{HII}\right>=0.3\pm0.4$~M$_\odot$~pc$^{-2}$, with a maximum of $\Sigma_{HII}=3.5$~M$_\odot$~pc$^{-2}$. While these estimates remain uncertain because of the number and sizes of HII regions within a single pixel, it shows that the total mass of the gas in HII regions is very small compared to the atomic and molecular gas components we have considered so far. Thus, we have decided not to account for the ionised gas mass to determine the relation between the attenuation and the optical depth.

\section{A first overview of the relation between the gas and the attenuation in resolved galaxies\label{sec:first-overview}}

\begin{figure}[!htbp]
\centering
\includegraphics[width=.7\columnwidth]{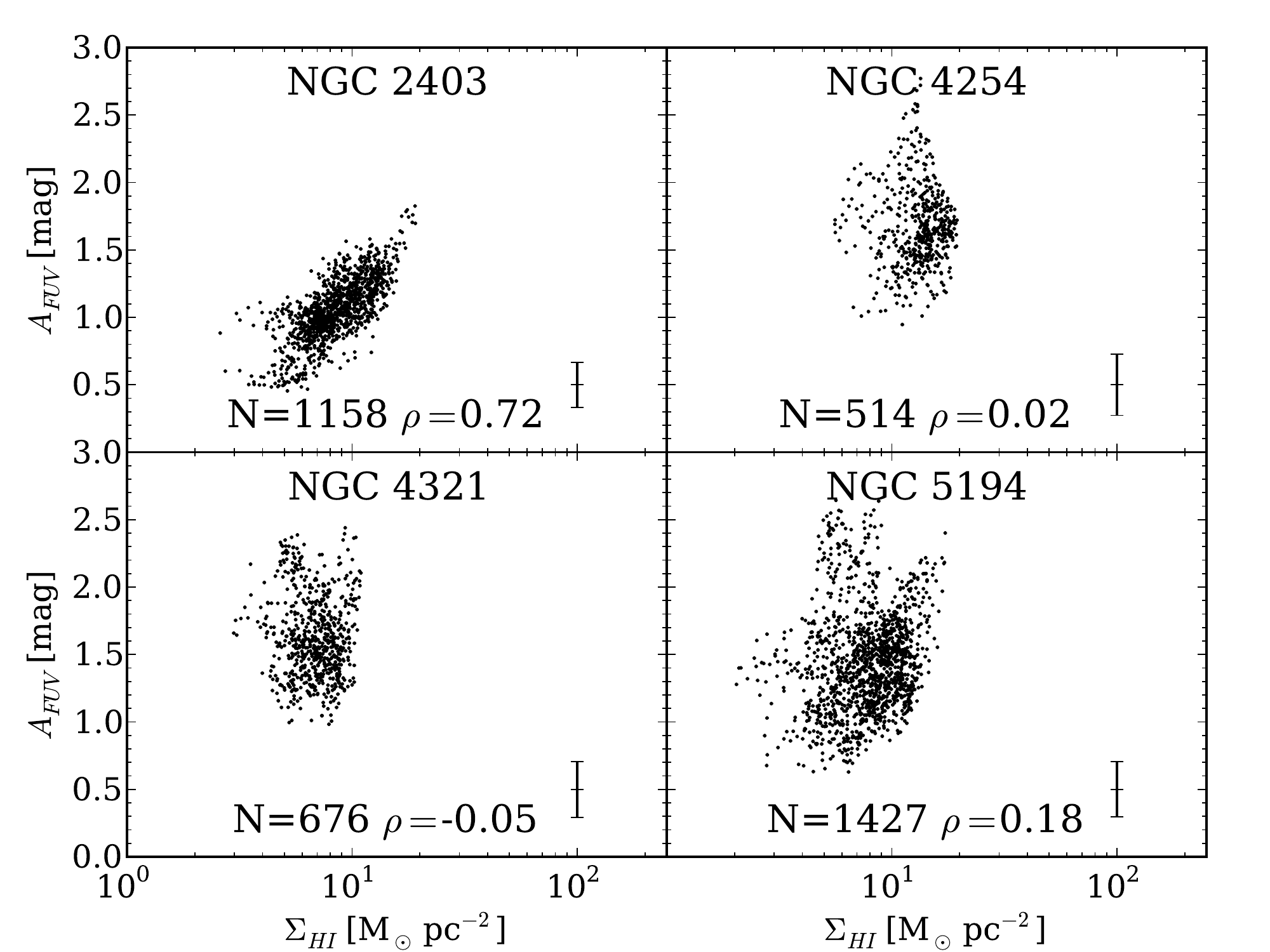}
\includegraphics[width=.7\columnwidth]{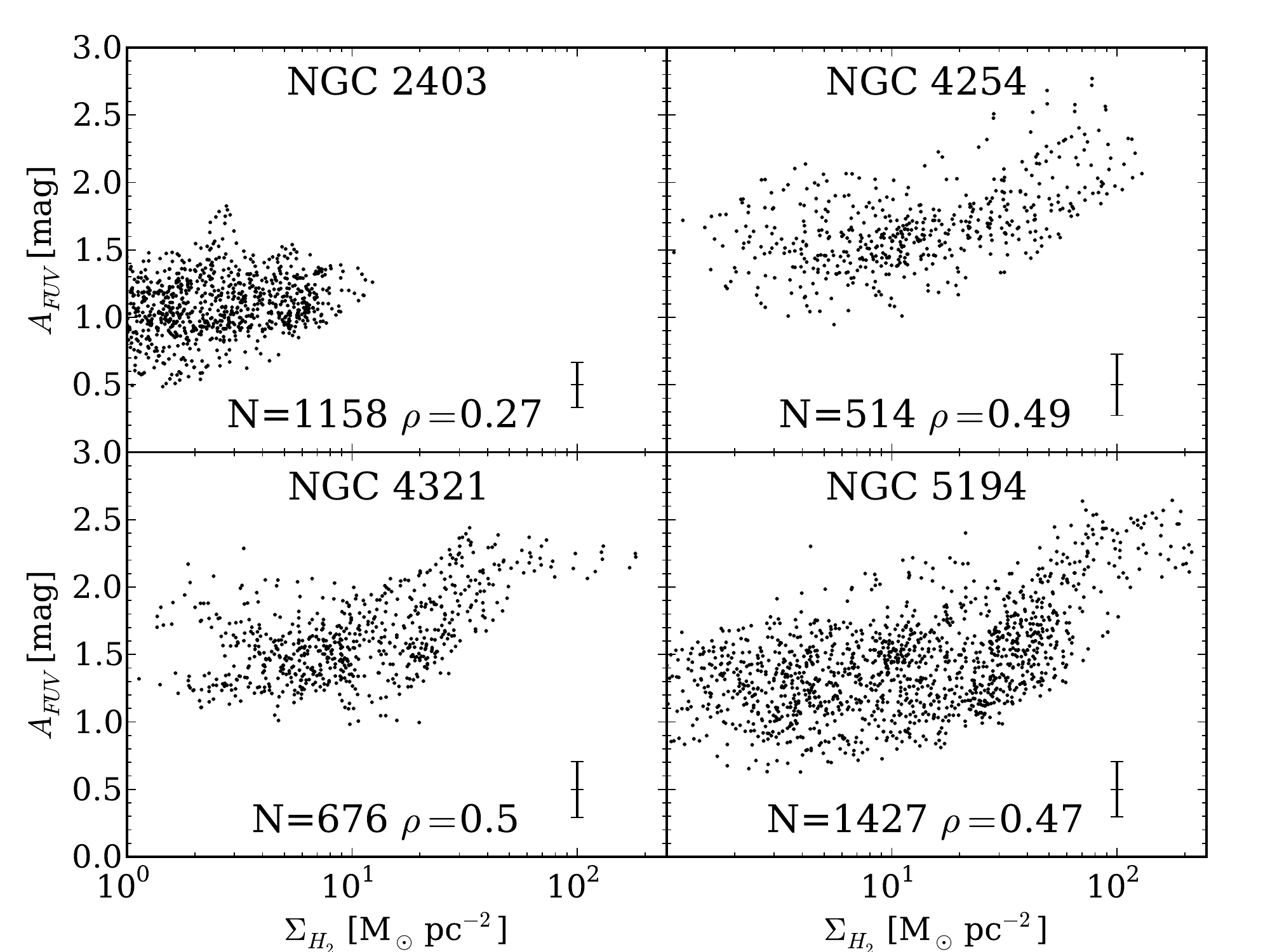}
\includegraphics[width=.7\columnwidth]{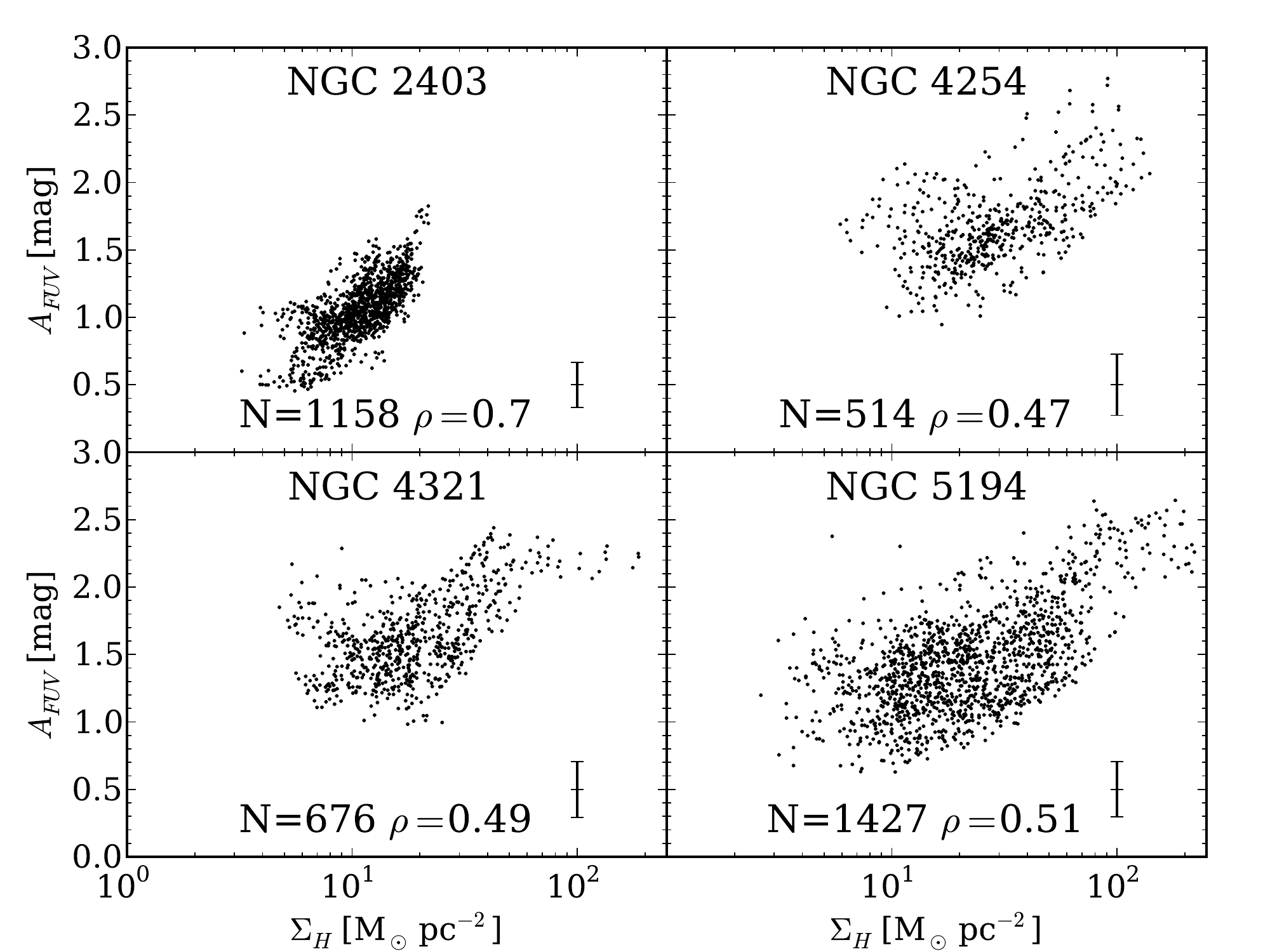}
\caption{Attenuation in the FUV band versus the atomic (top), molecular (middle), and total (bottom) gas mass surface density in units of M$_\odot$~pc$^{-2}$ for individual regions in galaxies (sample A). Each data point corresponds to a single pixel which has been detected at a 3--$\sigma$ level from the FUV to the FIR. No selection has been applied on CO and HI data. Data points with a poor fit ($\chi^2\ge2$) have also been removed. At the bottom of each panel we display the number of selected pixels N and the Spearman correlation coefficient $\rho$. Finally in the bottom right corner the median uncertainty on $A_{FUV}$ computed by CIGALE is indicated.}
\label{fig:AFUV-H}
\end{figure}

In Fig.~\ref{fig:AFUV-H} we plot for resolved galaxies the pixel--by--pixel relation between the FUV band attenuation derived in Sect.~\ref{ssec:AFUV} and the gas surface density derived in Sect.~\ref{ssec:NH}. 
We note that there is little correlation between the attenuation and the atomic gas in general ($|\rho|\le0.18$), except for NGC~2403 ($\rho=0.72$). Conversely, we observe mild correlations with the molecular gas for all galaxies ($\rho\ge0.47$) except for NGC~2403 ($\rho=0.27$). This difference between the atomic and molecular gas can be useful to understand where the attenuation takes place in galaxies as we will see hereafter. A good relation can also been seen when considering the total (HI+H$_2$) gas surface density for all galaxies ($\rho\ge0.47$).

Interestingly, we see that the relation between the attenuation and the gas column density follows 2 regimes. At high column density (typically beyond 10--20~M$_\odot$~pc$^{-2}$), there is a clear increase of the attenuation with the gas column density. Conversely, under this threshold, the attenuation seems to be rather independent from the gas column density and converges to a value circa 1.5~mag. This is best seen in the molecular gas, but can also be noted when the total gas column density is considered. We will examine the physical origin of this dual behaviour in Sect.~\ref{ssec:sigma-optdepth}.

At low molecular gas mass surface density, we see an increase of the scatter in $A_{FUV}$. This is at least in part due to larger relative uncertainties on the determination of $A_{FUV}$. Indeed, these regions tend to be on the outskirts of star--forming regions, with lower fluxes and therefore larger relative errors than in more active regions. This ultimately propagates to the determination of the FUV attenuation.

The reason why we see a correlation between CO and dust attenuation in some galaxies but not others may be related to changes in the dust heating among galaxies. The particular case of NGC~2403 is interesting in that it shows an opposite behaviour compared to the other elements of the sample. \cite{bendo2010b} found that in NGC~2403 the CO emission does not correlate well with tracers of cold dust when examined as small spatial scales, which may partly explain the absence of a good correlation between the molecular gas and the attenuation. Along this line, a recent study of \cite{bendo2012a} has shown that NGC~2403 has a thermal component that is not heated by the star forming regions. A similar analysis on NGC~4254 and NGC 4321 (Bendo et al. 2013, in prep.) does not find such a dust component. The recent results of \cite{mentuch2012a} on NGC~5194 also suggest that the dust is mainly heated by star forming regions. Hence, in NGC~4254, NGC~4321, and NGC~5194, the molecular gas and most of the dust attenuation is associated with the FIR emission near regions of star formation. In NGC~2403 however the dust attenuation associated with FIR emission is not in the same place as the molecular clouds, which causes the relation to break down. Though, since the FIR emission in NGC~2403 is at least partly from dust absorbing light in the diffuse interstellar medium (ISM), since the atomic gas is also in the diffuse ISM and not in star forming regions, the dust attenuation appears correlated with atomic gas. While studying further the relation between the star formation, and the attenuation is of utmost interest to improve our understanding of star formation across the Universe, this would go well beyond the scope of the present paper. We refer to the aforementioned papers for further details. We nevertheless briefly discuss the relation between the molecular gas fraction and the attenuation in appendix \ref{sec:molfrac}.

\section{Relation between the gas and the face--on optical depth in resolved galaxies\label{sec:rel-gas-tau}}

As we have shown earlier, in external galaxies the attenuation and the gas surface density are related in a non--trivial way. Indeed, whether a photon is affected or not by dust depends both on the relative geometry of the stars and the dust, but also on the line of sight of the observer. Unfortunately, no analytic model can fully grasp the impact of complex geometries of galaxies. They are bound to be coarse approximations. This may be worrisome as for a given quantity of dust some geometries will cause a more efficient attenuation than others, and edge--on galaxies will be more attenuated than face--on galaxies. If the attenuation is dependent on the inclination of the galaxy in a complex way, it is much easier to correct for it if we work on face--on optical depths. The overarching aim of this paper is to provide relations to compute the attenuation rather than estimates of the face--on optical depth per se, which are a means and not an end. Therefore we adopt a 2--step approach. First, we assume a given geometry to compute the relation between the face--on optical depth and the gas mass surface density (Sect.~\ref{sec:rel-gas-tau} and \ref{sec:comp-int}). In a second step, these relations can be applied on any object, assuming the same geometry, to retrieve the attenuation, as explained in Sect.~\ref{sec:recipe}. The use of the same geometry largely cancels out its initial impact on the determination of the relation between the attenuation and the face--on optical depth (appendix \ref{sec:impact-geo}).

In this section we present the stars--dust geometry we adopt (Sect.~\ref{ssec:choice-geom}) in order to compute the relation between the attenuation and the face--on optical depth (Sect.~\ref{ssec:sigma-optdepth}), and we examine the impact of the metallicity in Sect.~\ref{ssec:metal}.

\subsection{Geometry\label{ssec:choice-geom}}

\subsubsection{Choice of the geometry}

As described by \cite{disney1989a}, several geometries between the dust and the stars can be reasonably considered in galaxies, such as a simple dust screen, a sandwich model (a thin layer of dust embedded in a thick layer of stars), a slab (an infinite plane--parallel geometry in which stars and dust are perfectly mixed over the same scale), or even a clumpy ISM. While the screen model can be adapted for studies in the Milky Way, it is probably too simplistic to reproduce the complex relative distributions of stars and dust in external galaxies. The slab model is more realistic as gas is intimately mixed with stars in star--forming regions. In the sandwich model, the relative heights of the dust and the stars can be freely adjusted to take into account their different scales \citep[e.g.][]{boselli2003a}. Another common model is that of a clumpy ISM \citep[e.g.][]{witt1996a,witt2000a}. The clumps of dust can be modelled following a Poisson distribution, the source being partially attenuated by the presence of a variable number of clumps that all have the same size and optical depth \citep{natta1984a}.

More complex models, such as the one presented by \cite{charlot2000a} for instance, beyond defining a geometry also take into account the differential attenuation between stellar populations of different ages, as younger stars are still embedded in their dusty birth clouds whereas older ones have broken out of these clouds. Such models are by nature very dependent on the star formation history (SFH) of galaxies. However as we concentrate here exclusively on the FUV band where the youngest stellar populations dominate, such models do not present a significant advantage over simpler ones. To ensure this, we have carried out a supplementary run of CIGALE without differential attenuation. It yields FUV attenuations very close to when a differential attenuation is considered showing the FUV band is indeed dominated by one stellar population.

As stated earlier, the adopted geometry is a means to correct for the inclination and determine intrinsic relations between the optical gas and the gas surface density. As long as the geometry is fully consistent end--to--end, its choice is not critical. To determine the relation between the face--on optical depth and the gas mass surface density, we have decided to adopt the widely used slab geometry. We examine the impact of using different geometries in Sect.~\ref{ssec:impact-geom} and appendix~\ref{sec:impact-geo}.

For a slab geometry, if we assume an isotropic scattering the attenuation and the face--on optical depth are then related through the following equation adapted from \cite{natta1984a}:

\begin{equation}
A_\lambda=-2.5\log\left(\frac{1-\exp\left(-\sqrt{1-\omega_{\lambda}}\tau_{\lambda}/\cos i\right)}{\sqrt{1-\omega_{\lambda}}\tau_{\lambda}/\cos i}\right),\label{eqn:slab}
\end{equation}
with $A_\lambda$ the attenuation and $\tau_{\lambda}$ the face--on optical depth at wavelength $\lambda$, $i$ the inclination angle of the galaxy, and $\omega_{\lambda}$ the albedo at wavelength $\lambda$. The isotropic scattering is introduced through the term $\sqrt{1-\omega_\lambda}$. This is only a simple approximation. To take scattering fully into account radiative transfer models would be necessary \citep{baes2001a,pierini2004a,tuffs2004a,inoue2005a,rocha2008a}, which is much beyond the scope of the present article. Nevertheless, for the remainder of the paper we take $\omega_{FUV}=0.38$, following the values published by \cite{draine2003b} for a Milky Way extinction curve with $R_V=3.1$.

To compute the face--on optical depth from the attenuation, we have inverted Eq.~\ref{eqn:slab}:
\begin{equation}
 \tau_\lambda=\frac{\cos i}{\sqrt{1-\omega_\lambda}}\left[W\left(-10^{0.4A_\lambda}\times e^{-10^{0.4A_\lambda}}\right)+10^{0.4A_\lambda}\right],\label{eqn:slab-inv}
\end{equation}
with $W$ the Lambert $W$ function. The derivation of this relation is provided in appendix \ref{sec:inv-slab}.

\subsubsection{Impact of the inclination correction on the derived attenuation\label{ssec:impact-geom}}

\begin{figure}[!htbp]
\centering
\includegraphics[width=\columnwidth]{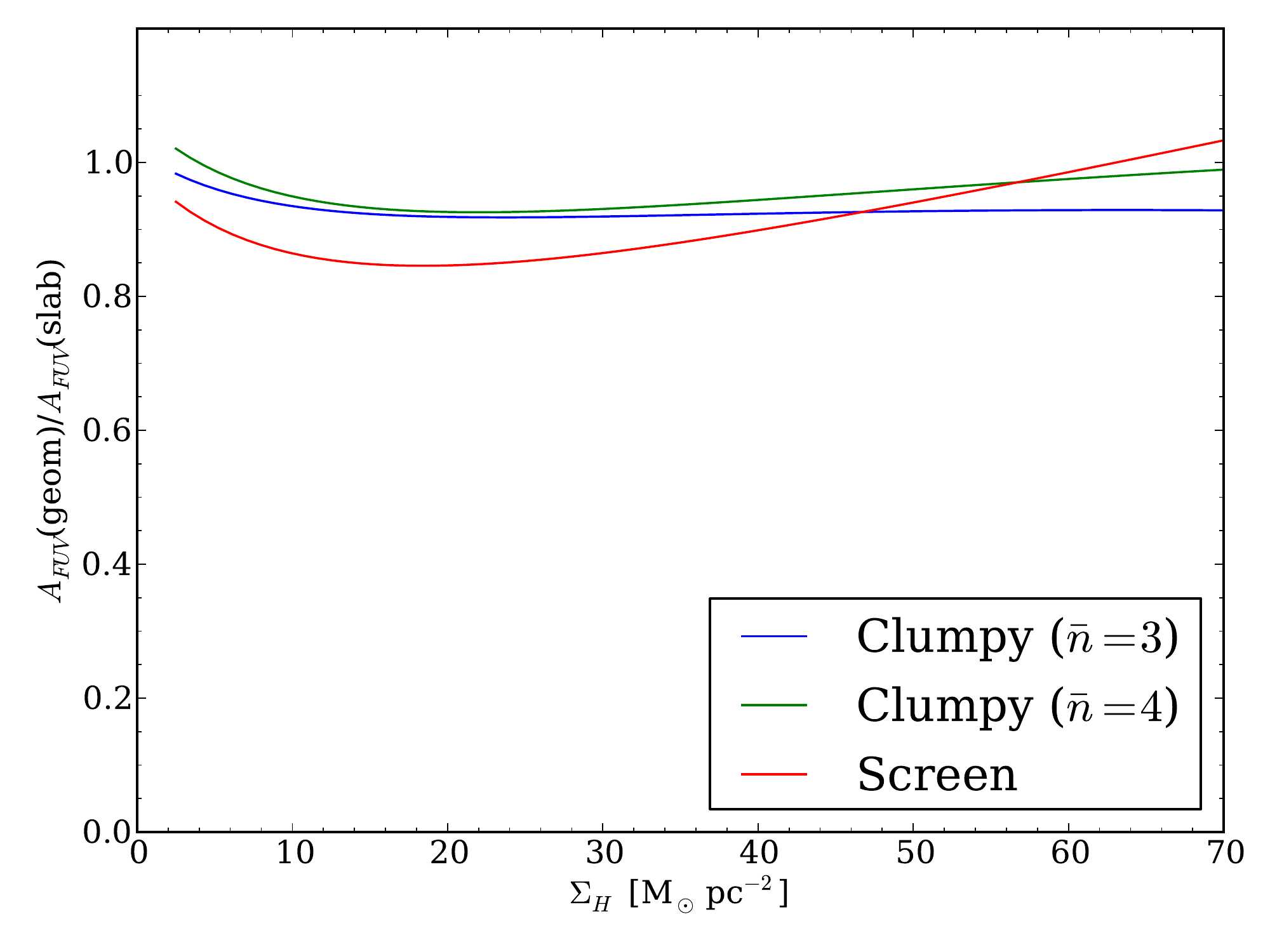}
\caption{Ratio of the FUV attenuation obtained from a relation for a given geometry to the attenuation for a slab geometry versus the gas surface density. We assume an inclination of 45$^\circ$, corresponding to the mean of datapoints in sample A, which is necessary to compare adequately with relations for clumpy geometries as the number of clumps can vary according to the inclination.}
\label{fig:comp-result}
\end{figure}

\begin{figure}[!htbp]
\centering
\includegraphics[width=\columnwidth]{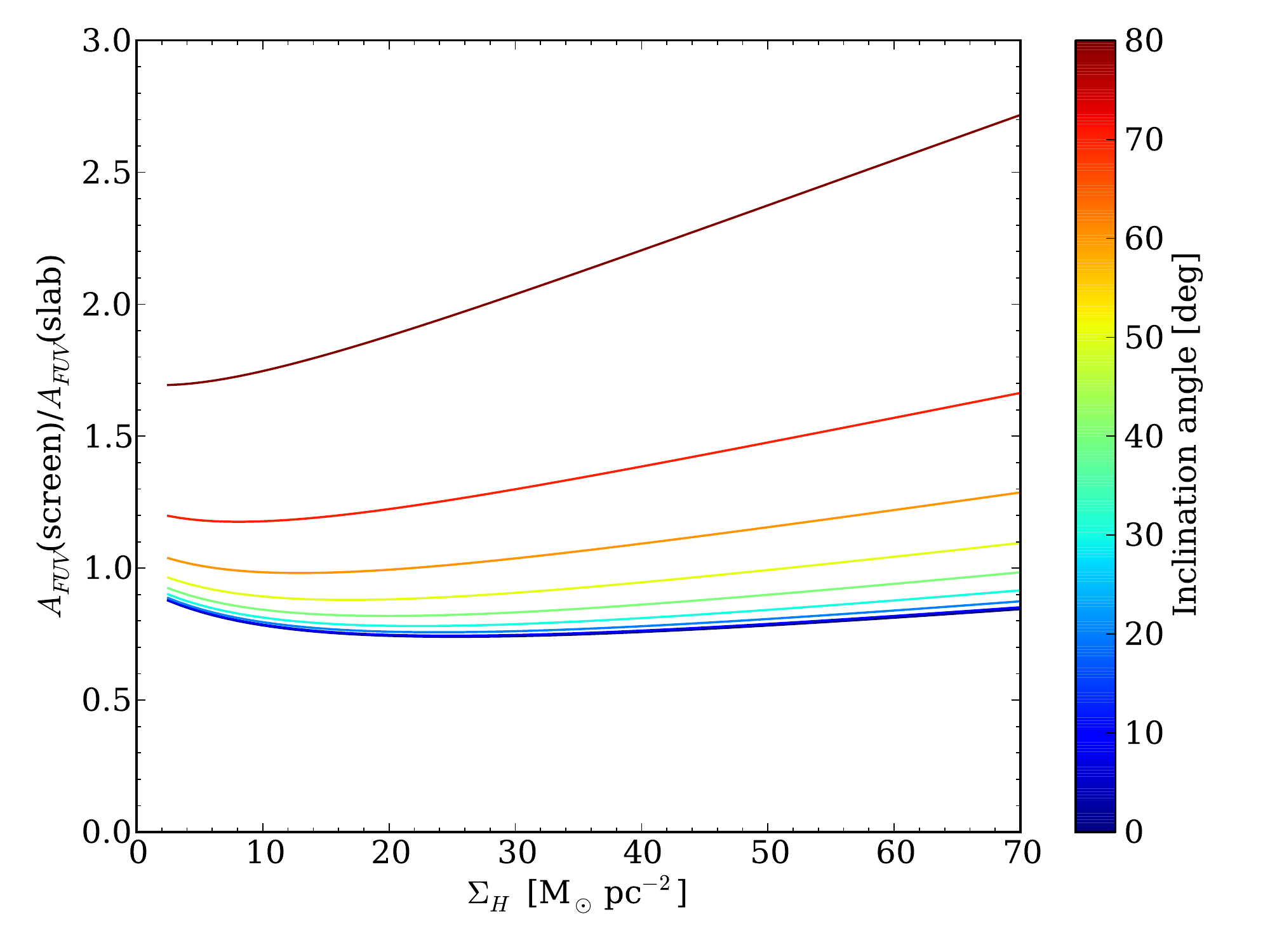}
\caption{Ratio of the FUV attenuation assuming screen and slab geometries. The inclination angle varies from 0$^\circ$ (blue) to 90$^\circ$ (red).}
\label{fig:effect-inc}
\end{figure}

As we will see later, the relations we determine to compute the face--on optical depth from the gas surface density depend significantly on the geometry. The reason is that a given attenuation corresponds to very different optical depths depending on the geometry. This is not necessarily a problem as long as the geometry is kept consistent. The application of these relations would yield identical attenuations if all galaxies were face--on and if there was no scatter around the relations. As our data deviate from these ideal conditions, to evaluate the actual impact of the geometry we present in Fig.~\ref{fig:comp-result} the ratio between the attenuation obtained assuming a given geometry and that obtained assuming a slab geometry, for the gas surface density spanned by sample A.
We see that while using relations other than for a slab geometry induces an offset, generally towards a lower attenuation, it is limited to 15\% at most, in case we consider an extreme, screen geometry. This shows that our original assumption that the geometry would only have a limited impact on the results is indeed validated. In Fig.~\ref{fig:effect-inc}, we show how the relative attenuations vary as a function of the inclination angle and the gas surface density, assuming slab and screen geometries.
We see that for inclinations no larger than 50$^\circ$, we get attenuations within 20\% from either a screen or a slab geometry. However the attenuations become increasingly discrepant for higher inclinations. This translates both the uncertainties due to the scatter around the relations, and probably the complexity of the geometry within galaxies compared to these simple models. We therefore warn of the greater uncertainties when deriving the attenuation from the face--on gas column density for highly inclined galaxies.

\subsection{Gas surface density and face--on optical depth\label{ssec:sigma-optdepth}}

\begin{figure}[!htbp]
\centering
\includegraphics[width=\columnwidth]{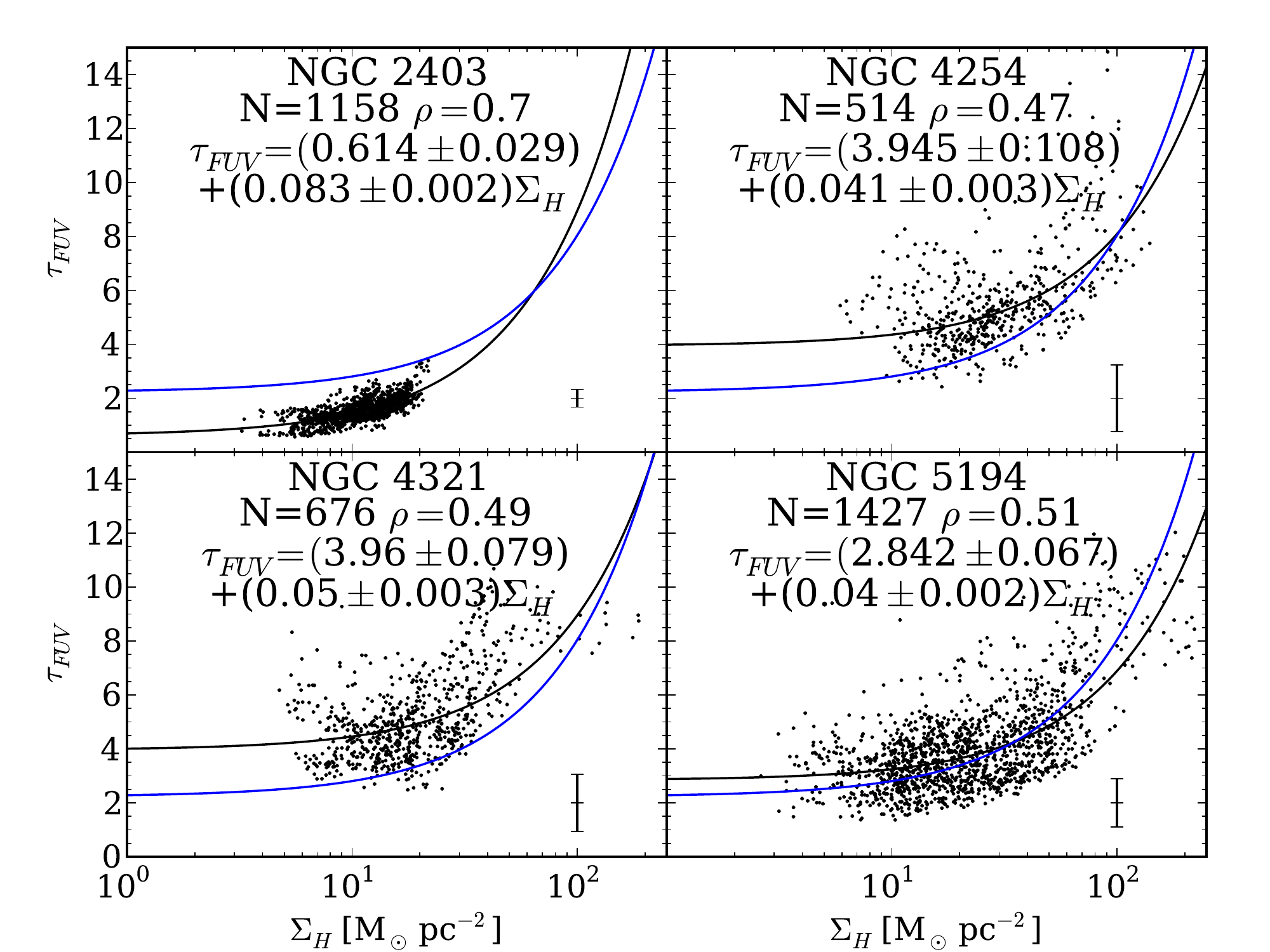}
\caption{Face--on optical depth in the FUV band versus the total gas mass surface density in units of M$_\odot$~pc$^{-2}$ for a slab geometry, and assuming a metallicity--independent $X_{CO}$ factor. The blue curves represent the best fit for the full sample whereas the black ones represent the best fit for individual galaxies for which the equation is given at the top of each panel. The uncertainties on $\tau_{FUV}$ are indicated by the error bar at the bottom right of each panel.}
\label{fig:tau-nh-slab}
\end{figure}

Following the choice of a slab geometry in the previous section, we have computed the face--on optical depth in the FUV band using Eq.~\ref{eqn:slab-inv}. To compute the gas column density, in this first step we do not take into account the impact of the metallicity which will be explored in detail in Sect.~\ref{ssec:metal}. We present in Fig.~\ref{fig:tau-nh-slab} the relation between the face--on optical depth and the gas surface density.
Combining data points from all regions of the galaxies in sample A, we find the following relation between the face--on optical depth and the gas surface density:

\begin{equation}
\tau_{FUV}=\left(2.226\pm0.040\right)+\left(0.058\pm0.001\right)\times \Sigma_H,\label{eqn:tau-sample-A}
\end{equation}
for $2.5\le \Sigma_H\le221.0$~M$_\odot$~pc$^{-2}$. The presence of a non--zero constant is surprising at first sight. A first explanation could be that there is a selection effect against FUV--bright, low--attenuation regions with a low gas surface mass density, as we require emission at a 3--$\sigma$ level in all bands. However, the inspection of selected regions shows that except for the most external regions, the entire extent of galaxies has been selected, within the limit of the observed field--of--view. We conclude that this is unlikely to be the reason for the observed levelling at low gas mass surface density. A more likely explanation would be that this is actually an effect due to the mixing of different regions because of the level of the resolution. Indeed, at a resolution of 0.5~kpc to 2.5~kpc (30\arcsec\ beamwidth) there is a blending between different regions. This affects differently the measure of the gas mass surface density and the attenuation. The computation of the resulting gas mass surface density is straightforward, being simply the average of all regions within the resolution element. However, the attenuation is strongly weighted towards the brightest regions in the FIR and in the FUV band. As a consequence, if we consider that a given resolution element actually contains 2 regions that have respective gas mass surface densities $s_1$ and $s_2$ covering the same area and with $s_1\gg s_2$, the measured gas surface density over the entire resolution element will be $\sim s_1/2$. If we assume that these regions have attenuations $a_1$ and $a_2$, if the first region is much brighter than the second one, the measured attenuation in the resolution element will the $\sim a_1$. We therefore see that such blending will differently affect the gas mass surface density and the attenuation. In other words, the latter is luminosity weighted whereas the former is governed by a simple geometrical mean over the surface of a pixel. This is even more evident if we consider that denser regions tend to form stars more actively, and then will drive the attenuation within the resolution element. The consequence is that if there is blending between different regions within a galaxy there will be a plateau at low gas mass surface density, which is what we see here. This means that the relations we derive in this paper should not be applied at a resolution more refined than that considered here. Conversely, this suggests that standard relations that do not take into account such blending should not be applied to unresolved galaxies or sufficiently large regions.

The second aspect that is to be noted is that there is a large scatter, both within galaxies and from one galaxy to another. The internal scatter can largely be explained by the uncertainty on the face--on optical depth. We have to note that this uncertainty is computed from the uncertainty on the FUV attenuation provided by CIGALE. When comparing the relation provided in Eq.~\ref{eqn:tau-sample-A} to relations derived for individual galaxies (Fig.~\ref{fig:tau-nh-slab}) there are clear differences. The slope is relatively similar for NGC~4254, NGC~4321, and NGC5194, but it is about twice as steep for NGC~2403. Conversely, NGC~2403 levels at a much lower face--on optical depth than average, whereas NGC~4254 and NGC~4321 level at a higher face--on optical depth. NGC~5194 is close to the value for the entire sample. This is probably due to intrinsic differences between galaxies in the sample. Another possibility would be that it is due to the physical resolution. However, after having carried out a similar study at a resolution of 64\arcsec\ we found that such a coarsening had no effect on the relative levels. Incidentally this suggests that beyond a certain resolution, the level of the face--on optical depth at low column density is left mostly unchanged. Indeed if the mixing within a resolution element is already representative of the content of the galaxy, increasing the mixing will not change the properties within a resolution element. 

Finally, we see that for NGC~4321, there is a visible saturation in the optical depth at high column density. This undoubtedly affects the fit. Performing a fit only when $\Sigma_H\le60$~M$_\odot$~pc$^{-2}$ yields a steeper slope for NGC~2403 than when all regions are considered but only change marginally the parameters for other galaxies.

\subsection{Impact of the metallicity\label{ssec:metal}}

A change in the metallicity can have dramatic effects on the measurement of the molecular gas reservoir of galaxies. The reason is that the molecular content is generally traced through quantities that are dependent on the metallicity. For instance the $X_{CO}$ factor mentioned in Sect.~\ref{ssec:NH} is expected to vary with the metallicity. Ignoring this effect naturally increases uncertainties on the gas mass derived from the CO emission. At lower metallicities CO lines trace an increasingly small fraction of the total molecular gas, directly increasing the value of the $X_{CO}$ factor, barring any other effect.

Several studies have been dedicated to quantifying the relation between the metallicity and the $X_{CO}$ factor. \cite{wilson1995a} in a seminal study found a sub--linear relation between the  $X_{CO}$ factor and the oxygen abundance. \cite{arimoto1996a} and \cite{boselli2002a} found nearly linear slope ($-1.01$ and $-1.00$, respectively). Recently, \cite{schruba2012a} found especially steep slopes between $-2.8$ and $-2.0$. Conversely, \cite{blitz2007a} or \cite{bolatto2008a} did not find any relation with the metallicity.

\begin{figure}[!htbp]
\centering
\includegraphics[width=\columnwidth]{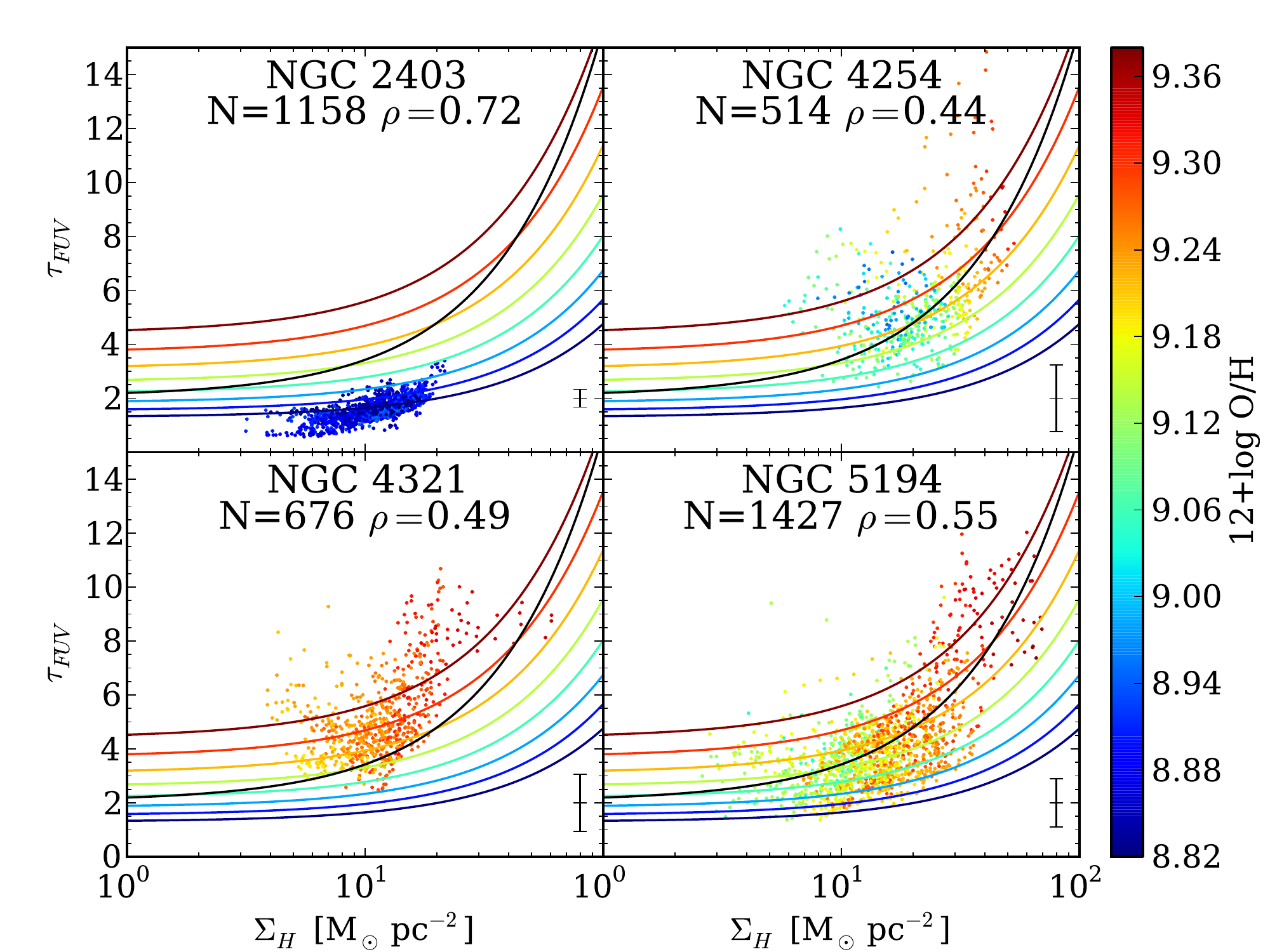}
\caption{Face--on optical depth in the FUV band versus the total gas mass surface density in units of M$_\odot$~pc$^{-2}$. The colour of each point indicates the oxygen abundance determined from the gradients published by \cite{zaritsky1994a}. The black line represents the best fit for the entire sample (Eq.~\ref{eqn:tau-sample-A-XCO-variable}) when taking into account a metallicity--dependent $X_{CO}$ factor (Eq.~\ref{eqn:boselli2002a}). The coloured lines represent the best fit for the entire sample (Eq.~\ref{eqn:tau-sample-A-XCO-variable-metal}) for different values of $12+\log O/H$, when fitting simultaneously the gas column density and the oxygen abundance. At the top of each panel the number of datapoints and the Spearman correlation coefficient is indicated.}
\label{fig:metal-H}
\end{figure}

To investigate the influence of the metallicity on the relation between the gas surface density and the face--on optical depth, we examine how this relation changes when we consider a metallicity--dependent $X_{CO}$ factor. To do so we adopt the relation derived by \cite{boselli2002a}, which is intermediate between the steep slopes found by \cite{schruba2012a}, and the lack of relation found by \cite{blitz2007a} or \cite{bolatto2008a}:
\begin{equation}
\label{eqn:boselli2002a}
 \log X_{CO}/X_{{CO}_{gal}}=8.92-1.01\times\left[12+\log O/H\right],
\end{equation}
with $X_{{CO}_{gal}}$, the standard Milky Way factor. To be consistent with the metallicity scale used by \cite{boselli2002a}, we adopt the metallicity gradients measured by \cite{zaritsky1994a} rather than the newer gradients published by \cite{moustakas2010a}. This prevents the introduction of biases due to differences between metallicity estimators. In Fig.~\ref{fig:metal-H}, we plot the relation between the total gas surface density and the face--on optical depth when using the relation given in Eq.~\ref{eqn:boselli2002a}.
Computing the relation between the gas surface density and the face--on optical depth, we find:

\begin{equation}
 \tau_{FUV}=\left(2.060\pm0.059\right)+\left(0.137\pm0.005\right)\times \Sigma_H\label{eqn:tau-sample-A-XCO-variable}.
\end{equation}
We note that there is a clear steepening of the relations with respect to those shown in Fig.~\ref{fig:tau-nh-slab}. This is expected because Eq.~\ref{eqn:boselli2002a} naturally yields a lower (respectively higher) $X_{CO}$ factor for $12+\log O/H>8.83$ (resp. $12+\log O/H<8.83$).

Beyond affecting the estimate of the molecular gas surface density, the metallicity also has a direct impact on the gas--to--dust mass ratio and therefore on the amount of dust attenuating the interstellar radiation. The simple relations we have derived do not fully capture the impact of the variation of the metallicity, in effect averaging over all the line--of--sights, objects, and metallicities. To take into account these variations, different methods can be used. A first, very na\"ive, method would be to correct the face--on optical depth computed as in Eq.~\ref{eqn:tau-sample-A-XCO-variable} for the difference of the gas--to--dust mass ratios between sample A and that of the observed object: $\tau_{FUV}^{corr}=\tau_{FUV}\left(\Sigma_H\right)\times\frac{\left[M_H/M_{dust}\right]_{sample~A}}{\left[M_H/M_{dust}\right]_{object}}$, with $\tau_{FUV}^{corr}$ the metallicity corrected face--on optical depth, $\tau_{FUV}\left(\Sigma_H\right)$ the face--on optical depth obtained from Eq.~\ref{eqn:tau-sample-A-XCO-variable}, and $M_H/M_{dust}$ the gas--to--dust mass ratio for sample A (subscript $sample~A$), and for the targeted object (subscript $object$). Indeed, as mentioned before, a variation of the gas--to--dust mass ratio will directly affect the quantity of dust affecting the emission. This ratio is evidently strongly dependent on the metallicity. Therefore if is important that the metallicity of the sample the relation was derived on is homogeneous and well--known. With a metallicity spread over more than 0.5~dex in sample A, the former criterion is certainly not fulfilled as there are significant variations of the gas--to--dust mass ratio from one region to another, and from one galaxy to another. A better way to take the metallicity into account is to derive relations between the face--on optical depth and the gas surface density, parameterising them explicitly on the oxygen abundance. With this method, we find:

\begin{equation}
\begin{split}
 \tau_{FUV}=\left[\left(1.926\pm0.034\right)+\left(0.051\pm0.002\right)\times \Sigma_H\right]\\
\times10^{\left(0.947\pm0.024\right)\left(\left[12+\log O/H\right]-9.00\right)}.\label{eqn:tau-sample-A-XCO-variable-metal}
\end{split}
\end{equation}
The normalisation of the oxygen abundance, here $9.00$, is arbitrary. We have chosen this value because it is close to the typical values yielded by the oxygen abundance estimator of \cite{zaritsky1994a} for nearby spiral galaxies. This relation is necessarily dependent on the way the metallicity has been determined as different methods give different estimates. One can use the relations provided by \cite{kewley2008a} to convert the metallicity computed from one method to the \cite{zaritsky1994a} method. Note that because of the metallicity and the gas surface density are not completely independent quantities in galaxies (regions with a higher gas surface density tend to be more metal rich than regions with a lower gas column density), great care must be taken to interpret the coefficients of the best fit related to the gas surface density and the metallicity. Examining the variance--covariance matrix of the best fit parameters, we find that the correlation coefficient between the coefficients related to these quantities is $r=-0.63$.

\begin{figure}[!htbp]
\centering
\includegraphics[width=\columnwidth]{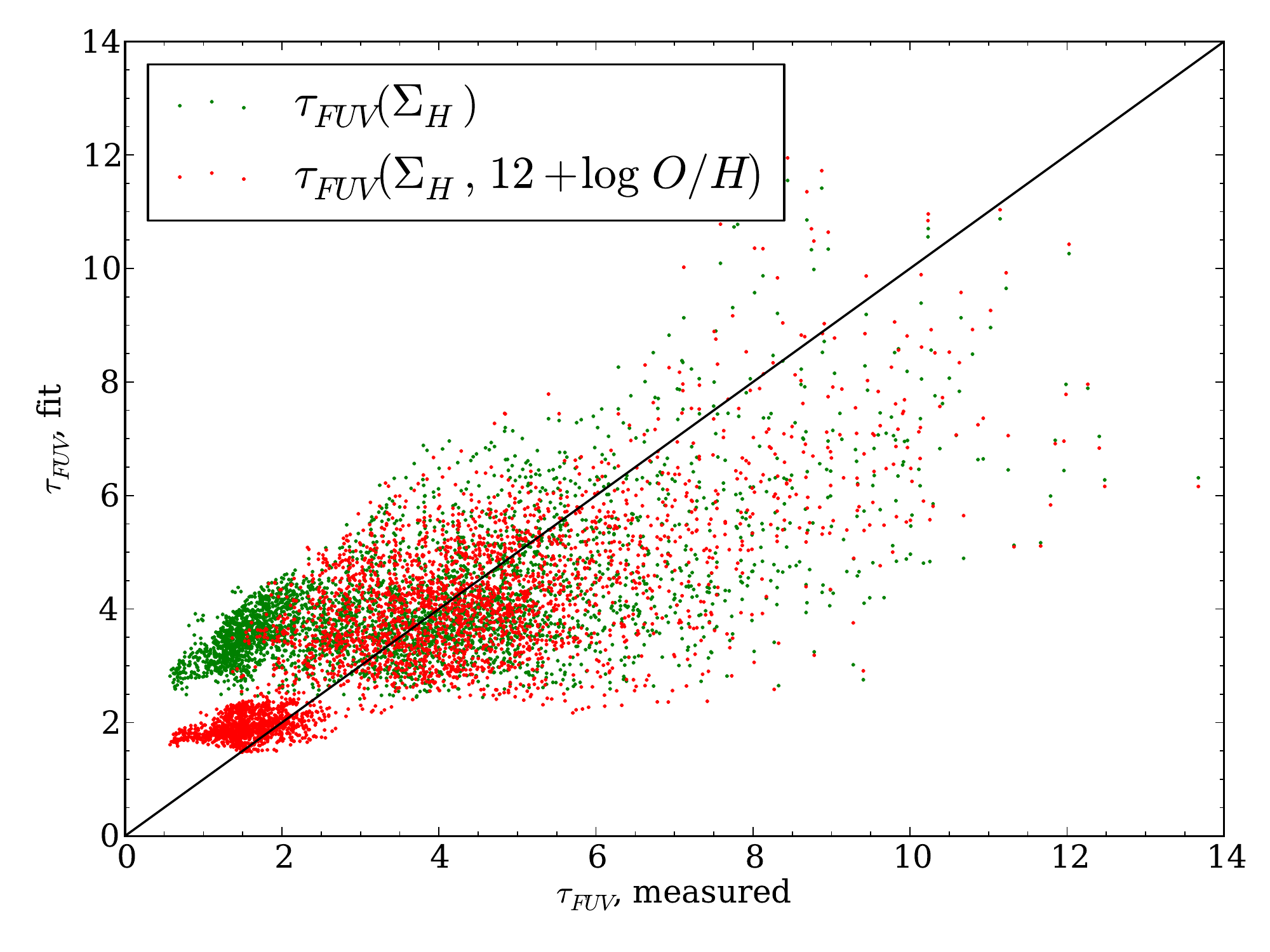}
\caption{FUV face--on optical depth computed with Eq.~\ref{eqn:tau-sample-A-XCO-variable} taking into account the gas surface density (green dots), and Eq.~\ref{eqn:tau-sample-A-XCO-variable-metal} taking into account the gas surface density and the metallicity (red dots), versus the measured face--on optical depth from the attenuation provided by CIGALE and an assumed geometry. The black line represents to 1--to--1 relation.}
\label{fig:res-fit-tau-slab}
\end{figure}

In Fig.~\ref{fig:res-fit-tau-slab} we show the face--on optical depth computed with these relations versus the face--on optical depth derived from the attenuation provided by CIGALE and an assumed slab geometry.
Taking into account the metallicity brings as expected substantial improvements. The most notable change is due to the regions belonging to NGC~2403 which have a systematically lower metallicity than other galaxies in the sample. The regions belonging to NGC~2403 correspond to the concentration of points in the lower left corner in Fig.~\ref{fig:res-fit-tau-slab}. If the impact of the metallicity is not taken into account, Eq.~\ref{eqn:tau-sample-A-XCO-variable} would systematically overestimate the optical depth. Eq.~\ref{eqn:tau-sample-A-XCO-variable-metal} brings a strong improvement, yielding optical depths much closer to the ones obtained from CIGALE attenuations.

The relative difference of the face--on optical depth goes from:
\begin{equation}
\left<\frac{\Delta\tau_{FUV}}{\tau_{FUV,\ measured}}\right>=0.43\pm0.77,
\end{equation}
with $\Delta\tau_{FUV}=\tau_{FUV,\ fit}-\tau_{FUV,\ measured}$ when considering only the gas surface density, to:
\begin{equation}
\left<\frac{\Delta\tau_{FUV}}{\tau_{FUV,\ measured}}\right>=0.11\pm0.39,
\end{equation}
when also taking into account the metallicity. This means that the uncertainty on the face--on optical depth is of the order of $\sim50$\%. This hints at the importance of additional, undetermined, parameters, and it shows the inherent complexity of determining the attenuation, and by extension the face--on optical depth, from the gas column density.

For the remaining of this article we will always consider a metallicity dependent $X_{CO}$ factor unless specifically noted otherwise.

\section{Comparison with unresolved galaxies\label{sec:comp-int}}

Due to their more complex, large scale geometries, gas distribution, and stellar populations, the relation between the gas surface density and the face--on optical depth in unresolved galaxies is expected to be more challenging to determine. It is important to verify whether the relations previously obtained provide reliable results for unresolved galaxies as it conditions their applicability. One of the first attempts to connect the attenuation and the gas surface density in unresolved galaxies was made by \cite{xu1997a}. They found weak relations between the attenuation in the B band and the face--on optical depth. More recent studies have yielded similar results \citep{boissier2004a,boissier2007a}.

\begin{figure}[!htbp]
\centering
\includegraphics[width=.7\columnwidth]{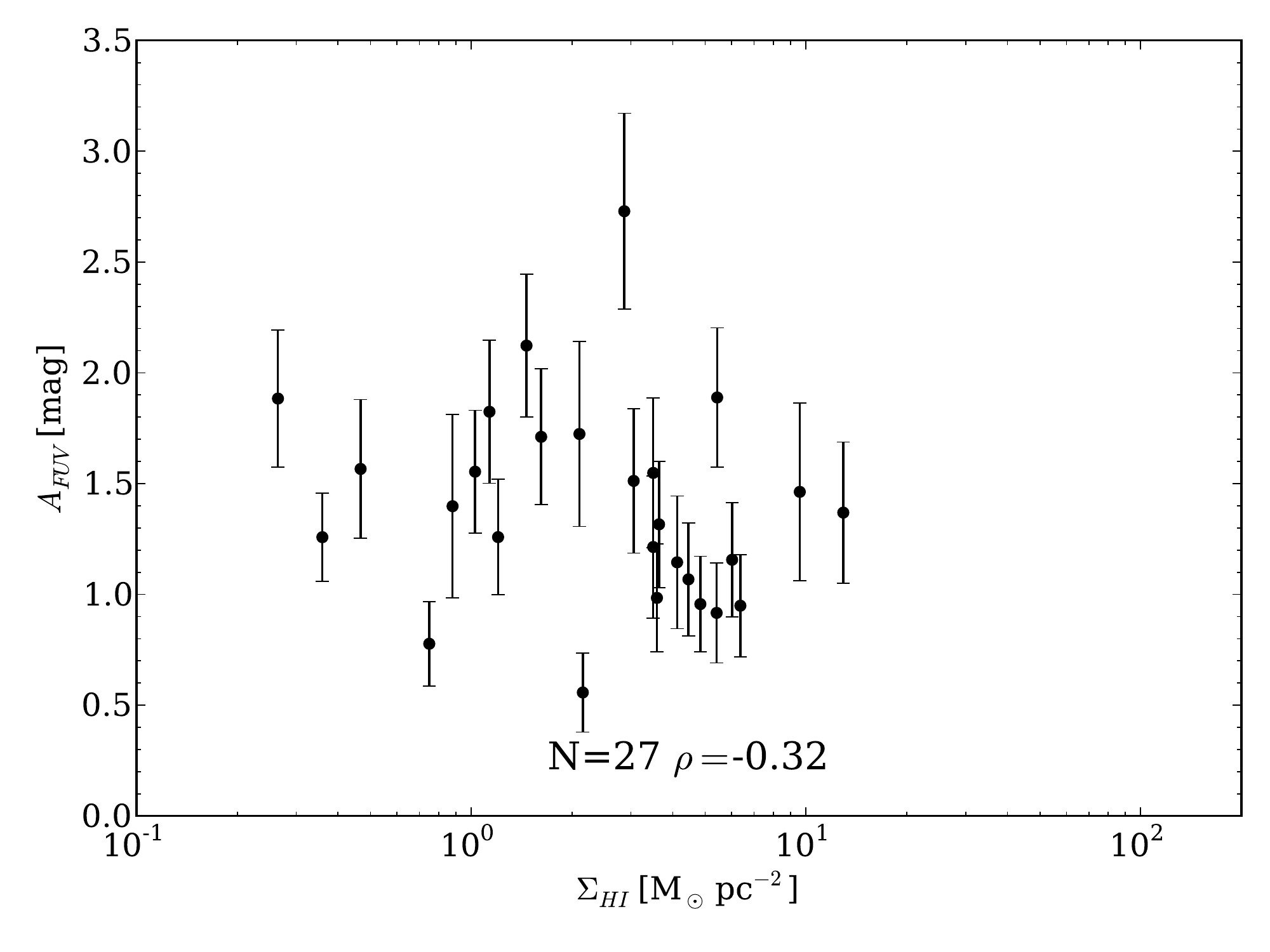}
\includegraphics[width=.7\columnwidth]{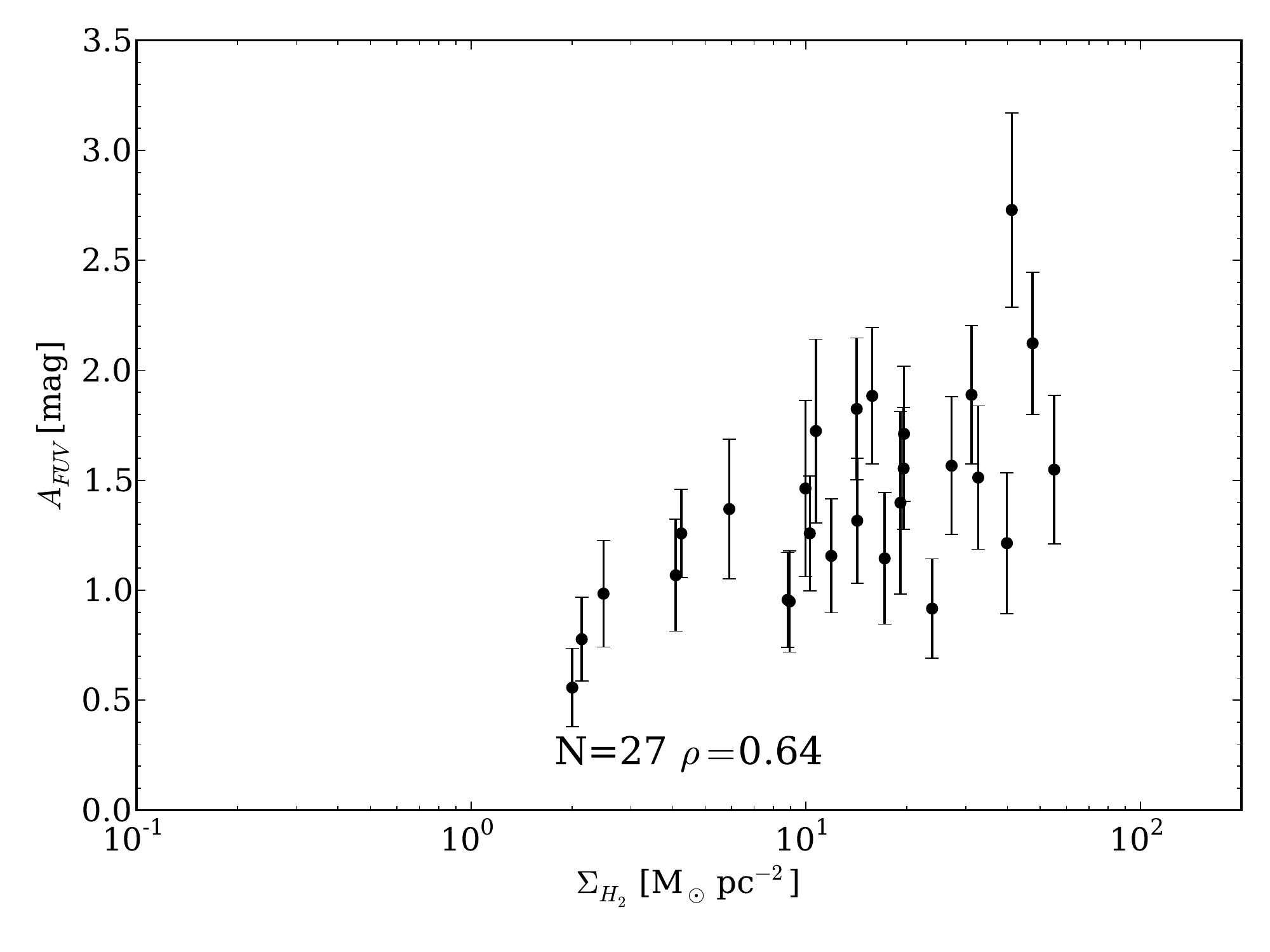}
\includegraphics[width=.7\columnwidth]{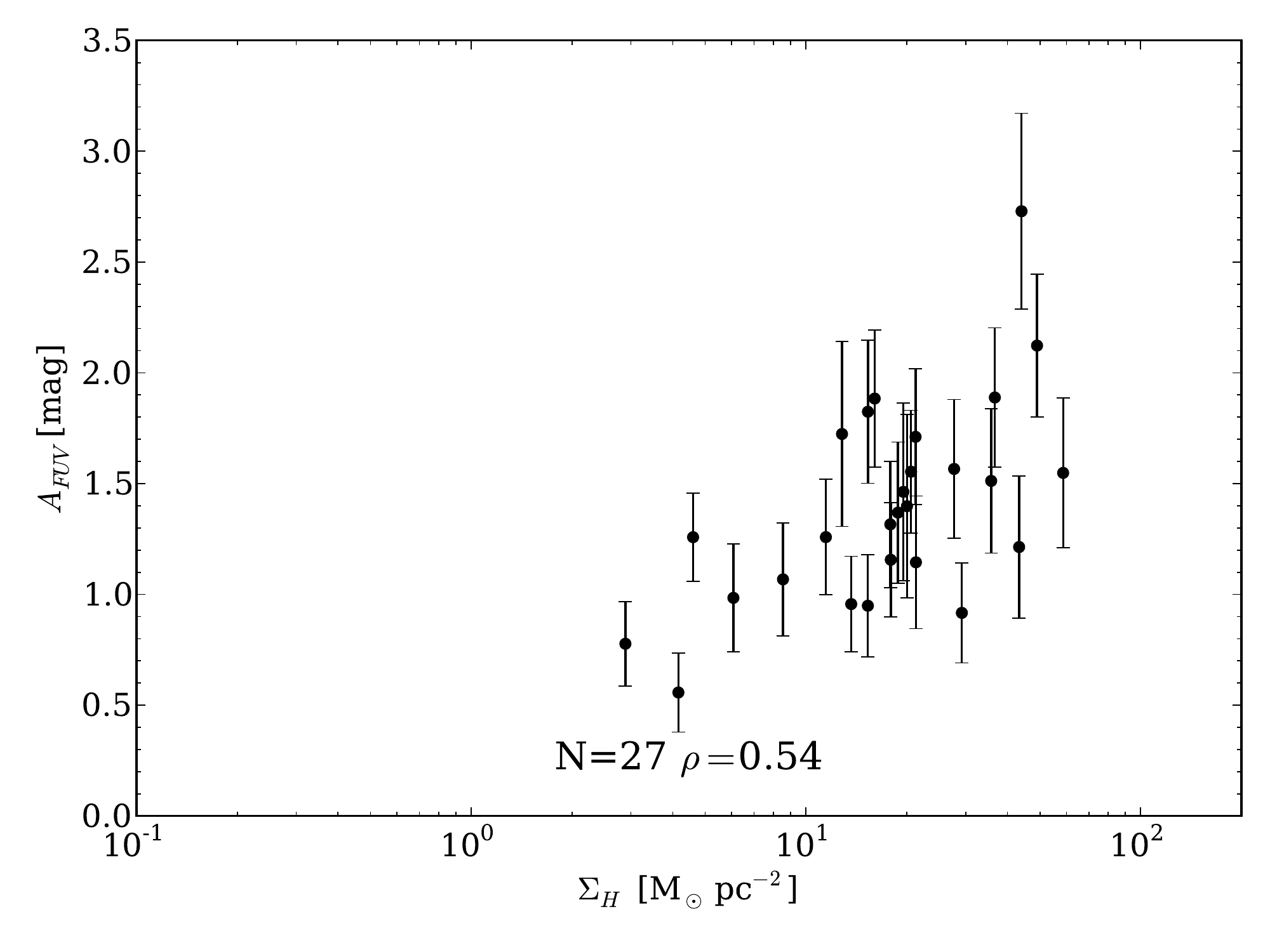}
\caption{Same as Fig.~\ref{fig:AFUV-H} for sample B.}
\label{fig:AFUV-H-int}
\end{figure}

A first inspection of the relation between the FUV attenuation and the gas surface density in sample B is shown in Fig.~\ref{fig:AFUV-H-int}.
There is a weak anti--correlation between the attenuation and the atomic gas mass surface density ($\rho=-0.33$). Conversely, there is a clear correlation with both the molecular ($\rho=0.64$) and the total gas content ($\rho=0.54$). This is actually due to the fact that some galaxies have a high molecular fraction. If attenuation is indeed linked to denser regions, then it is expected that galaxies with relatively little HI due to the high molecular fraction will be more attenuated.

\begin{figure}[!htbp]
\centering
\includegraphics[width=\columnwidth]{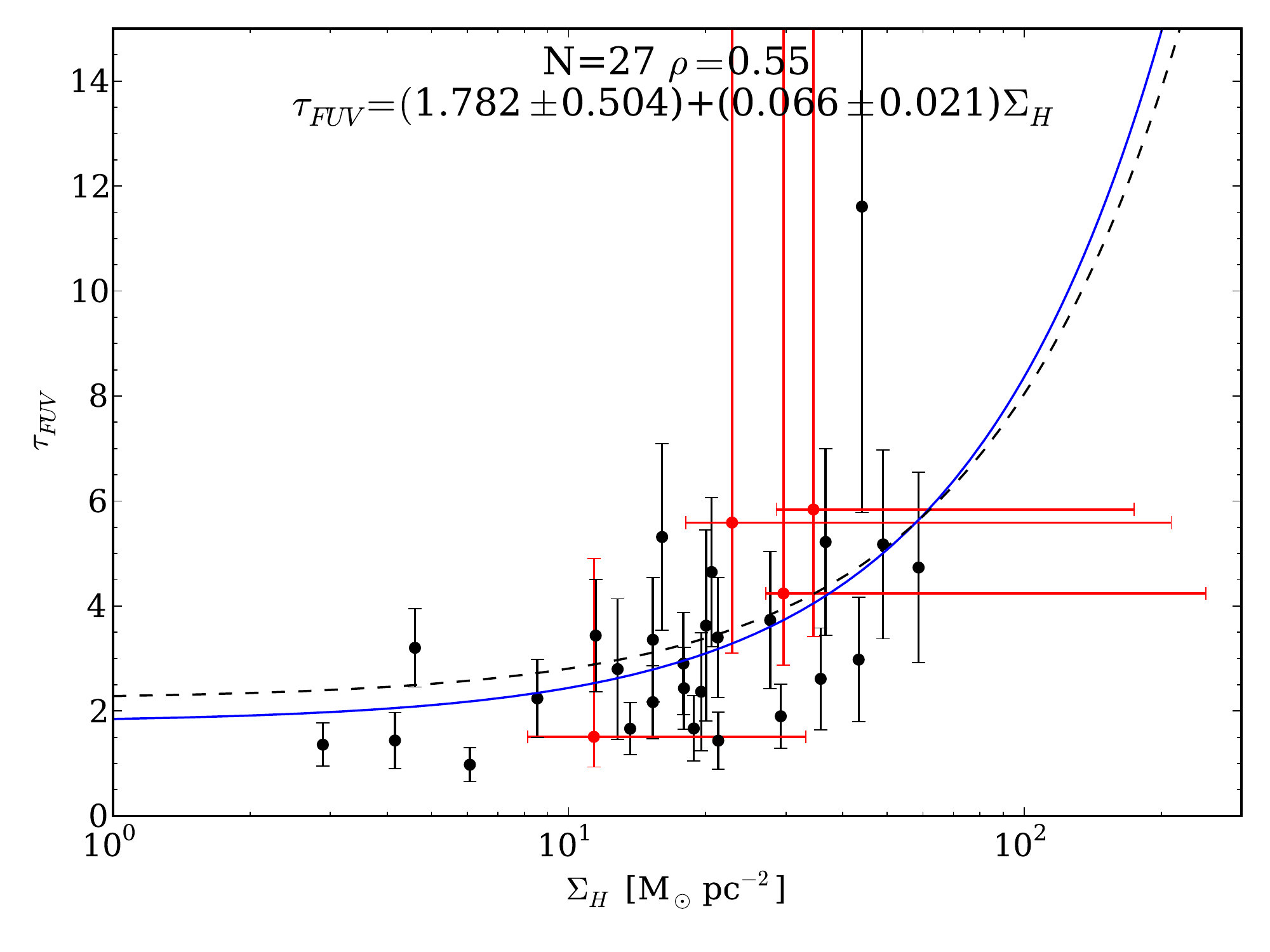}
\caption{Relation between the FUV face--on optical depth and the gas surface density for unresolved galaxies (black circles with error bars). The blue line corresponds to the best linear fit for the unresolved sample. The corresponding relation is indicated at the top of the figure, along with the number of elements and the Spearman correlation coefficient. The black dashed line corresponds to the relation for resolved galaxies when assuming a solar metallicity (Eq.~\ref{eqn:tau-sample-A}). The absence of metallicity estimates for a number of galaxies prevents us from using a metallicity--dependent $X_{CO}$ factor. The red dots represent the 4 galaxies in sample A, with the bars indicating the full dynamical range probed by each of them.}
\label{fig:NH-tau-int}
\end{figure}

In Fig.~\ref{fig:NH-tau-int} we show the relation between the face--on optical depth and the gas surface density.
In spite of the scatter and the limited size of the sample, for most of the galaxies the relation derived for resolved galaxies provides a reasonable estimate of the face--on optical depth. The relations derived for both samples prove to be very similar. This suggests that both resolved and unresolved galaxies follow relatively similar laws. Had they been markedly different, the derived relations would have shown obvious differences. This is especially important as sample B is made of only 27 galaxies and the scatter is large. A larger sample would be particularly useful to ascertain this result. Ultimately, the similarity between the relations suggests that the relations derived from resolved galaxies can be reasonably applied to unresolved galaxies.

To compare the dynamical range covered by resolved galaxies versus unresolved ones, we have computed the global face--on optical depth for each galaxy in sample A. To do so, 1) we correct each pixel for the FUV attenuation using the estimate from CIGALE, 2) we sum the attenuation corrected flux in each pixel, 3) we compute the global FUV attenuation of each galaxy comparing the summed corrected and uncorrected FUV fluxes, and finally 4) assuming a slab geometry we compute the face--on optical depth. The red dots in Fig.~\ref{fig:NH-tau-int} represent these integrated values and the bars around them the dynamical range covered by individual pixels within each of these galaxies. We see that our sample is well within the envelope described by sample B. Note that the mean gas surface densities for the 4 galaxies in sample A determined this way is much more secure than the values derived for sample B as they do not depend on assumptions on the distribution of the gas in galaxies.

\section{Practical methods to compute the attenuation\label{sec:recipe}}

In this section we present some methods to compute the attenuation at all wavelengths in galaxies from the relations presented above, at kpc scales for resolved galaxies and for unresolved galaxies as a whole. The computation of the attenuation can be broken into several individual steps which we describe in detail hereafter:

\begin{enumerate}
 \item The mean gas surface density must be computed for each element (either an unresolved galaxy or an individual subregion) in the sample, taking into account both the atomic and molecular gas.
 \begin{enumerate}
   \item If either component is missing, it can be compensated for by assuming a certain molecular fraction.
   \item In the case of unresolved galaxies, the computation of the gas surface density will be especially dependent on the actual distribution of the gas in the galaxy because this determines the area it should be averaged upon. We recommend averaging over 3 scale--lengths as we mentioned in Sect.~\ref{sssec:NH-sample-B}: $r_{HI}=3\times0.61\times r_{25}$ \citep{bigiel2012a}, and $r_{H_2}=3\times0.2\times r_{25}$ \citep{leroy2008a,lisenfeld2011a}, with $r_{25}$ the optical radius at 25~mag~arcsec$^{-2}$.
 \end{enumerate}
 \item From the gas surface density, the attenuation in the FUV band should be computed using one of the relations provided in this paper.
 \begin{enumerate}
  \item If an estimate of the metallicity is available, we strongly recommend the use of a relation explicitly dependent on the metallicity (Eq.~\ref{eqn:tau-sample-A-XCO-variable-metal}). This is important as the metallicity affects the gas--to--dust mass ratio which has a direct consequence on the optical depth. We would like to stress as already mentioned in Sect.~\ref{ssec:metal} that for consistency, the oxygen abundance estimator of \cite{zaritsky1994a} has been adopted throughout this paper. It is known for yielding rather high values compared to other estimators. In case estimates of the metallicity have been carried out with another estimator, we recommend either to recompute it with the \cite{zaritsky1994a} estimator, or to convert it with the formulas published in \cite{kewley2008a}.
  \item If no estimate of the metallicity is available, we recommend the use of a relation parameterised only on the gas surface density (Eq.~\ref{eqn:tau-sample-A-XCO-variable}). We caution that these relations have been derived on local galaxies that are relatively metal rich on average, $8.8\lesssim12+\log O/H\lesssim9.4$ with the \cite{zaritsky1994a} estimator, which could lead to large biases in the estimate of the optical depth if applied without correction to lower metallicity objects.
 \end{enumerate}
 \item From the FUV face--on optical depth, the FUV attenuation can be computed assuming a given geometry (Eq.~\ref{eqn:slab} for a slab geometry).
  \begin{enumerate}
   \item As the relations we have derived between the gas surface density and the face--on optical depth depend on the geometry we initially assumed, it is of utmost importance to adopt the same geometry\footnote{If another geometry is adopted, please contact the lead author of this paper to obtain updated relations applicable to this geometry.}.
   \item For consistency, we recommend the use the value of the albedo published by \cite{draine2003b}.
  \end{enumerate}
 \item Finally, from the attenuation in the FUV band, it is possible to compute the attenuation at any wavelength assuming an attenuation curve. In case of a starburst galaxy, the starburst attenuation curve \citep{calzetti1994a,calzetti2000a} can be applied. For high redshift galaxies, the recently derived attenuation curve of \cite{buat2011b} can also be used.
\end{enumerate}

\section{Comparison with relations from the literature\label{sec:comp-literature}}

In this section we examine how the relations we have obtained compare to some of the relations published in the literature.

In the Milky Way it is possible to directly measure the quantity of gas towards individual objects and the extinction undergone by these objects. In this case, relations between the optical depth and the gas surface density take the form of a linear relation between these 2 quantities, and have an explicit dependency on the attenuation curve and the gas--to--dust mass ratio. \cite{guiderdoni1987a} expressed this as:

\begin{equation}
 \tau_\lambda=\frac{1}{1.086}\times\frac{A_\lambda}{A_V}\times\frac{A_V}{E(B-V)}\times\frac{E(B-V)}{N_H}\times \left(\frac{Z}{Z_\odot}\right)^s\times\Sigma_H,\label{eqn:guiderdoni}
\end{equation}
with $\Sigma_H$ the face--on gas surface density, $E(B-V)/N_H$ the quantification of the reddening, which is generally calibrated in the Milky Way with a screen geometry, $Z$ the metallicity, $Z_\odot$ the solar metallicity, and $s$ a wavelength--dependent parameter to take into account the variation of the shape and the amplitude of the reddening with the metallicity. Following \cite{guiderdoni1987a}, we take $s=1.35$ in the FUV band. If we assume an extinction curve with $R_V=3.1$ and $E(B-V)/N_H=1.81\times10^{-2}$~mag~(M$_\odot$~pc$^{-2}$)$^{-1}$ from \cite{guver2009a}, in the $V$ band and for a solar metallicity it reduces to:
\begin{equation}
 \tau_V=0.052\times \Sigma_H,
\end{equation}
which is close to the widely used relation of \cite{guiderdoni1987a} and \cite{devriendt1999a}:
\begin{equation}
 \tau_V=0.059\times \Sigma_H.
\end{equation}
We have to note that at very high column density, $R_V$ may be larger as can be seen in dense molecular clouds. For instance in the Perseus cloud, \cite{foster2013a} measured an increase from $R_V=3$ for $A_V=2$, to $R_V=5$ for $A_V=10$. At such attenuations, no significant quantity of UV photon could escape. However, on the scale of the current study, it should concern only a small fraction of UV photons. Otherwise it should be noted that CIGALE fits would fail catastrophically, which is not the case. We therefore think that the adopted value of $R_V=3.1$ is reasonable under the present conditions.

Using such relations directly has several drawbacks to trace star formation in galaxies. First of all they are normalised to the V band which is a poor tracer of star formation. Computing the optical depth in the FUV, which is critical to compute attenuated and non--attenuated SFR, requires not only to assume an extinction curve (typically Milky Way, Large Magellanic Cloud, or Small Magellanic Cloud depending on the metallicity of the object), but also to take into account the mixing of different populations which are differently affected by the attenuation. Also, these relations are calibrated on individual lines of sight in the Milky Way or in the Magellanic Clouds assuming a screen geometry, but not on large scale regions or even unresolved galaxies that have more complex stars--dust geometries.

\begin{figure}[!htbp]
\centering
\includegraphics[width=\columnwidth]{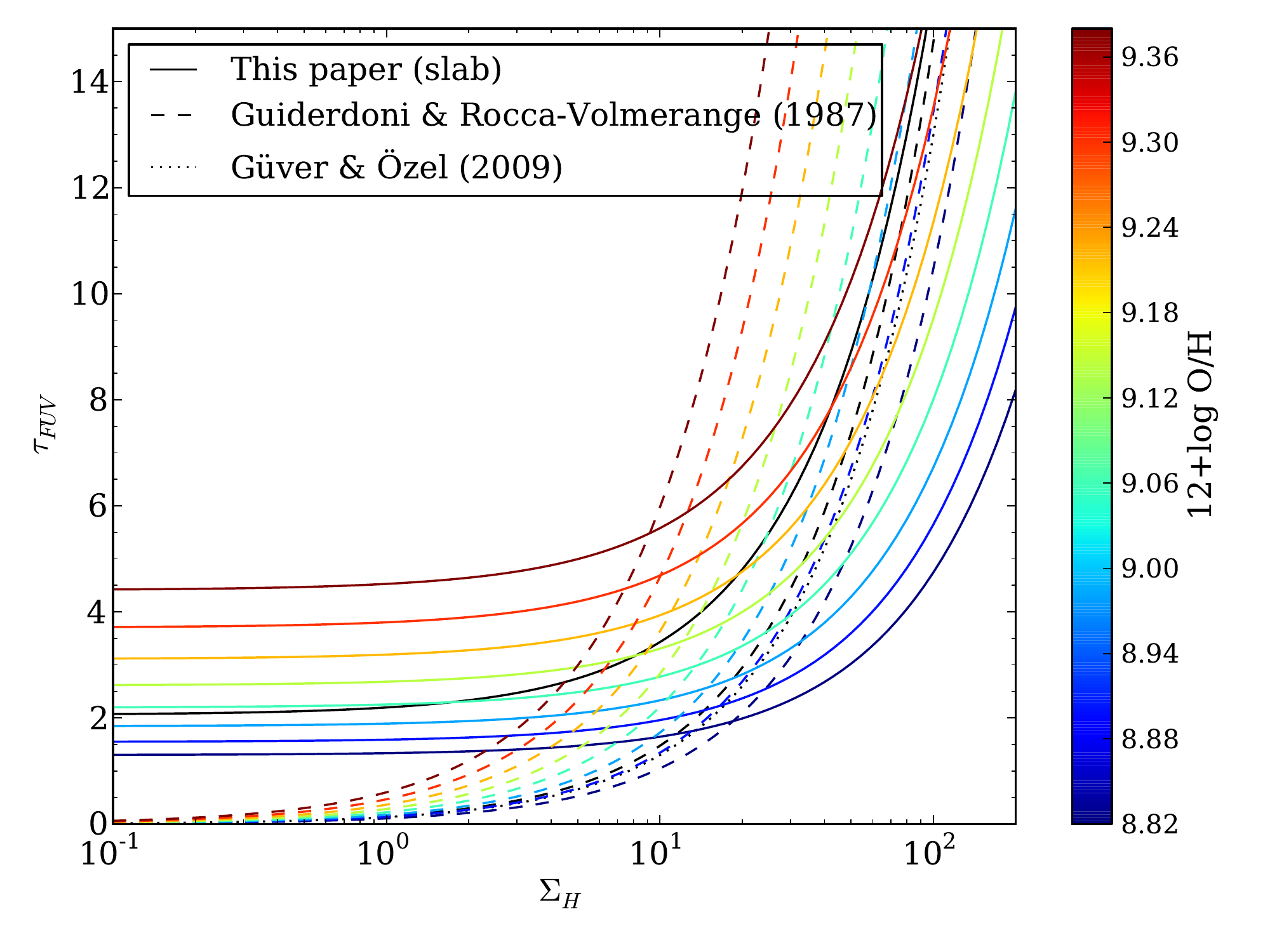}
\caption{Comparison of the relations between the face--on optical depth and the gas surface density obtained in this paper (solid lines, Eq.~\ref{eqn:tau-sample-A-XCO-variable} and \ref{eqn:tau-sample-A-XCO-variable-metal}), with the relation from \cite{guiderdoni1987a} (dotted line), and with the relation from \cite{guver2009a} (dash--dot line). The coloured lines show metallicity--dependent relations according to the oxygen abundance scale show on the right. For clarity, we have plotted the \cite{guver2009a} relation only in the case of a solar metallicity.}
\label{fig:comp-rel}
\end{figure}

To compare these relations to the ones we have obtained at kpc scales, we assume an extinction curve with $\tau_{FUV}/\tau_{V}=2.5$ which approximately corresponds to a Milky Way--like extinction curve. Steeper extinction curves such as the LMC or the SMC would yield larger FUV optical depths. We present the results in Fig.~\ref{fig:comp-rel}.
There are several striking differences that must be noted. First of all, as discussed earlier at low gas surface density, the relations derived in this paper are asymptotically constant in the typical range $1.5\lesssim\tau_{FUV}\lesssim4.5$. The relations from \cite{guiderdoni1987a} or \cite{guver2009a} on the contrary fall linearly to $\tau_{FUV}=0$ as $\Sigma_H\rightarrow0$. This comes from the fact that these relations are calibrated along a set of individual line--of--sights in the Local Group, with an actual geometry that is close to the assumed screen geometry, without any mixing between different regions. Conversely the relations derived in this paper are based on large regions in galaxies and naturally, undergo a strong mixing between high and low gas surface density regions that have a different face--on optical depth (Sect.~\ref{ssec:sigma-optdepth}). The difference at low column density shows in particular the expected effect of mixing compared to classically used relations. In any case, at zero metallicity $\tau_{FUV}=0$ for all relations. The second feature we can note is that as the gas surface density increases, the metallicity--dependent relation of \cite{guiderdoni1987a} yields, for the same metallicity, higher optical depths. Once again, the empirical relations obtained in this paper more naturally take into account the complexity of the galaxies, which we see makes dust less efficient at attenuating stellar radiation than the relation of \cite{guiderdoni1987a}.

\section{Impact on galaxy formation and evolution\label{sec:impact}}

The study of the LFs is one of the main tools to understand galaxy formation and evolution, either through observations or models. It is therefore important to understand the exact impact of the new relations we have derived in this paper on LFs output from SAMs. As mentioned in Sect.~\ref{sec:introduction}, UV LF obtained from hydrodynamical simulations can be sensitive to the way the attenuation is computed \citep{finlator2006a,fontanot2009a} To understand the impact of different gas--attenuation prescriptions on models, we examine here how such relations would impact UV and IR LFs derived from cosmological simulations, as these are among the main targets to apply such relations. To do so we exploit the catalogue of galaxies of \cite{guo2011a} which is based on the Millennium \citep{springel2005a} and Millennium--II \citep{boylan2009a} simulations.

\subsection{Sample and observational parameters\label{ssec:simu-params}}

Before computing LFs we need 1) to select a sample of galaxies from the simulated catalogue, and 2) derive their observational parameters such as the mean gas surface density or the UV and IR luminosities. We have to stress that the objective is to test the relative effect of different attenuation--gas surface density relations but not to constitute highly accurate LFs in an absolute way, therefore the UV and IR luminosities will be built from relations widely used in the literature, that we will describe hereafter. The reproducibility of local LFs is discussed in appendix~\ref{sec:reproduce-LF}.

As a first step, we select all star--forming galaxies ($\mathrm{SFR>0}$) at $z=0$ from the \cite{guo2011a} catalogue, which yields a sample of 708005 galaxies out of a full catalogue of 13191859 galaxies.

The \cite{guo2011a} catalogue provides for each galaxy the cold gas mass in units of $10^{10}$ M$_\odot$/$h$, and the size of the gas disk in units of Mpc/$h$, which is defined as 3 times the scale length, assuming an exponential disk, which is similar to the way we have determined the disk sizes for sample B. We assume $h=0.73$ to follow the parameters used for the simulation \citep{boylan2009a}. We compute the mean gas surface density simply by dividing the cold gas mass by the disk area.

To compute the attenuation in the FUV band, we use the relations between the face--on optical depth and the gas surface density determined for sample A. The inclination of each galaxy is drawn randomly from a uniform distribution between 0$^\circ$ and 90$^\circ$. As the catalogue of \cite{guo2011a} does not directly provide an estimate of the metallicity, we compute the oxygen abundance using the stellar mass--metallicity relation of \cite{kewley2008a} adapted for the \cite{zaritsky1994a} scale.

Ideally, the FUV luminosity could be determined from the SFH of each galaxy. Unfortunately, the sampling in redshift is too sparse to trace accurately the variations of the SFH over the sensitivity timescale of the FUV. As a consequence, to compute the FUV luminosity, we assume that the FUV contamination from previous star formation episodes is small and therefore that it traces the current star formation. Following \cite{murphy2011a}, if we assume that galaxies have been forming stars at a constant rate over the last 100~Myr and follow a \cite{kroupa2001a} IMF, we find:
\begin{equation}
 L(FUV)=5.89\times10^{9}\times \mathrm{SFR},
\end{equation}
with $L(FUV)$ the FUV luminosity in $L_\odot$, and the SFR in M$_\odot$~yr$^{-1}$. The use of another relation, such as the one of \cite{salim2007a} for instance, would yield slightly different FUV luminosities but would not affect the results. The IR luminosity can be determined from a relation between the UV--to--IR luminosity ratio (IRX), and the attenuation. Several of such relations have been published based on local star--forming galaxies. They yield very similar results \citep{boquien2012a}. We adopt the formula determined by \cite{buat2005a}:

\begin{equation}
 A_{FUV}=-0.0333x^3+0.3522x^2+1.1960x+0.4967,
\end{equation}
with $x\equiv\log L(IR)/L(FUV)$. To retrieve the value of $L(IR)$, we simply invert this equation numerically. Galaxies being mostly optically thin in the mid-- and far--IR, the IR emission does not depend on the inclination of the galaxy. We therefore determine the value of $L(IR)$ for an $A_{FUV}$ calculated for a face--on observer.

\subsection{Impact of relations between the face--on optical depth and the gas surface density}

\begin{figure}[!htbp]
\centering
\includegraphics[width=\columnwidth]{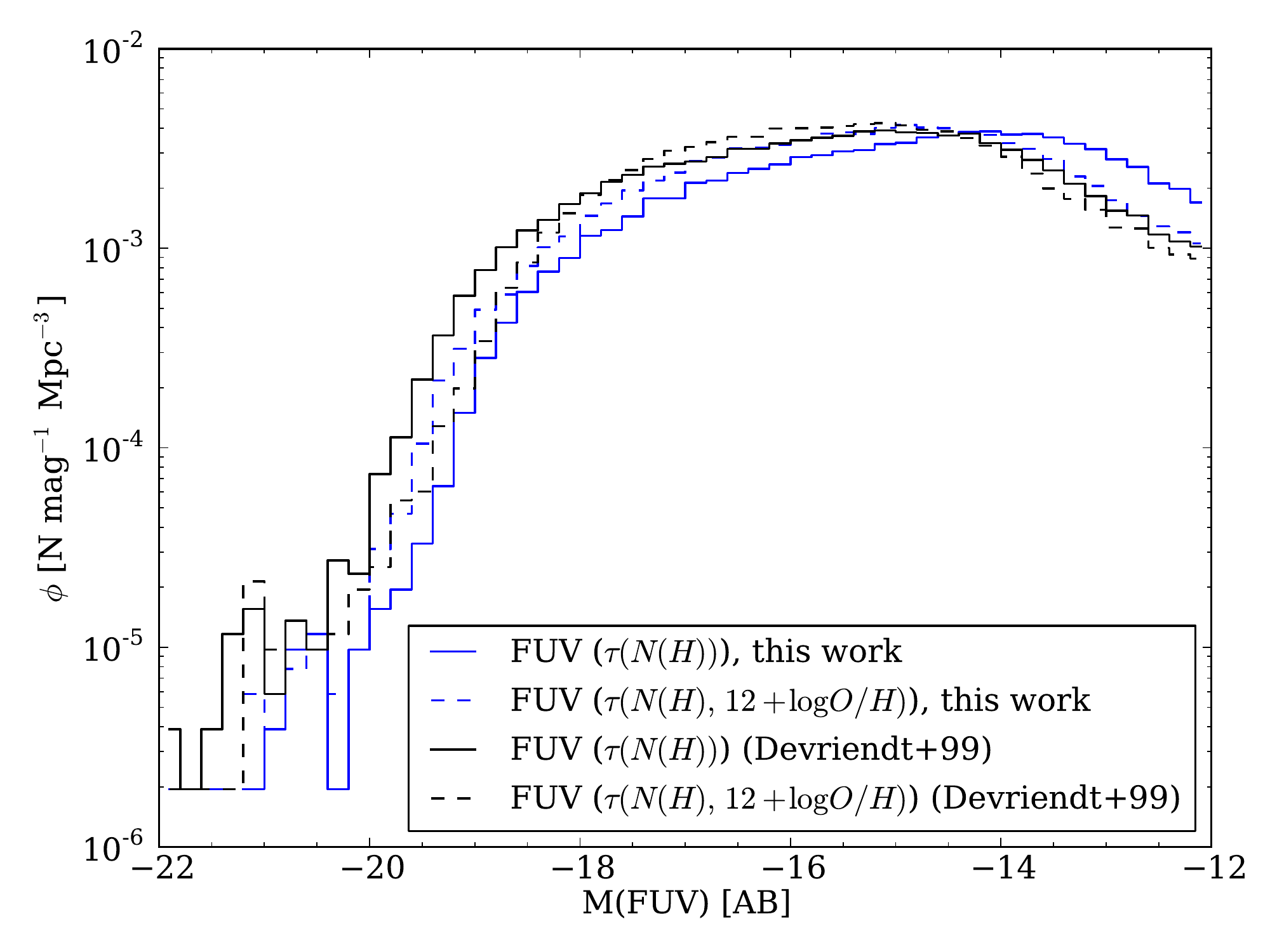}
\includegraphics[width=\columnwidth]{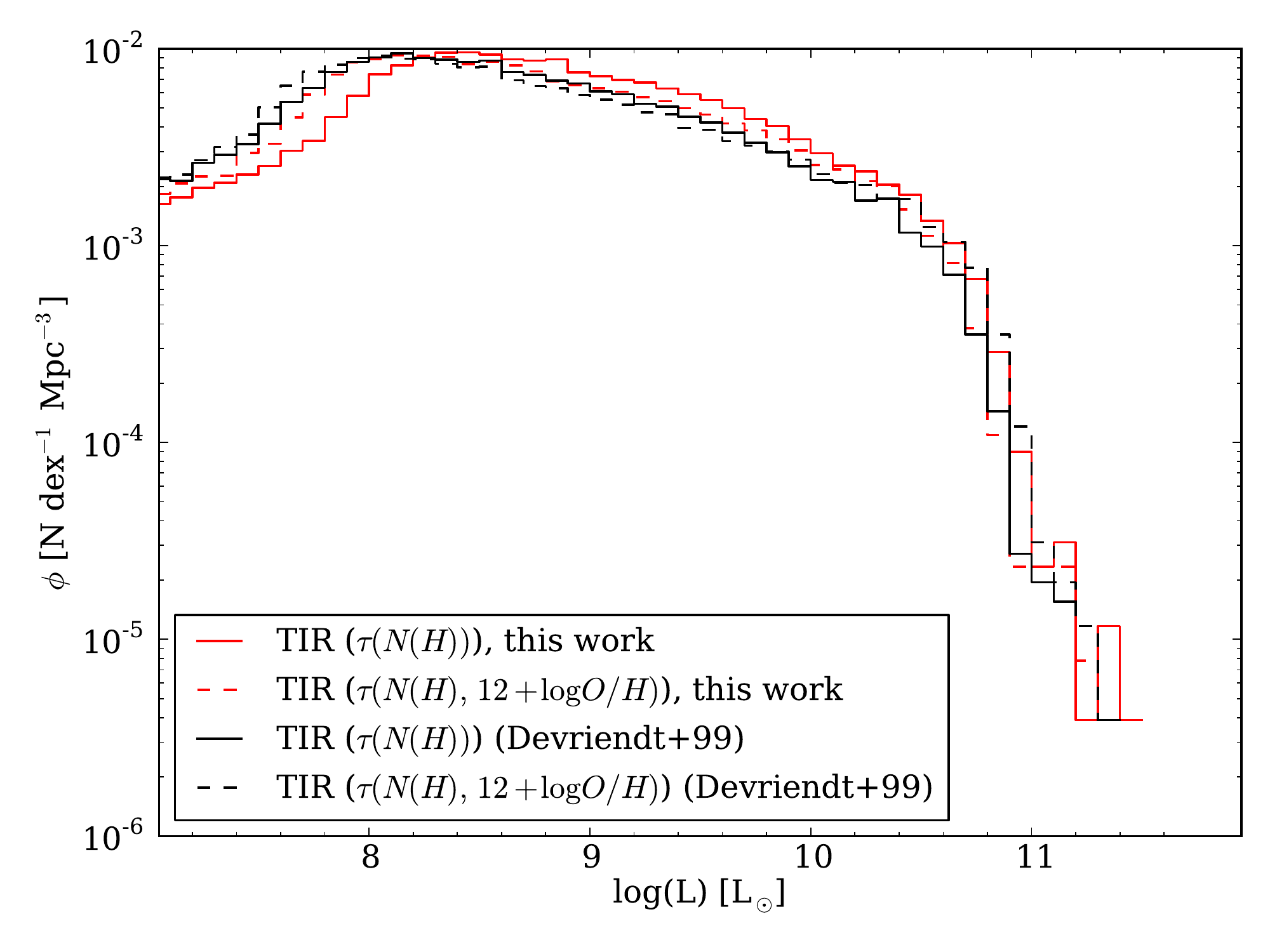}
\caption{Top: FUV LFs derived following the method outlined in Sect.~\ref{ssec:simu-params}. The blacks lines represent the LFs when using the gas surface density versus attenuation relation of \cite{guiderdoni1987a,devriendt1999a} whereas the blue lines represent the LFs when using the relations presented in the current paper, assuming a slab geometry. For both relations, the solid line represent a relation parameterised only on the gas surface density whereas the dashed lines represent relations paramaterised both on the gas mass surface density and the oxygen abundance. Bottom: IR LF, with the red lines representing the use of the relations derived in this paper.}
\label{fig:lumfunc}
\end{figure}

To evaluate the impact of different relations between the face--on optical depth and the gas surface density, we compare in Fig.~\ref{fig:lumfunc} the FUV and IR LFs derived for the relation we derived for sample A (Eq.~\ref{eqn:tau-sample-A-XCO-variable} and \ref{eqn:tau-sample-A-XCO-variable-metal}) to the one presented in \cite{guiderdoni1987a,devriendt1999a}. Note that in this section we select only galaxies that have $8.5<\log M_\star<11$ because the stellar mass--metallicity relation of \cite{kewley2008a} is valid only in this range, reducing the catalogue to 54716 objects. This does not impact our results as we only examine here the relative variations on LFs induced by different relations.

We note several striking features. First of all, the use of a metallicity--dependent estimator can make a large difference. For instance when selecting bright galaxies that have $\mathrm{M(FUV)}<-18$, taking the metallicity into account systematically lowers the counts in each bin by $-0.19\pm0.22$~dex on average for the relation of \cite{guiderdoni1987a,devriendt1999a} and increases by $0.21\pm0.14$~dex for our relations (Eq.~\ref{eqn:tau-sample-A-XCO-variable} and \ref{eqn:tau-sample-A-XCO-variable-metal}). For fainter galaxies with $\mathrm{M(FUV)}\geq-18$, even if the number counts are left nearly unchanged, respectively $-0.00\pm0.01$~dex and $-0.00\pm0.08$~dex, there is a change of behaviour around $\mathrm{M(FUV)=-14.5}$. Fainter than this threshold the number counts are lower with the metallicity--dependent relation and higher when brighter than this threshold. A difference in the number counts between metallicity--independent and metallicity--dependent relations is seen for the IR LFs. This shows the importance of taking the metallicity explicitly into account, especially when studying the bright end of the LFs. When considering different metallicity--dependent relations between the gas surface density and the attenuation, we see that there are again significant differences that affect the overall LFs. The peak difference for UV bright galaxies reaches nearly 0.6~dex. Conversely in the IR, the difference in number counts remains small.

To ensure that these results are not affected by the 0.2~mag width of the bins (smaller bins are more affected by random sampling effects), we have also examined the FUV LFs using a bin width of 0.5~mag. This size is similar to that of the observed FUV LF published by \cite{wyder2005a}. We find that the previous results remain unaffected by an increase of the FUV bin width to 0.5~mag.

These results are in line with those obtained by \cite{fontanot2009a} and show the significant impact the relation between the gas surface density and the face--on optical depth can have on the predicted LFs with peak differences in counts reaching about 1 dex.

\section{Summary and conclusions\label{sec:conclusion}}

In this paper we have studied a sample of 4 resolved galaxies, and a sample of 27 unresolved galaxies in order to determine new but still simple relations between the gas surface density and the face--on optical depth at kpc scales, while also taking into account the effect of the metallicity.

To do so, we have determined the gas mass surface density from HI and CO observations. In parallel, we have determined the face--on optical depth in two steps.
\begin{enumerate}
 \item We have determined the attenuation for each pixel in the resolved galaxies and for each unresolved galaxy using the CIGALE SED fitting code, which is based on an energy balance principle between the attenuation in the UV--optical domain and the emission of the dust in the FIR.
 \item From the attenuation, we have determined the face--on optical depth assuming a slab geometry.
\end{enumerate}
We have then determined new relations between the gas surface density, and the face--on optical depth in resolved galaxies, taking into account the metallicity indirectly by way of only metallicity--dependent $X_{CO}$ and gas--to--dust mass ratios, or also adding an explicit parameterisation. We have found that for galaxies with an inclination no larger than $50^\circ$, the assumptions made on the geometry have little impact on the derived attenuation. For larger inclinations, deviations over 30\% can be found, depending on the geometry. We can summarise our main findings as following:
\begin{enumerate}
 \item The attenuation seems to primarily take place in regions that have a high molecular fraction, which is consistent with the fact that star--forming regions are more attenuated than more quiescent ones.
 \item The attenuation and the gas column density are affected differently by the averaging between different regions within a given resolution element in galaxies. This mixing effect strongly affects the relation between the face--on optical depth and the gas surface density, in particular in regions with a low surface density. This shows that oft--used relations calibrated in the Milky Way are not adapted for use on large areas in galaxies or even entire galaxies.
 \item Unsurprisingly the impact of the metallicity is very significant, even for nearby spiral galaxies whose metallicity range is not extreme. As such it should be taken into account whenever possible.
 \item Relations between the gas surface density and the face--on optical depth derived on resolved galaxies are able to reproduce the trend for unresolved galaxies despite the scale difference of these objects and the intrinsic difficulty to measure reliably the total gas surface density in unresolved galaxies.
 \item There is both a significant scatter around the best fits and galaxy--to--galaxy variations. This shows that there are additional physical parameters governing the relation between the face--on optical depth and the gas surface density.
 \item The new relations have a visible impact on LFs from the Millennium--II simulation when compared to the oft--used relations of \cite{guiderdoni1987a,devriendt1999a}. This is especially visible on the bright end, and could therefore have a significant impact on the study of galaxy evolution through the use of large cosmological simulations combined with SAMs.
\end{enumerate}

Finally, to compute the attenuation from the gas surface density in practice, we recommend to apply the method laid out in Sect.~\ref{sec:recipe}. We remind that if another geometry is assumed, the lead authors of this paper should be contacted in order to obtain relations adapted to this geometry.

\begin{acknowledgements}
{\it Herschel} is an ESA space observatory with science instruments provided by European-led Principal Investigator consortia and with important participation from NASA.

SPIRE has been developed by a consortium of institutes led by Cardiff University (UK) and including Univ. Lethbridge (Canada); NAOC (China); CEA, LAM (France); IFSI, Univ. Padua (Italy); IAC (Spain); Stockholm Observatory (Sweden); Imperial College London, RAL, UCL-MSSL, UKATC, Univ. Sussex (UK); and Caltech, JPL, NHSC, Univ. Colorado (USA). This development has been supported by national funding agencies: CSA (Canada); NAOC (China); CEA, CNES, CNRS (France); ASI (Italy); MCINN (Spain); SNSB (Sweden); STFC (UK); and NASA (USA).

The Millennium--II Simulation database used in this paper and the web application providing online access to them were constructed as part of the activities of the German Astrophysical Virtual Observatory.

This research has made use of the NASA/IPAC Extragalactic Database (NED) which is operated by the Jet Propulsion Laboratory, California Institute of Technology, under contract with the National Aeronautics and Space Administration. 

The research leading to these results has received funding from the European Community's Seventh Framework Programme (/FP7/2007-2013/) under grant agreement No 229517.
\end{acknowledgements}

\bibliographystyle{aa}
\bibliography{article}

\appendix

\section{The combination of FUV and dust emission as a proxy to the attenuation\label{sec:afuv-measured-with-irx}}
\begin{figure}[!htbp]
\centering
\includegraphics[width=\columnwidth]{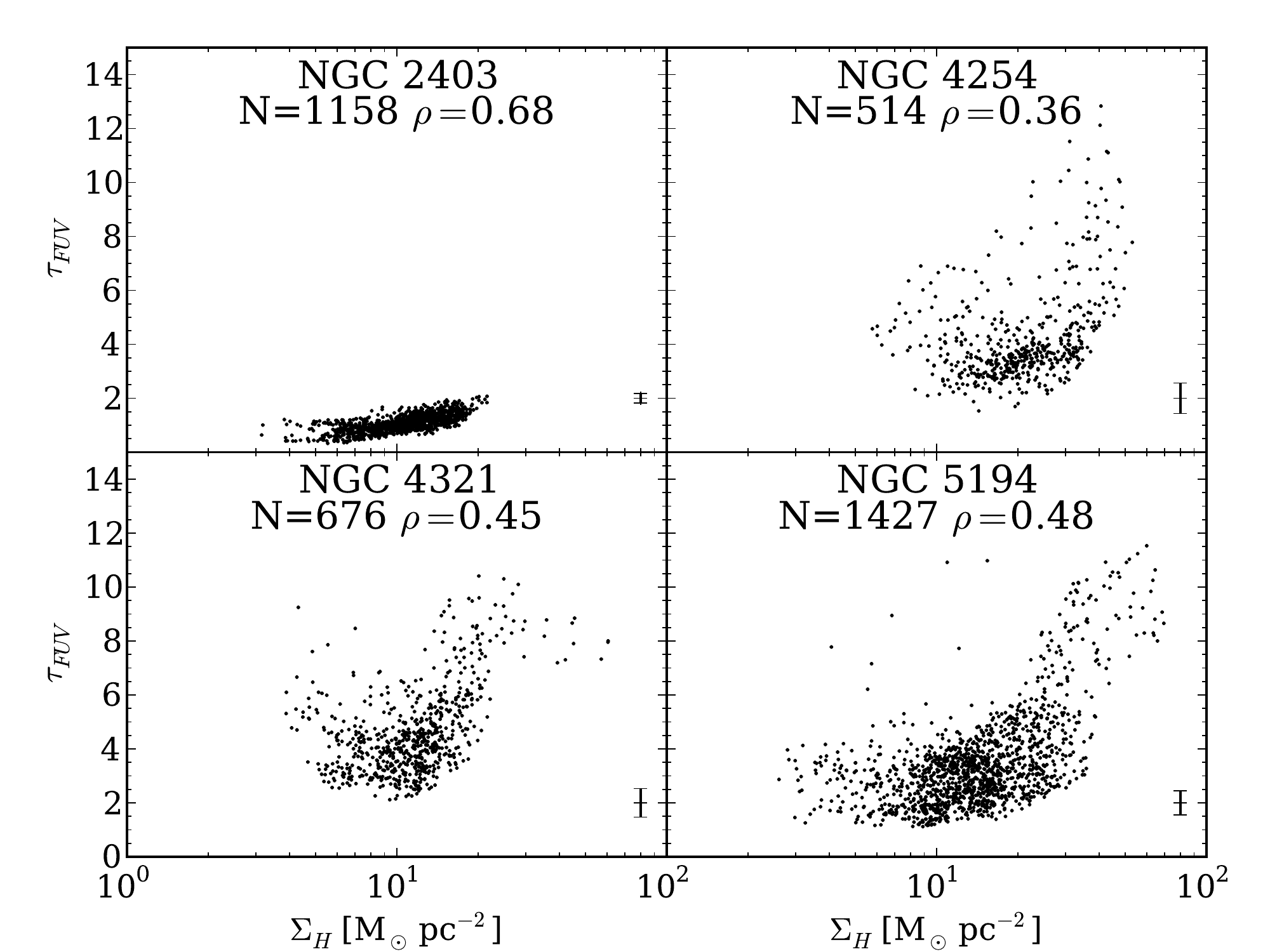}
\caption{Face--on optical depth in the FUV band determined from the FUV and dust emission \citep{hao2011a} versus the atomic total gas mass surface density in units of M$_\odot$~pc$^{-2}$.}
\label{fig:AFUV-H-att-hao}
\end{figure}

While we have shown in Sect.~\ref{sssec:phys-params} that the determination of the FUV attenuation with CIGALE is reliable, we examine here whether and how the results we have obtained are affected by using an alternative method to compute the attenuation and therefore the face--on optical depth. Several relations exist in the literature to quantify the attenuation from the FUV and the emission of the dust. These relations have generally been derived from the observation of unresolved star--forming galaxies \citep{cortese2008a,buat2011a,hao2011a}. \cite{boquien2012a} have also derived such a relation but on subregions in star--forming galaxies using CIGALE. Even though it yields attenuations close to that of previous relations, we do not consider it here as it has also been determined using CIGALE. Among the relations from the literature, \cite{boquien2012a} found that the one provided by \cite{hao2011a} yields the largest difference with the attenuation determined with CIGALE in resolved galaxies: $\left<\Delta A_{FUV}\right>=0.156\pm0.056$~mag. We therefore adopt this relation as it will be indicative of the maximum deviation that could be induced by such an attenuation tracer compared to SED modelling. In Fig.~\ref{fig:AFUV-H-att-hao}, we present the relations between the gas surface density and the FUV face--on optical depth determined from the formula published in \cite{hao2011a}, assuming a slab geometry.
We see that using an alternative method to determine the attenuation leaves the relations qualitatively unchanged. After verification, the relations provided by \cite{cortese2008a} and \cite{buat2011a} show little qualitative difference compared to when using the relation of \cite{hao2011a}. However the lower attenuations yielded by the relation of \cite{hao2011a} naturally affect the quantitative relations between the face--on optical depth and the gas surface density:
\begin{equation}
 \tau_{FUV}=\left(2.353\pm0.127\right)+\left(0.109\pm0.009\right)\times\Sigma_H,
\end{equation}
and 
\begin{equation}
\begin{split}
 \tau_{FUV}=\left[\left(1.859\pm0.095\right)+\left(0.029\pm0.004\right)\times \Sigma_H\right]\\
\times10^{\left(1.168\pm0.085\right)\left(\left[12+\log O/H\right]-9.00\right)}.
\end{split}
\end{equation}
These relations yield smaller optical depths. Averaging over the entire sample, we obtain $\left<\Delta\tau_{FUV}/\tau_{FUV}\right>=0.02\pm0.04$ when taking only the gas surface density into account, and $\left<\Delta\tau_{FUV}/\tau_{FUV}\right>=0.10\pm0.07$ and when taking into account both the gas surface density and the metallicity. The amplitude of these shifts is not larger than the error bars on the optical depths deduced from the uncertainties on the FUV attenuation provided by CIGALE. Therefore the use of different methods to determine the attenuation should only have a limited impact on the derived optical depth, with offsets of the order of 10\%.

\section{Dust emission as a proxy to the gas surface density\label{sec:gas-measured-with-dust}}
If the 21~cm emission is a direct tracer of the atomic gas, as mentioned earlier, the molecular gas is generally traced indirectly through CO emission. An alternative way to measure the total gas content in a galaxy is through the emission of the dust. Indeed, as gas and dust are intimately intertwined, assuming a given gas--to--dust mass ratio, one can retrieve the gas mass from the dust mass. Both methods have different biases. Whether the gas mass determination method affects our results or not is important to estimate their reliability.

\begin{figure}[!htbp]
\centering
\includegraphics[width=\columnwidth]{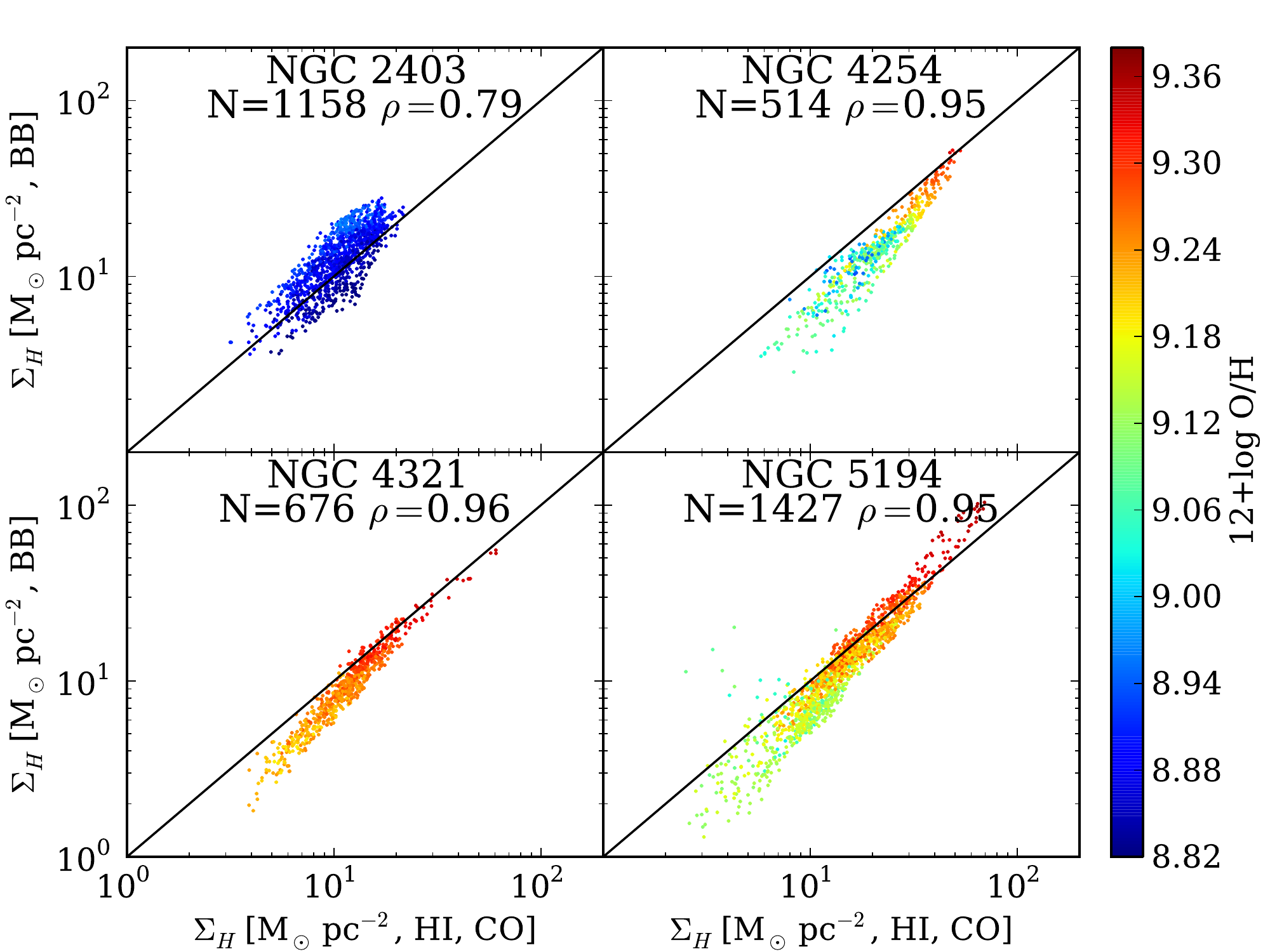}
\caption{Total gas surface density derived from a modified black body (Eq.~\ref{eqn:BB}) fit of the FIR data from 70~$\mu$m to 350~$\mu$m versus the total gas surface density derived from the combination of HI and CO observations. The colour of each dot indicates the metallicity indicated by the scale on the right. The black lines show the 1--to--1 relation between these quantities.}
\label{fig:comp-mass}
\end{figure}

One common method to determine the dust mass is to fit the FIR emission with a modified black body \citep[e.g.][ and many others]{boselli2002a,james2002a,eales2010a,eales2012a}. It is described in Eq.~\ref{eqn:BB}:
\begin{equation}
 f_\nu=\frac{1}{4\pi D^2}\times\frac{8\pi h\nu^3}{c^2}\times\frac{1}{\exp\left(h\nu/kT\right)-1}\times\kappa\left(\frac{\nu}{\nu_\kappa}\right)^\beta\times M_{dust},\label{eqn:BB}
\end{equation}
with $f_\nu$ the flux density, $D$ the distance of the object, $h$ the Planck constant, $\nu$ the frequency, $c$ the speed of light in vacuum, $k$ the Boltzmann constant, $T$ the temperature of the dust, $\kappa$ the dust emissivity taken at frequency $\nu_\kappa$, $\beta$ the emissivity index, and $M_{dust}$ the dust mass. Following \cite{draine2003a} we adopt $\kappa=0.192$~m$^2$~kg$^{-1}$ at $\nu_\kappa=857$~GHz (350~$\mu$m). The value of $\beta$ is the subject of an intense debate in the literature. A value comprised between 1.5 and 2.0 is adequate for star--forming galaxies. Even though $\beta$ can vary from region to region \citep{smith2012a}, \cite{boselli2012a} found that $\beta=1.5$ best reproduces the FIR properties of the star--forming galaxies in the HRS sample. We therefore adopt their value. Note however that this emissivity has been determined from a dust model that assumes $\beta=2$ rather than $\beta=1.5$, which may induce a bias in the dust mass. We determine the dust mass by fitting Eq.~\ref{eqn:BB} to the data at 70~$\mu$m, 250~$\mu$m, and 350~$\mu$m, using the {\sc curve\_fit} function in the {\sc scipy} library \citep{scipy}, which is based on a Levenberg--Marquardt algorithm. This combination of bands allows us to obtain a good handle on both the dust temperature because these bands straddle the FIR emission peak, and on the dust mass as the 350~$\mu$m band does not show a strong dependence upon temperature variations. In the case of NGC~2403, the 70~$\mu$m emission comes from a distinctly different dust component rather than the dust emitting at longer wavelength. As a consequence, the dust mass within this galaxy will be underestimated. In Fig.~\ref{fig:comp-mass}, we compare the gas mass estimated from the dust mass, versus the one derived from HI and CO emission. In both cases we use the metallicity--dependent conversion factors of \cite{boselli2002a} which assumes a gas--to--dust mass ratio of 160 for $12+\log O/H=8.91$.
We see that there is a tight correlation between the total gas surface density derived with a modified black body and from the combination of HI and CO observation as shown by the high value of the Spearman correlation coefficient for NGC~2403 ($\rho=0.79$), NGC~4254 ($\rho=0.95$), NGC~4321 ($\rho=0.96$), and NGC~5194 ($\rho=0.95$). We observe that the surface densities obtained using either method are similar even though there are some small offsets, generally towards lower masses except for NGC~2403. However they are small compared to the full dynamical range. These relations will be discussed in detail in Boquien et al. (2013, in prep.).

\begin{figure}[!htbp]
\centering
\includegraphics[width=\columnwidth]{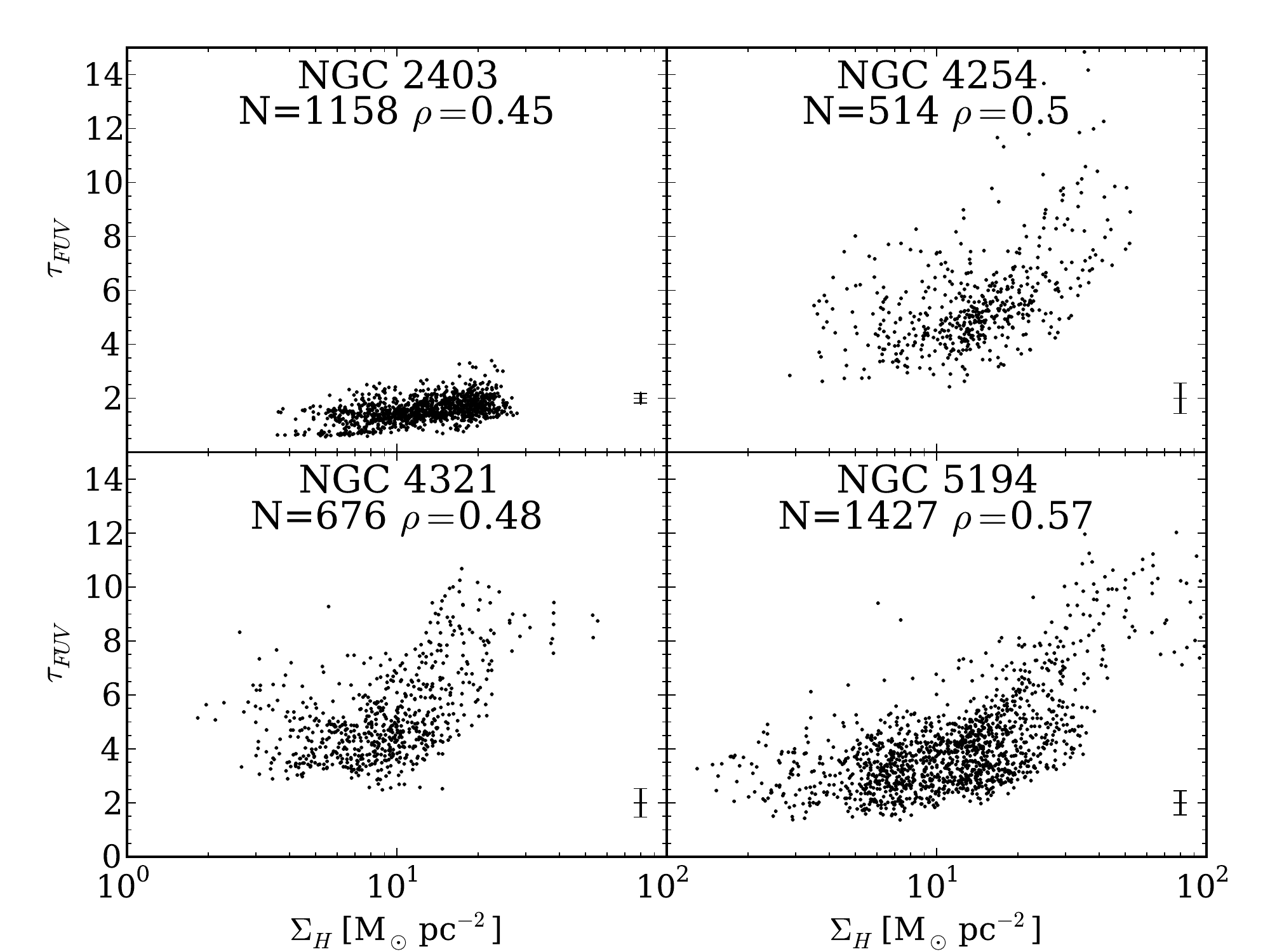}
\caption{Face--on optical depth in the FUV band versus total gas surface density determined by fitting the FIR emission with a modified black body as described in Eq.~\ref{eqn:BB}.}
\label{fig:NH-dust}
\end{figure}

To verify whether determining the total gas surface density affects the results, we plot in Fig.~\ref{fig:NH-dust} the relation between the face--on optical depth and the total gas surface density derived from a modified black body fit.
When comparing to Fig.~\ref{fig:metal-H}, we see that the results are very similar. The Spearman correlation coefficients for NGC 4254, NGC~4321, and NGC~5194 yield values close to what is found when combining CO and HI observations. However, in the case of NGC~2403 there is a significant decrease from $\rho=0.72$ to $\rho=0.45$. This could be due to the lower metallicity of this galaxy ($\left<12+\log O/H\right>=8.90$) compared to the rest of the sample ($\left<12+\log O/H\right>=9.20$). As such there could be a significant quantity of ``dark'' molecular gas not traced by CO. Quantitatively, the relations we derive are close to Eq.~\ref{eqn:tau-sample-A-XCO-variable} and \ref{eqn:tau-sample-A-XCO-variable-metal}:
\begin{equation}
 \tau_{FUV}=\left(2.297\pm0.059\right)+\left(0.092\pm0.004\right)\times\Sigma_H,
\end{equation}
and 
\begin{equation}
\begin{split}
 \tau_{FUV}=\left[\left(2.206\pm0.032\right)+\left(0.029\pm0.002\right)\times \Sigma_H\right]\\
\times10^{\left(1.048\pm0.025\right)\left(\left[12+\log O/H\right]-9.00\right)}.
\end{split}
\end{equation}
These relations yield similar optical depths. Averaging over the entire sample, we obtain $\left<\Delta\tau_{FUV}/\tau_{FUV}\right>=0.09\pm0.05$ when taking only the gas surface density into account, and $\left<\Delta\tau_{FUV}/\tau_{FUV}\right>=-0.02\pm0.06$ and when taking into account both the gas surface density and the metallicity. The amplitude of these shifts is not larger than the error bars on the optical depths deduced from the uncertainties on the FUV attenuation provided by CIGALE. Therefore the use of different methods to determine the gas surface density should only have a limited impact on the derived face--on optical depth, with offsets of the order of 10\%.

\section{Computation of the HI surface density\label{sec:computation-HI}}
In Sect.~\ref{ssec:NH} we provide a relation to compute the HI mass surface density. It is derived in the following way. First we consider this well known relation linking $N_{HI}$, the HI column density in atoms~cm$^{-2}$, to the brightness temperature $T_B$ in K:
\begin{equation}
 N_{HI}=1.823\times10^{15}\sum_iT_{B,i}\Delta V,
\end{equation}
with $\Delta V$ in m~s$^{-1}$. Considering that 1~M$_\odot$~pc$^{-2}=1.248\times10^{20}$~atoms~cm$^{-2}$, we have:
\begin{equation}
\Sigma_{HI}=1.461\times10^{-5}\sum_iT_{B,i}\Delta V.\label{eq:sigmaH-TB}
\end{equation}
The brightness temperature and the flux density per beam are related through the following equation:
\begin{equation}
S=\frac{2kT_B}{\lambda^2}\iint_{mb}Pd\Omega,
\end{equation}
with $P$ the angular power pattern of the beam, $d\Omega$ the solid angle, and $mb$ denoting the main beam. For a gaussian beam with a full width half maximum size $\Delta\alpha$ and $\Delta\delta$ along the major and minor axes, it easily comes that
\begin{equation}
 \iint_{mb}Pd\Omega=\frac{\pi}{4\ln2}\Delta\alpha\Delta\delta.
\end{equation}
Combining the 2 previous equations we get:
\begin{equation}
T_B=\frac{S\lambda^2\times2\ln2}{k\pi\Delta\alpha\Delta\delta}.
\end{equation}
Inserting this relation into equation \ref{eq:sigmaH-TB}:
\begin{equation}
 \Sigma_{HI}=\frac{\lambda^2\times1.461\times10^{-5}\times2\ln2}{k\pi\Delta\alpha\Delta\delta}\sum_iS_i\Delta V.
\end{equation}
Assuming that $\Delta\alpha$ and $\Delta\delta$ are expressed in arcsec, $S_i$ in Jy per beam, and $\Delta V$ in m~s$^{-1}$, a simple numerical application yields:
\begin{equation}
 \Sigma_{HI}=\frac{8.85}{\Delta\alpha\Delta\delta}\sum_iS_i\Delta V.
\end{equation}

\section{Impact of different CO transitions\label{sec:impact-CO-transitions}}
As explained earlier, to trace the molecular gas this study relies on the CO(2--1) transition. Yet different transitions may yield different results. Indeed the intensity of the emission of the various CO transitions not only depends on the abundance of CO but also on the local physical conditions of the molecular gas, such as the density or the temperature. For a given quantity of CO, a high level transition will trace more closely the denser and warmer part of the gas. Disentangling between these effects is particularly challenging and would require a dedicated study of the sample with appropriate data. Here we therefore do not attempt to separate these parameters. We rather examine whether and how using different transitions would affect our results.

\begin{figure}[!htbp]
 \centering
 \includegraphics[width=\columnwidth]{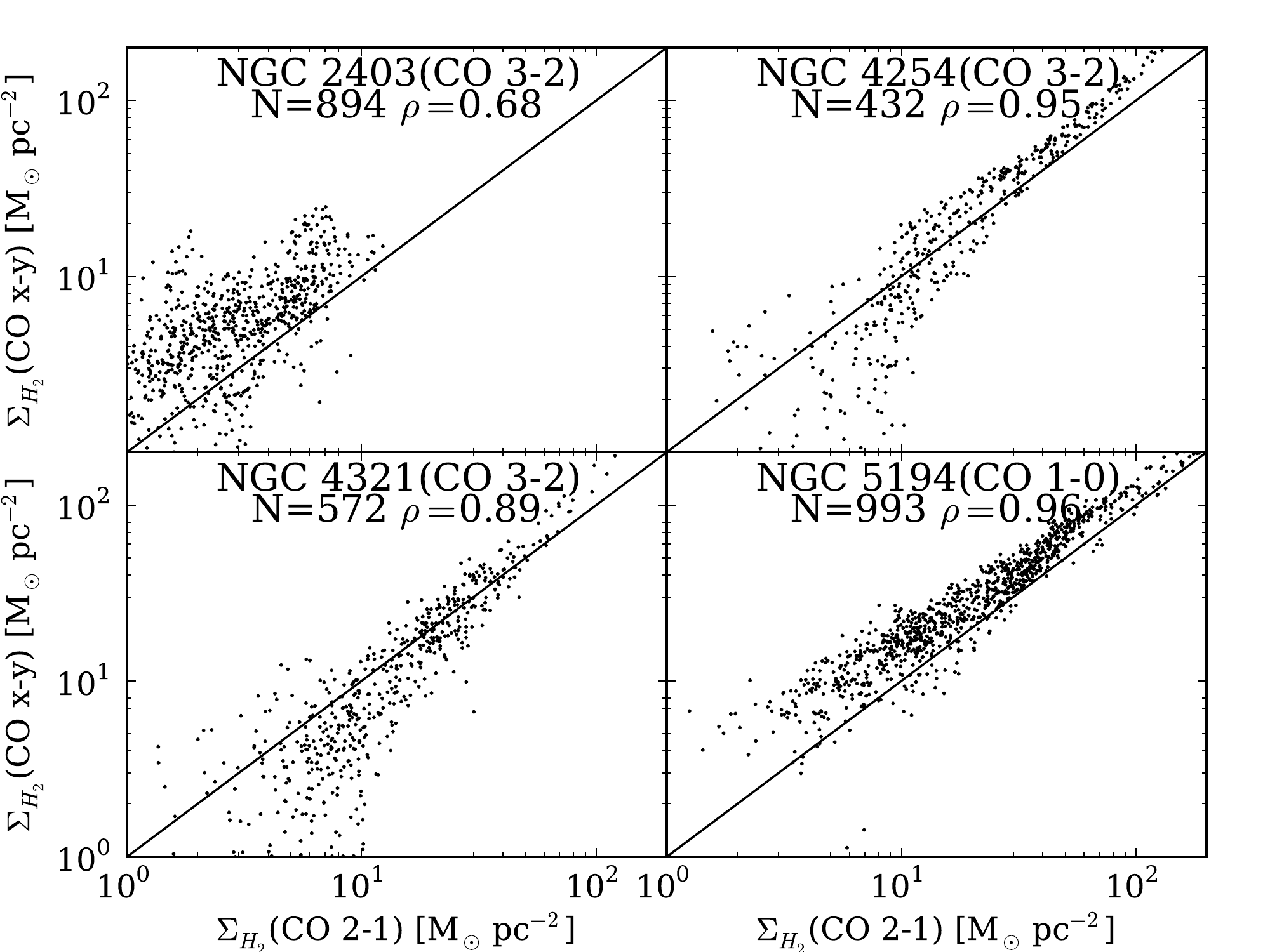}
 \caption{Molecular gas surface densities computed from CO(3--2) observations (CO(1--0) for NGC~5194) versus the molecular gas surface densities computed from the CO(2--1) transition. All CO lines have been scaled to that of the CO(1--0) in order to apply the same $X_{CO}$ factor. The black line represent the 1--to--1 relation where both lines yield the same molecular gas surface densities.\label{fig:comp-CO}}
\end{figure}

\begin{figure}[!htbp]
 \centering
 \includegraphics[width=\columnwidth]{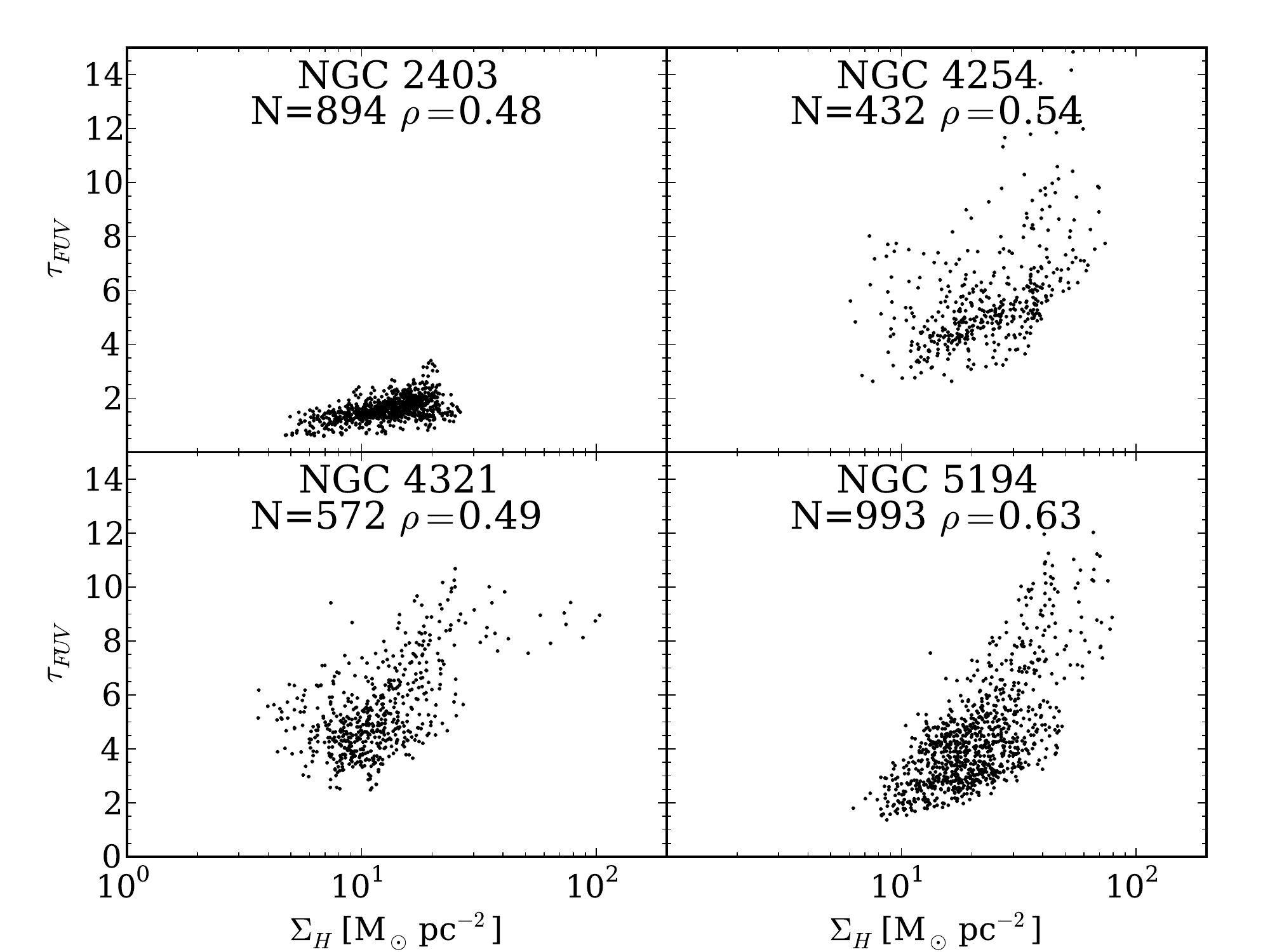}
 \caption{Same as for Fig.~\ref{fig:AFUV-H} with the molecular gas surface density determined from the CO(3--2) line (CO(1--0) for NGC~5194).\label{fig:NH-altCO}}
\end{figure}

To complement our CO(2--1) data, we need additional transitions at a resolution that is similar or better than 30\arcsec. This is a particularly limiting factor because CO(1--0) observations have a resolution twice as coarse as that of CO(2--1) for a given instrument. For higher order transitions, such as CO(3--2), observations are difficult to carry out as they require optimal atmospheric conditions. Nevertheless, 3 galaxies in our sample have adequate CO(3--2) maps to further our study. As part of the JCMT Nearby Galaxies Legacy Survey \citep{wilson2009b}, \cite{bendo2010b} published a CO(3--2) map of NGC~2403, and \cite{wilson2009b} maps of NGC~4254 and NGC~4321. For NGC~5194, \cite{koda2011a} have published a high resolution single--dish CO(1--0) map which we adopt. Additional CO(1--0) observations of several galaxies in the sample have been published by \cite{kuno2007a}. Unfortunately, their 15\arcsec\ beam is not Nyquist--sampled which makes the maps not suitable for a pixel--by--pixel analysis. In Fig.~\ref{fig:comp-CO} we present the molecular gas surface density derived from the baseline CO(2--1) observations and the CO(3--2) ones (CO(1--0) for NGC~5194). Note that as we divided our CO(2--1) intensities by a factor 0.89 to scale them to the CO(1--0) line. We also scale the CO(3--2) observations to CO(1--0) by dividing the intensities by a factor 0.34 \citep{wilson2009b}. 
There is visibly a strong correlation between CO(2--1) and other CO lines, with the Spearman correlation coefficient ranging from $\rho=0.68$ for NGC~2403 to $\rho=0.96$ for NGC~5194. In Fig.~\ref{fig:NH-altCO} we plot the relation between the face--on optical depth and the gas mass surface density.
We see that qualitatively the use of different CO transitions does not affect the global trends between these quantities and that the qualitative relations remain unchanged. We note however that there are much fewer datapoints for NGC~5194, which makes the flattening at low surface densities less evident. After inspection of the images, these are regions on the outskirts of the galaxy which are not in the field--of--view of the CO(1--0) map and have therefore been eliminated from the analysis. We should note that as commented by \cite{wilson2009b,bendo2010b}, there is a significant scatter in the ratio between the CO(3--2) line and lower CO transitions. It has been suggested that the CO(3--2) line are more closely associated with molecular gas fueling star formation than lower order lines, which may contribute to the pixel--to--pixel scatter.

Quantitatively, the relations we derive are close to Eq.~\ref{eqn:tau-sample-A-XCO-variable} and \ref{eqn:tau-sample-A-XCO-variable-metal}:
\begin{equation}
 \tau_{FUV}=\left(2.259\pm0.063\right)+\left(0.099\pm0.004\right)\times\Sigma_H,
\end{equation}
and 
\begin{equation}
\begin{split}
 \tau_{FUV}=\left[\left(1.997\pm0.033\right)+\left(0.033\pm0.002\right)\times \Sigma_H\right]\\
\times10^{\left(1.041\pm0.024\right)\left(\left[12+\log O/H\right]-9.00\right)}.
\end{split}
\end{equation}
These relations yield similar optical depths. Averaging over the entire sample, we obtain $\left<\Delta\tau_{FUV}/\tau_{FUV}\right>=0.08\pm0.05$ when taking only the gas surface density into account, and $\left<\Delta\tau_{FUV}/\tau_{FUV}\right>=0.05\pm0.04$ and when taking into account both the gas surface density and the metallicity. The amplitude of these shifts is not larger than the error bars on the optical depths deduced from the uncertainties on the FUV attenuation provided by CIGALE. Therefore the use of different CO transitions to determine the gas surface density should only have a limited impact on the derived face--on optical depth, with offsets of the order of 10\%.

\section{Relation with the molecular fraction\label{sec:molfrac}}
\begin{figure}[!htbp]
\centering
\includegraphics[width=\columnwidth]{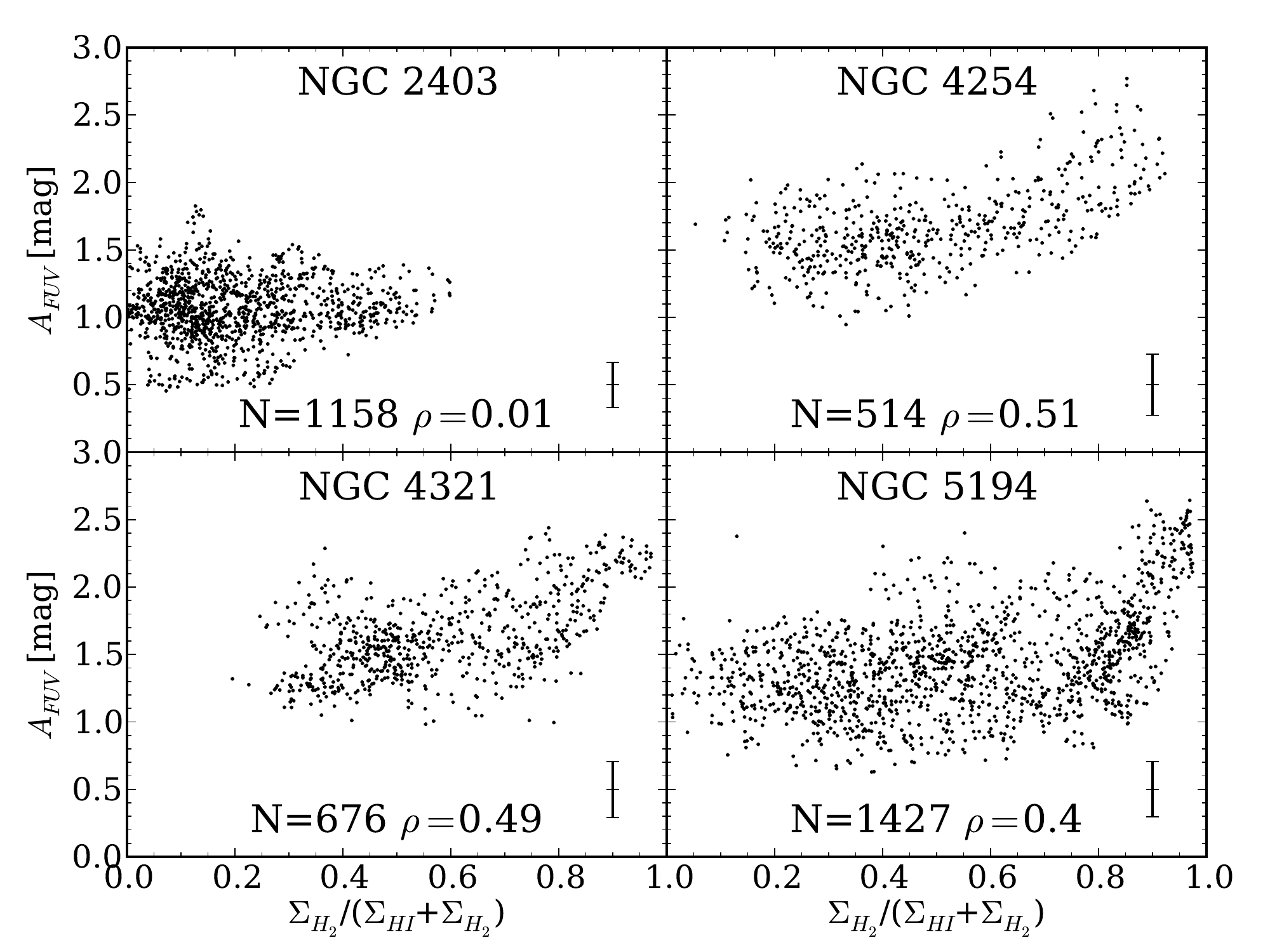}
\caption{Relation between the molecular gas fraction and the attenuation. There is a clear trend, with increasing molecular fraction the attenuation increases, with the exception of NGC~2403 which shows no trend. This is consistent with the trends observed in Fig.~\ref{fig:AFUV-H}. The number of elements and the Spearman correlation coefficient are indicated at the bottom of each panel whereas the uncertainty on $A_{FUV}$ is shown in the bottom right corner.}
\label{fig:molfrac}
\end{figure}

Whether the attenuation chiefly takes place in HI or H$_2$ dominated regions gives us an interesting insight into radiation transfer in galaxies. In Fig.~\ref{fig:molfrac} we show the FUV attenuation versus the molecular fraction.
The regions that have a low molecular fraction generally correspond to peripheral regions. In the case of NGC~2403 we do not see any trend of the attenuation with the molecular fraction. Conversely the other 3 galaxies show 2 regimes. Up to a molecular fraction of 0.6--0.8, there is little to no trend, the attenuation remaining constant. Beyond this limit, the attenuation strongly increases with the molecular fraction. This suggests that where the molecular fraction is high, the attenuation is strongly linked to the quantity of molecular gas. At lower molecular fraction it is possible that the atomic gas plays a larger role in the attenuation. However, there could also be mixing effects if the FUV attenuation is driven by regions more closely linked to molecular gas even in low density regions. We will examine this possibility in more detail in Sect.~\ref{ssec:sigma-optdepth}.

\section{Impact of the geometry\label{sec:impact-geo}}
\subsection{Attenuation--optical depth relations}

For this study, we have adopted a slab geometry. However, as mentioned in Sect.~\ref{ssec:choice-geom}, other geometries can also be considered. First of all, we examine the case of a sandwich model. Following \cite{boselli2003a}, the relation between the face--on optical depth $\tau_\lambda$ and the attenuation is:
\begin{dmath}
A_\lambda=-2.5\log\left(\left[\frac{1-\zeta_{\lambda}}{2}\right]\left(1+\exp\left(\sqrt{1-\omega_{\lambda}}\tau_{\lambda}/\cos i\right)\right)
+\left[\frac{\zeta_{\lambda}}{\sqrt{1-\omega_{\lambda}}\tau_{\lambda}/\cos i}\right]\left(1-\exp\left(\sqrt{1-\omega_{\lambda}}\tau_{\lambda}/\cos i\right)\right)\right),\label{eqn:sandwich}
\end{dmath}
with $\zeta$ the height ratio between the dust and stars layers. \cite{boselli2003a} provide the following calibration based on observations:

\begin{equation}
 \zeta_\lambda=1.0867-5.501\times10^5\lambda,
\end{equation}
with $\lambda$ the wavelength in m. For the FUV band at 151.6~nm, $\zeta\simeq1.0033$. When $\zeta=1$, Eq.~\ref{eqn:sandwich} is identical to the slab case (Eq.~\ref{eqn:slab}). As we limit the computation of the face--on optical depth to the FUV band (see Sect.~\ref{sec:recipe} for the extension to other wavelengths), we will not discuss the sandwich geometry further.

We also consider the case where dust is concentrated into clumps that are distributed following a Poisson distribution. In this case the relation between the attenuation and the optical depth is:

\begin{equation}
 A_\lambda=-2.5\log\left(\exp\left[-\bar{n}\left(1-e^{-\sqrt{1-\omega_{\lambda}}\tau_{\lambda_c}}\right)\right]\right),\label{eqn:clumpy}
\end{equation}
with $\bar{n}$ the average number of clouds along the line of sight, and $\tau_{\lambda_c}$ the optical depth of an individual cloud.

\begin{figure}[!htbp]
\centering
\includegraphics[width=\columnwidth]{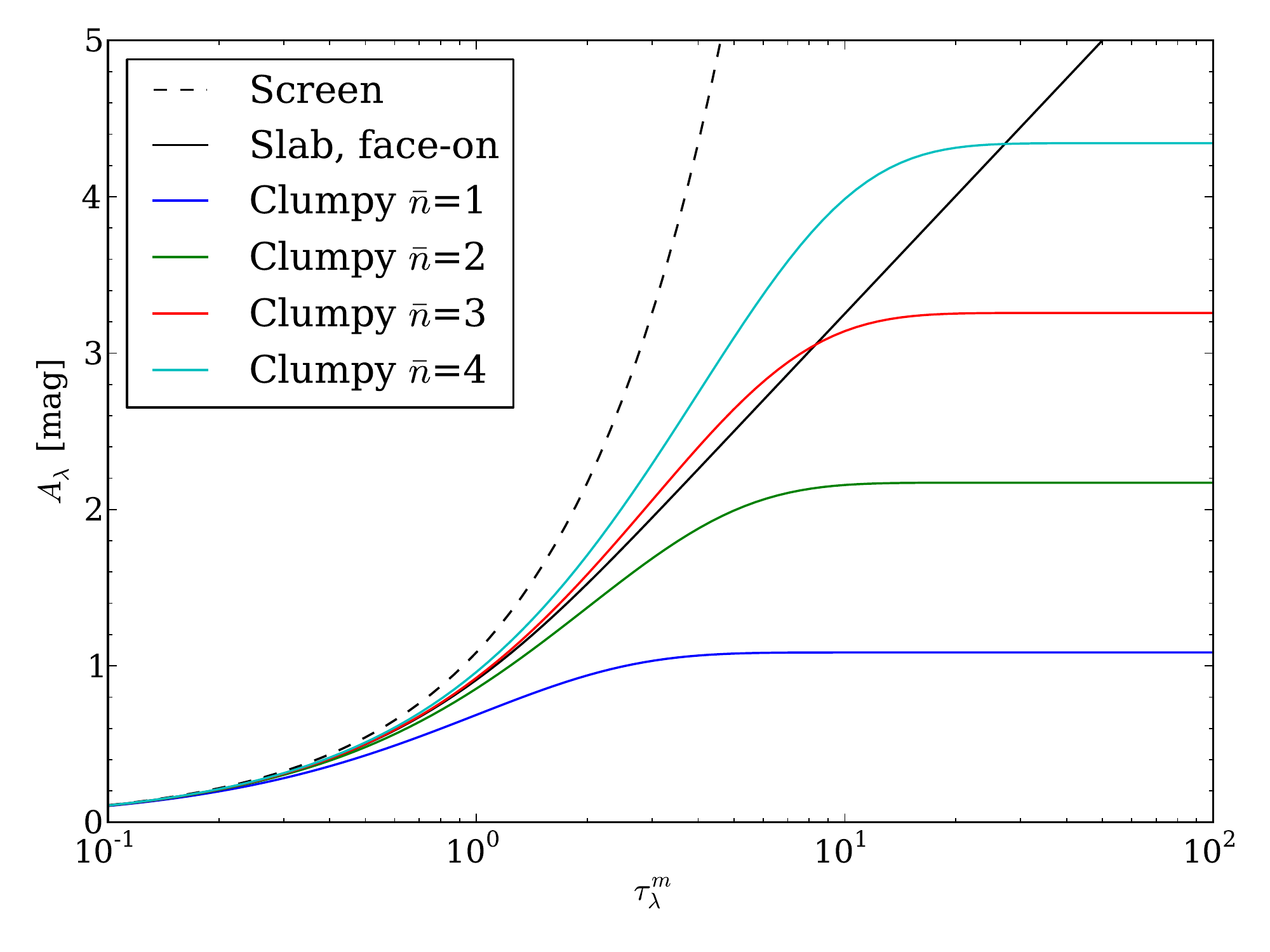}
\caption{Attenuation versus the mean optical depth $\tau_\lambda^m$ at a given wavelength $\lambda$, for a face--on galaxy with a slab geometry (solid black line), and a clumpy geometry assuming a mean number of clouds on the line of sight of 1 (blue), 2 (green), 3 (red), or 4 (cyan). A simple screen geometry is also shown for reference (dashed black line). For simplicity we have not taken into account the effect of the albedo in this plot, that is we have assumed that $\omega_\lambda=0$.}
\label{fig:comp-geo}
\end{figure}

In Fig.~\ref{fig:comp-geo}, we compare the effective attenuation for various geometries, versus the mean optical depth $\tau_\lambda^m$.
At low mean optical depths, all geometries yield the same attenuation. This means that the actual distribution of the dust, slab, clumpy, or even screen has little impact on the effective attenuation. However from an optical depth of unity, there is a strong divergence. For a slab geometry (solid black line in Fig.~\ref{fig:comp-geo}), $A_\lambda\propto\log\tau_\lambda^m$ whereas in the case of a clumpy distribution, $A_\lambda\propto\bar{n}$, when $\tau_\lambda^m\rightarrow+\infty$. As a consequence in the latter case the resulting attenuation curve is more and more grey as the optical depth increases. As we see, in the case of a distribution of very opaque clumps, the maximum attenuation is directly determined by the number of clumps and not by the optical depth of individual clouds.

\subsection{Gas surface density and the optical depth for different geometries\label{ssec:sigma-optdepth-appendix}}

We compute the relation between the gas surface density and the face--on optical depth for screen, and clumpy geometries to understand the impact of the assumed geometries on these relations.

If we assume that all galaxies in the sample have the same average number of clumps along the line of sight, we need $\bar{n}\ge3$ to be able to reproduce the observations. This is slightly larger than what was found by \cite{holwerda2007b} for a sample of 14 disk galaxies in the nearby Universe ($\bar{n}=2.6$). Considering $\bar{n}=3$ and $\bar{n}=4$, we find the following relations between the optical depth and the gas surface density:

\begin{equation}
 \tau_{FUV}=\left(1.455\pm0.022\right)+\left(0.065\pm0.002\right)\times \Sigma_H,\label{eqn:tau-sample-A-clump-3}
\end{equation}
and
\begin{equation}
\begin{split}
 \tau_{FUV}=\left[\left(1.361\pm0.017\right)+\left(0.038\pm0.001\right)\times \Sigma_H\right]\\
\times10^{\left(0.448\pm0.014\right)\left(\left[12+\log O/H\right]-9.00\right)}.
\end{split}
\end{equation}
for $\bar{n}=3$, and
\begin{equation}
 \tau_{FUV}=\left(1.464\pm0.018\right)+\left(0.052\pm0.001\right)\times \Sigma_H.\label{eqn:tau-sample-A-clump-4}
\end{equation}
and
\begin{equation}
\begin{split}
 \tau_{FUV}=\left[\left(1.364\pm0.014\right)+\left(0.029\pm0.001\right)\times \Sigma_H\right]\\
\times10^{\left(0.427\pm0.013\right)\left(\left[12+\log O/H\right]-9.00\right)}.
\end{split}
\end{equation}
for $\bar{n}=4$.

When we compare to Eq.~\ref{eqn:tau-sample-A-clump-3} or Eq.~\ref{eqn:tau-sample-A-clump-4}, Eq.~\ref{eqn:tau-sample-A-XCO-variable} yields very different optical depths. For instance, if we assume $\Sigma_H=10$~M$_\odot$~pc$^{-2}$, $\tau_{FUV}=3.426$ for a slab geometry and $\tau_{FUV}=1.988$ for a clumpy geometry with $\bar{n}=4$, reaching $\tau_{FUV}=15.715$ and $\tau_{FUV}=6.704$ respectively for $\Sigma_H=100$~M$_\odot$~pc$^{-2}$. This exemplifies strong the model--dependence of such relations. In order to avoid biases, it is therefore particularly important to use the same geometry when computing the attenuation as the one that was assumed for the relation between the gas surface density and the optical depth.

\section{Inversion of the relation between the attenuation and the face--on optical depth for a slab geometry\label{sec:inv-slab}}
In Sect.~\ref{ssec:choice-geom} we introduced the Lambert W function to express the face--on optical depth as a function of the attenuation (Eq.~\ref{eqn:slab-inv}). We indicate here how we have derived this equation and how the Lambert W function has been introduced. First of all, let's consider the equation we want to invert:
\begin{equation}
 A_\lambda=-2.5\log\left(\frac{1-\exp\left(-\sqrt{1-\omega_{\lambda}}\tau_{\lambda}/\cos i\right)}{\sqrt{1-\omega_{\lambda}}\tau_{\lambda}/\cos i}\right).
\end{equation}
Let $\tau'=\sqrt{1-\omega_\lambda}\tau_\lambda/\cos i$. We easily get:
\begin{equation}
 e^{-\tau'}=1-10^{-0.4A_\lambda}\tau',
\end{equation}
\begin{equation}
 -10^{0.4A_\lambda}\times e^{-\tau'}=\tau'-10^{0.4A_\lambda}
\end{equation}
\begin{equation}
 -10^{0.4A_\lambda}\times e^{-\tau'+10^{0.4A_\lambda}}\times e^{-10^{0.4A_\lambda}}=\tau'-10^{0.4A_\lambda}
\end{equation}
\begin{equation}
 -10^{0.4A_\lambda}\times e^{-10^{0.4A_\lambda}}=\left(\tau'-10^{0.4A_\lambda}\right)\times e^{\tau'-10^{0.4A_\lambda}}.
\end{equation}
Using the $W$ Lambert function, which gives the invert relation of $f\left(w\right)=we^w$, this is equivalent to:
\begin{equation}
 \tau'-10^{0.4A_\lambda}=W\left(-10^{0.4A_\lambda}\times e^{-10^{0.4A_\lambda}}\right).
\end{equation}
Redeveloping $\tau'$ and rearranging slightly we obtain the final relation:
\begin{equation}
 \tau_\lambda=\frac{\cos i}{\sqrt{1-\omega_\lambda}}\left[W\left(-10^{0.4A_\lambda}\times e^{-10^{0.4A_\lambda}}\right)+10^{0.4A_\lambda}\right].
\end{equation}

\section{On the reproducibility of local LFs\label{sec:reproduce-LF}}

As stated before, the aim of this exercise is not to reproduce exactly the LFs of nearby galaxies but to examine the impact of different gas surface density attenuation relations. Indeed, several intrinsic aspects of the simulations tend to incur biases affecting the LFs.

\begin{figure}[!htbp]
\centering
\includegraphics[width=\columnwidth]{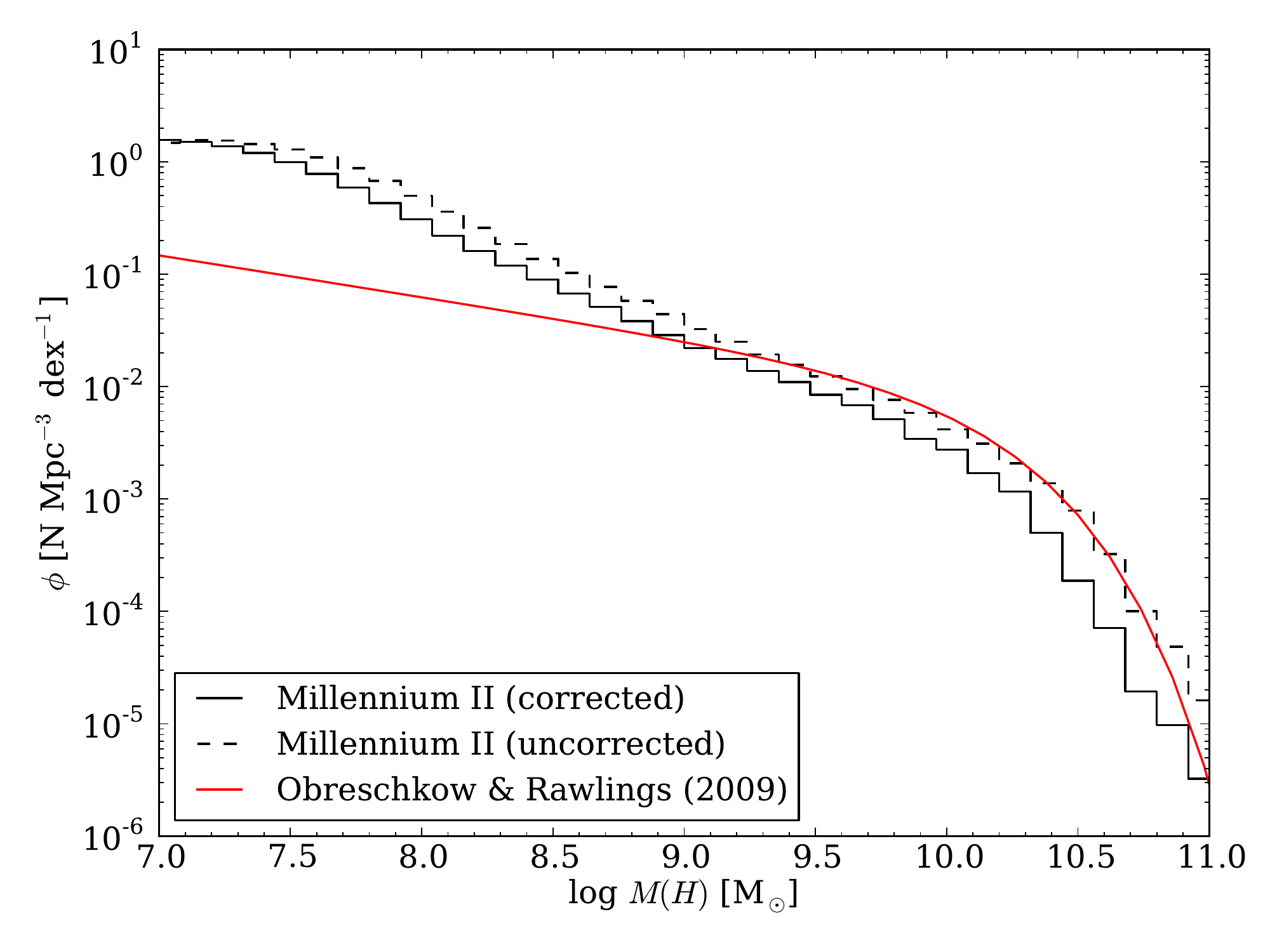}
\caption{Mass function of the cold gas mass of the Millennium--II simulation (black lines), uncorrected (dashed line), and corrected (solid) for the mass excess. For comparison the observed cold gas mass function derived by \cite{obreschkow2009a} is shown in red. All functions have been rescaled to $h=0.73$, following the Millennium--II simulation parameters \citep{boylan2009a}.}
\label{fig:Hfunc}
\end{figure}

A first well known problem is that semi--analytic relations can yield too much gas compared to what is observed in the local universe. To correct for this effect, following \cite{obreschkow2009b}, we rescale the gas mass in each galaxy so that the total cold gas mass density at $z=0$ is consistent with the value derived by \cite{obreschkow2009a}: $\Omega_{gas}^{obs}=6.0\times10^{-4}$, with $\Omega_{gas}^{obs}$ normalised to the critical density for closure:
\begin{equation}
 \rho_{crit}=\frac{3H^2}{8\pi G},
\end{equation}
$H$ being the Hubble constant, and $G$ the gravitational constant. For $H=73$~km~s$^{-1}$~Mpc$^{-1}$, $\rho_{crit}=1.0\times10^{-26}$~kg~m$^{-3}$. We then rescale the cold gas masses from the simulation by multiplying them with the factor $\xi^{-1}$, with $\xi$ defined as:
\begin{equation}
 \xi=\frac{\rho_{gas}^{simu}}{\Omega_{gas}^{obs}\rho_{crit}},
\end{equation}
with $\rho_{gas}^{simu}$ the mean gas mass density of the simulation over the entire volume. We find that $\rho_{gas}^{simu}=9.9\times10^{-30}$~kg~m$^{-3}$, leading to $\xi=1.5$. This is very similar to the factor 1.45 found by \cite{obreschkow2009b} for the \cite{delucia2007a} catalogue which was based on the Millennium simulation. In Fig.~\ref{fig:Hfunc}, we show the Millennium--II cold gas mass luminosity. 
At high mass, the model slightly underestimates the cold gas mass function compared to the observed on by \cite{obreschkow2009a}. Conversely there is a very strong excess of galaxies with a lower mass of gas, up to about 1 dex for $\log M(H)=7.0$.

\begin{figure}[!htbp]
\centering
\includegraphics[width=\columnwidth]{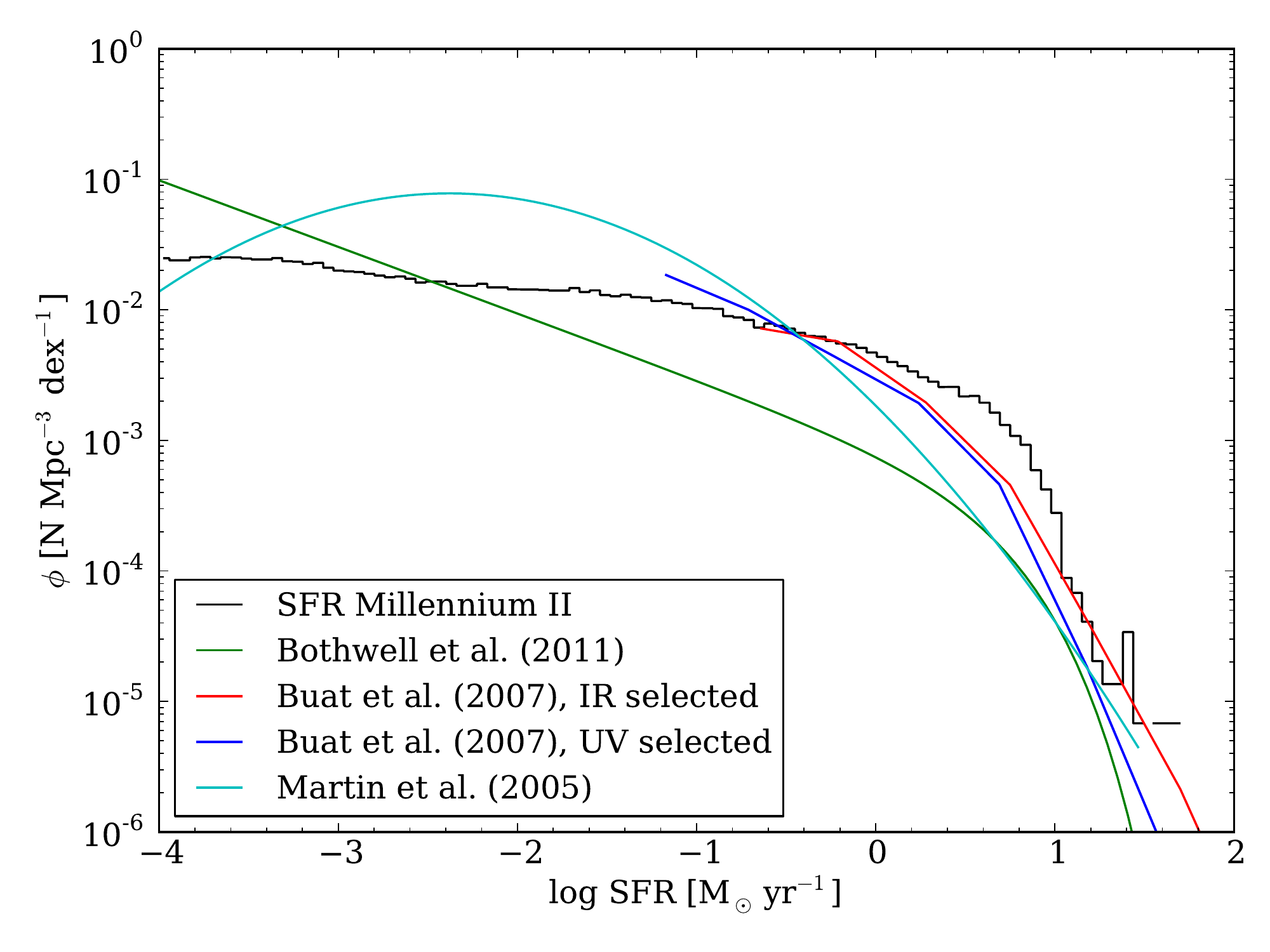}
\caption{SFR function of the Millennium--II simulation (black line). For comparison, the \cite{martin2005a} function (cyan), the IR (red) and UV (blue) selected \cite{buat2007a} functions, and the \cite{bothwell2011a} functions observed in the local universe have also been plotted. All functions have been scaled to $h=0.73$ and to a \cite{chabrier2003b} IMF.}
\label{fig:SFRfunc}
\end{figure}

The SFR distribution function, which is not affected by the attenuation, can also be compared to the observed ones in the local Universe. In Fig.~\ref{fig:SFRfunc}, we show the observed SFR distribution functions of \cite{martin2005a}, \cite{buat2007a}, and \cite{bothwell2011a}.
It turns out that the simulated SFR distribution function does not convincingly reproduce any of them. At the same time, they also have significant differences from another. The difference is the smallest at relatively high SFR, but the difference rapidly increases under SFR=10~M$_\odot$~yr$^{-1}$, at the level of the ``knee'' of the distribution. For a $\mathrm{SFR<1}$~M$_\odot$~yr$^{-1}$, the simulated SFR distribution function presents a much flatter slope than observed ones. At the same time, \cite{guo2011a} remarked that the Millennium--II simulation produces a large fraction of non--star forming dwarf galaxies compared to what is observed. This dearth of star--forming galaxies combined with the effect of the SFH -- dwarf galaxies tend to have a more bursty SFH than large spirals which form stars at a steady rate -- could explain such a flattening of the slope. At optical and near--infrared wavelengths which are increasingly less sensitive to the exact SFH, this problem has a much weaker impact. \cite{guo2011a} showed that the Millennium--II simulations can reproduce convincingly the SDSS LFs derived by \cite{blanton2005a} in the $g'$, $r'$, $i'$, and $z'$ bands. However, in the star--formation tracing bands, these departures from the observed SFR distribution functions indicate that because we derive the extinction--free UV luminosity of each galaxy from the SFR as we saw in Sect.~\ref{ssec:simu-params}, we cannot compare directly our simulated UV and IR LFs with the observed ones. This is the reason why we do not attempt to carry out a comparison between our LFs and the observed ones.
\end{document}